\documentclass[10pt, conference, a4paper]{IEEEtran}
\usepackage{etoolbox}
\usepackage{cite}
\usepackage{amsmath,amssymb,amsfonts}
\usepackage{algorithm,algorithmic}
\usepackage{graphicx}
\usepackage{textcomp}
\usepackage{xcolor}
\usepackage{enumerate,enumitem}
\usepackage{url}
\usepackage{color}
\usepackage{bm}
\usepackage{subfigure}

\DeclareMathOperator*{\argmax}{\arg\!\max}
\DeclareMathOperator*{\argmin}{\arg\!\min}
\newtheorem{Thm}{Theorem}
\newtheorem{Lem}{Lemma}

\newtheorem{Prob}{Problem}
\newtheorem{Rem}{Remark}

\definecolor{green_wq}{rgb}{0.47,0.67,0.19}
\definecolor{purple_wq}{rgb}{0.49,0.18,0.56}
\newcommand{\eqa}{\overset{(a)}{=}}
\newcommand{\eqb}{\overset{(b)}{=}}
\newcommand{\eqc}{\overset{(c)}{=}}
\newcommand{\eqd}{\overset{(d)}{=}}
\newcommand{\eqe}{\overset{(e)}{=}}
\newcommand{\eqf}{\overset{(f)}{=}}
\newcommand{\eqg}{\overset{(g)}{=}}
\newcommand{\eqh}{\overset{(h)}{=}}
\newcommand{\eqi}{\overset{(i)}{=}}

\newcommand{\lea}{\overset{(a)}{\le}}

\newcommand{\lec}{\overset{(c)}{\le}}
\newcommand{\led}{\overset{(d)}{\le}}

\newcommand{\lef}{\overset{(f)}{\le}}

\newcommand{\gea}{\overset{(a)}{\ge}}
\newcommand{\geb}{\overset{(b)}{\ge}}
\newcommand{\gec}{\overset{(c)}{\ge}}

\newcommand{\gee}{\overset{(e)}{\ge}}

\newcommand{\llc}{\overset{(c)}{<}}

\newcommand{\gld}{\overset{(d)}{>}}

\newcommand{\Jwqm}{\textcolor{black}}

\newcommand{\TCOMr}{\textcolor{black}}
\newcommand{\TCOMb}{\textcolor{black}}
\newcommand{\TCOMc}{\textcolor{black}}

\newcommand{\TCOMbr}{\textcolor{black}}
\newcommand{\TCOMg}{\textcolor{black}}

\newcommand{\DAIm}{\textcolor{black}}

\newcommand{\CUIb}{\textcolor{black}}

\newcommand{\REPLYb}{\textcolor{black}}

\newcommand{\SecREPLYb}{\textcolor{black}}
\newcommand{\SecREPLYc}{\textcolor{black}}

\allowdisplaybreaks[4]
\makeatletter
\renewcommand{\maketag@@@}[1]{\hbox{\m@th\normalsize\normalfont#1}}
\patchcmd{\@makecaption}
  {\scshape}
  {}
  {}
  {}
\makeatletter
\patchcmd{\@makecaption}
  {\\}
  {.\ }
  {}
  {}
\makeatother
\IEEEoverridecommandlockouts
\begin{document}

\title{Optimization-based Block Coordinate Gradient Coding for Mitigating Partial Stragglers in Distributed Learning}

\author{\IEEEauthorblockN{Qi Wang,\ Ying Cui,\ Chenglin Li,\ Junni Zou,\ Hongkai Xiong}\\
%\IEEEauthorblockA{\textit{Dept. of EE,\ Shanghai Jiao Tong University,\ China}}
\thanks{The authors are with Shanghai Jiao Tong University, China.
This paper was presented in part at IEEE GLOBECOM 2021~\cite{Qi_GLOBECOM}.}
}

\maketitle
%\vspace{-1cm}
\begin{abstract}
% ===================================================================
%\DAIm{Gradient-based methods are popular in distribution optimization for large-scale problems in machine learning. 
%For gradient-based distributed optimization, stragglers  significantly affect the overall computation efficiency.}
%This paper considers a distributed computation system consisting of one master and $N$ workers, characterized by a general partial straggler model and focuses on solving a general large-scale machine learning problem with $L$ model parameters using gradient coding. 
%Existing gradient coding schemes introduce identical redundancy in coded local partial derivatives corresponding to all $L$ coordinates and hence cannot effectively utilize the computing capabilities of partial stragglers \DAIm{since incomplete computation results from stragglers are wasted}.
% ===================================================================
Gradient coding schemes effectively mitigate full stragglers in distributed learning by introducing identical redundancy in coded local partial derivatives corresponding to all model parameters.
However, they are no longer effective for partial stragglers as they cannot utilize incomplete computation results from partial stragglers.
This paper aims to design a new gradient coding scheme for mitigating partial stragglers in distributed learning.
Specifically, we consider a distributed system
consisting of one master and $N$ workers, characterized by a general partial straggler model and focuses on solving a general large-scale machine learning problem with $L$ model parameters using gradient coding.
First, we propose a coordinate gradient coding scheme with $L$ coding parameters representing $L$ possibly different diversities for the $L$ coordinates, which generates most gradient coding schemes.
Then, we consider the minimization of the expected overall runtime and the maximization of the completion probability with respect to the $L$ coding parameters for coordinates, which are challenging discrete optimization problems.
To reduce computational complexity, we first transform each to an equivalent but much simpler discrete problem with $N \ll L$ variables representing the partition of the $L$ coordinates into $N$ blocks, each with identical redundancy.
This indicates an equivalent but more easily implemented block coordinate gradient coding scheme with $N$ coding parameters for blocks.
Then, we adopt continuous relaxation to further reduce computational complexity.
For the resulting minimization of expected overall runtime, we develop an iterative algorithm of computational complexity $\mathcal{O}(N^2)$ to obtain an optimal solution 
and derive two closed-form approximate solutions both with computational complexity $\mathcal{O}(N)$.
For the resultant maximization of the completion probability, we develop an iterative algorithm of computational complexity $\mathcal{O}(N^2)$ to obtain a stationary point and
derive a closed-form approximate solution with computational complexity $\mathcal{O}(N)$ at a large threshold.
Finally, numerical results show that the proposed solutions significantly outperform existing coded computation schemes and their extensions.

\end{abstract}

%\newpage

\section{Introduction}
Due to the explosion in the numbers of samples and features of modern datasets, it is generally impossible to train a model by solving a large-scale machine learning problem on a single node. This challenge naturally leads to distributed learning in a master-worker distributed computation system. 
\SecREPLYb{Recently, distributed learning (or federated learning) has been actively investigated \cite{FedAvg, Huang, xu2021learning, Ye_CC, Li_YC}.}
%In such a distributed computation system, 
Due to various factors such as insufficient power, contention of shared resources, imbalanced work allocation and network congestions \cite{dean2013tail, straggler_detection_1}, some processing nodes may be slower than others or even fail from time to time.
These nodes, referred to as \emph{stragglers}, can significantly affect the overall computation efficiency.
Generally speaking, there are two commonly used straggler models. One is the \emph{full (persistent) straggler} model where stragglers are unavailable permanently\cite{ mat_mat_Yu_polynomial, mat_mat_Dutta_PolyDot_Codes , GC_Tandon_exact, GC_approx_Tandon_approx, approx_GC_Bitar}. 
The other is the \emph{partial (non-persistent) straggler} model where stragglers are slow but can conduct a certain amount of work\cite{mat_vec_Lee_speeding_up, mat_vec_LT_Codes, mat_vec_Dutta_Shot_Dot_Codes, mat_vec_Gunduz_speeding_up,   mat_vec_Draper_hierarchical_ISIT, mat_mat_Draper_hierarchical_TIT, mat_mat_Lee_high_dimensional, mat_mat_Yu_polynomial, mat_mat_Draper_exploitation, GC_WH_improving_RS_Codes, GC_Min_Ye_communication, GC_SA_near_optimal, GC_Gunduz_clustering, GC_Gunduz_dynamic_clustering_J, approx_mat_vec_Draper_anytime, approx_mat_vec_Coded_Sequential}. 
The partial straggler model is more general than the full straggler model, as the former with each worker's computing time following a Bernoulli distribution degenerates to the latter.

Recently, several coded distributed computation techniques, including \emph{coded computation}~\cite{mat_vec_Lee_speeding_up, mat_vec_LT_Codes, mat_vec_Dutta_Shot_Dot_Codes, mat_vec_Gunduz_speeding_up,   mat_vec_Draper_hierarchical_ISIT, mat_mat_Lee_high_dimensional, mat_mat_Yu_polynomial, mat_mat_Dutta_PolyDot_Codes , mat_mat_Draper_exploitation,  mat_mat_Draper_hierarchical_TIT, GC_Tandon_exact, GC_approx_Tandon_approx, GC_WH_improving_RS_Codes, GC_Min_Ye_communication, GC_SA_near_optimal, GC_Gunduz_clustering, GC_Gunduz_dynamic_clustering_J, heterogeneous_new} and \emph{approximate coded computation}~\cite{approx_mat_vec_Draper_anytime, approx_mat_vec_Coded_Sequential, GC_approx_Tandon_approx, approx_GC_Bitar, approx_GC_Charles}, have been proposed to mitigate the effect of stragglers in training the model via gradient descent (or stochastic gradient descent) algorithms \cite{survey_ng}.\footnote{\REPLYb{Coded distributed computation techniques have also been applied in 
%other domains such as heterogeneous networks~\cite{heterogeneous}, 
secure and private computing~\cite{secure_and_private}, distributed optimization~\cite{approx_mat_vec_Coded_Sequential}, federated learning~\cite{federated}, blockchains~\cite{blockchain}, timely computing~\cite{timely_computing}, etc., which are not the focus of this paper.}} 
The common idea is to enable robust collaborative computation of a gradient (or stochastic gradient) for reducing the overall computation time in the presence of stragglers.
Note that approximate coded computation schemes achieve faster computation speeds with higher accuracy losses than coded computation schemes.
\DAIm{In both coded computation and approximate coded computation,}
matrix multiplications (called \emph{coded matrix multiplication}) \cite{mat_vec_Lee_speeding_up, mat_vec_LT_Codes, mat_vec_Dutta_Shot_Dot_Codes,  mat_vec_Gunduz_speeding_up,   mat_vec_Draper_hierarchical_ISIT, mat_mat_Lee_high_dimensional, mat_mat_Yu_polynomial, mat_mat_Dutta_PolyDot_Codes , mat_mat_Draper_exploitation,  mat_mat_Draper_hierarchical_TIT, approx_mat_vec_Draper_anytime, approx_mat_vec_Coded_Sequential, heterogeneous_new} and calculations of gradients in general forms (called \emph{gradient coding}) \cite{GC_Tandon_exact, GC_approx_Tandon_approx, GC_WH_improving_RS_Codes, GC_Min_Ye_communication, GC_SA_near_optimal, GC_Gunduz_clustering, GC_Gunduz_dynamic_clustering_J, approx_GC_Bitar, approx_GC_Charles} are investigated.
Specifically, in \cite{mat_mat_Yu_polynomial, mat_mat_Dutta_PolyDot_Codes, mat_vec_Lee_speeding_up, mat_vec_LT_Codes, mat_vec_Dutta_Shot_Dot_Codes, mat_mat_Lee_high_dimensional,  approx_mat_vec_Draper_anytime,approx_mat_vec_Coded_Sequential, heterogeneous_new}, coded submatrix-submatrix products corresponding to all data blocks have identical redundancy, and in \cite{GC_Tandon_exact, GC_approx_Tandon_approx, approx_GC_Bitar,GC_WH_improving_RS_Codes, GC_Min_Ye_communication, GC_SA_near_optimal}, coded local partial derivatives corresponding to all coordinates have identical redundancy.
Note that with identical redundancy, the computing capabilities of partial stragglers cannot be effectively utilized since incomplete computation results from stragglers are wasted.
To avoid wasting incomplete computation results, in \cite{mat_vec_Gunduz_speeding_up, mat_vec_Draper_hierarchical_ISIT,mat_mat_Draper_hierarchical_TIT,   mat_mat_Draper_exploitation,GC_Gunduz_clustering, GC_Gunduz_dynamic_clustering_J}, diverse redundancies are introduced in coded computation schemes, and incomplete computation results such as coded submatrix-submatrix products corresponding to some data blocks \cite{mat_mat_Yu_polynomial, mat_mat_Dutta_PolyDot_Codes, mat_vec_Lee_speeding_up, mat_vec_LT_Codes, mat_vec_Dutta_Shot_Dot_Codes, mat_mat_Lee_high_dimensional,  approx_mat_vec_Draper_anytime,approx_mat_vec_Coded_Sequential} and coded local partial derivatives corresponding to some coordinates \cite{GC_Tandon_exact, GC_approx_Tandon_approx, approx_GC_Bitar,GC_WH_improving_RS_Codes, GC_Min_Ye_communication, GC_SA_near_optimal} are utilized.
Note that in \cite{mat_vec_Gunduz_speeding_up,    mat_mat_Draper_exploitation,GC_Gunduz_clustering, GC_Gunduz_dynamic_clustering_J}, coding parameters determining the amount of redundancies are artificially fixed, which may limit the performances of coded computation schemes.
To address the limitation, in \cite{mat_vec_Draper_hierarchical_ISIT,mat_mat_Draper_hierarchical_TIT}, the optimization of coding parameters for coded matrix multiplication is formulated, and an efficient solution for an approximated problem is obtained.
%Nevertheless, the optimization of coding parameters for gradient coding, remains open.
Nevertheless, 
%\DAIm{since coding parameters represent different meanings in coded matrix multiplication and gradient coding, the optimization method of coding parameters for coded matrix multiplication cannot be readily applied to optimize parameters for gradient coding.}
the optimization of coding parameters for gradient coding, \CUIb{which is significantly different from that of coded matrix multiplication,} still remains open.

To shed some light, this paper investigates optimization-based gradient coding schemes, which are applicable to a broader range of applications. 
Specifically, we consider a distributed computation system consisting of one master and $N$ workers, characterized by a general partial straggler model with the computing times of workers independent and identically distributed (i.i.d.) according to an arbitrary distribution, and focus on solving a general large-scale machine learning problem with $L$ model parameters using gradient descent methods.
Our considered general partial straggler model includes the one under the shifted exponential distribution \cite{mat_vec_Lee_speeding_up, mat_vec_LT_Codes, mat_vec_Dutta_Shot_Dot_Codes, mat_vec_Gunduz_speeding_up,   mat_vec_Draper_hierarchical_ISIT, mat_mat_Draper_hierarchical_TIT, mat_mat_Lee_high_dimensional, mat_mat_Yu_polynomial, mat_mat_Draper_exploitation, GC_WH_improving_RS_Codes, GC_Min_Ye_communication, GC_SA_near_optimal, GC_Gunduz_clustering, GC_Gunduz_dynamic_clustering_J, approx_mat_vec_Draper_anytime, approx_mat_vec_Coded_Sequential} as special cases.
Besides, our results also apply to mini-batch stochastic gradient descent methods.
Our detailed contributions are summarized below.
\begin{itemize}
\item 
We propose a coordinate gradient coding scheme with $L$ coding parameters, one for each coordinate, to maximally diversify the redundancies in coded local partial derivatives corresponding to all $L$ coordinates.  
It is worth noting that it systematically generates \TCOMb{existing gradient coding schemes \cite{GC_Tandon_exact,GC_WH_improving_RS_Codes}} by allowing introduced redundancies for all coordinates to be different.
\item 
We formulate the minimization of the expected overall runtime with respect to the $L$ coding parameters for coordinates.
The problem is a challenging stochastic optimization problem with a large number of ($L$) discrete variables.
\TCOMb{First}, we transform the original problem with $L$ variables to an equivalent but much simpler problem with $N \ll L$ variables \TCOMb{representing $L$ possibly different diversities for the $L$ coordinates}, by characterizing the optimality properties. 
This indicates that we can optimally partition the $L$ coordinates into $N$ blocks, each with identical redundancy.
\TCOMb{Then, we} adopt continuous relaxation with a negligible approximation error at $N \ll L$ to further reduce computational complexity.
\TCOMb{Next}, we develop an iterative algorithm of computational complexity $\mathcal{O}(N^2)$ to obtain an optimal solution of the relaxed \TCOMb{convex} stochastic problem using the stochastic projected subgradient method.
We also obtain two closed-form approximate solutions of the relaxed \TCOMb{convex} stochastic problem with computational complexity $\mathcal{O}(N)$ by solving its two deterministic approximations.
Furthermore, we show that the expected overall runtimes of the two low-complexity approximate solutions and the minimum overall runtime have sub-linear multiplicative gaps in $N$.
\item 
We formulate the maximization of the completion probability with respect to the $L$ coding parameters for coordinates.
The problem is a challenging (deterministic) optimization problem \TCOMb{with a large number of ($L$) discrete variables and} a large number of ($\Omega(2^N)$) summands \TCOMb{in the objective function}.
Similarly, we transform the original problem to an equivalent but much simpler problem with $N \ll L$ variables  \TCOMb{representing $L$ possibly different diversities for the $L$ coordinates} and focus on solving the continuous relaxation of the equivalent problem with a negligible approximation error at $N \ll L$.
%\TCOMb{This indicates that we can optimally partition the $L$ coordinates into $N$ blocks, each with identical redundancy.}
Then, we develop an iterative algorithm of computational complexity $\mathcal{O}(N^2)$ to obtain a stationary point of the relaxed \TCOMb{non-convex} problem using the stochastic successive convex approximation (SSCA) method.
Besides, we obtain a closed-form approximate solution of the relaxed problem \TCOMb{with computational complexity $\mathcal{O}(N)$} at a large threshold.
% which has computational complexity $\mathcal{O}(N)$.
\item 
Numerical results show that the proposed solutions significantly outperform the optimized version of the gradient coding scheme in \cite{GC_Tandon_exact} and two extensions of the coded matrix multiplication scheme in \cite{mat_vec_Draper_hierarchical_ISIT}.
Numerical results also show the impacts of the system parameters on the performances \TCOMb{of} the proposed solutions and the close-to-optimal performances of the approximate solutions.
\end{itemize}

\TCOMb{To the best of our knowledge, this is the first work that optimizes the redundancies for gradient coding to effectively utilize the computing capabilities of partial stragglers.}
\DAIm{The preliminary version of this paper appeared as \cite{Qi_GLOBECOM}. 
In this paper, we additionally consider the minimization of the completion probability consisting of theoretical analysis and numerical results.}

\noindent\emph{\textbf{Notation}}

Throughout this paper, 
$\mathbb{R}$   denotes the set of real numbers,
$\mathbb{R}_+$ denotes the set of positive real numbers,
$\mathbb{N}$   denotes the set of natural numbers, 
$\mathbb{N}_+$ denotes the set of positive integers,
and $[N]$ denotes set $\{1,\cdots,N\}$, for any $N\in\mathbb{N}$.
We also use calligraphic capitalized letters, e.g., $\mathcal{K}$, to denote sets.  
$|\mathcal{K}|$ denotes the cardinality of set $\mathcal{K}$.
For random quantities, we use upper case italic letters, e.g., $T$, for scalars, upper case non-italic bold letters, e.g., $\mathbf{T}$, for vectors.
For deterministic quantities, we use lower case italic letters, e.g.,  $t$, for scalars, lower case bold letters, e.g., $\mathbf{t}$, for vectors.
$x_i$ denotes the $i$-th coordinate of $\mathbf{x}$.
$I(\cdot)$ denotes the indicator function.
$\mathbf{1}_{m}$ denotes the $m \times 1$ identity vector and
$\mathbf{1}_{m \times n}$ denotes the $m \times n$ identity matrix.
$\binom{n}{k_0,k_1,\cdots,k_{r-1}} \triangleq \frac{n!}{k_0! k_1! \cdots k_{r-1}!}$ denotes the multinomial coefficient, where $\sum_{i=0}^{r-1} k_i = n$.

\section{System Setting}
\label{sec:System_Setting}

%\begin{figure}
%\begin{center}
%  {\resizebox{5cm}{!}{\includegraphics{master_worker.pdf}}}
%%  \vspace{-2mm}
%         \caption{Master-worker distributed computation system.
%         \vspace{-6mm}
%         }\label{fig:system_master_worker}
%\end{center}
%\end{figure}

%As illustrated in Fig.~\ref{fig:system_master_worker}, 
We consider a master-worker distributed computation system which consists of one master and $N$ workers all with  computation and communication capabilities~\cite{GC_Tandon_exact,GC_approx_Tandon_approx,approx_GC_Charles,Maity_approx,Wang_approx_1,approx_Wang_fundamental,approx_GC_Bitar,Kadhe_approx,GC_Min_Ye_communication,GC_WH_improving_RS_Codes,mat_vec_Lee_speeding_up,mat_mat_Lee_high_dimensional,mat_mat_Yu_polynomial,mat_mat_Draper_exploitation,mat_vec_Draper_hierarchical_ISIT,mat_mat_Draper_hierarchical_TIT, mat_vec_Gunduz_speeding_up,GC_Gunduz_clustering}.
Let $[N]\triangleq\{1,\cdots,N\}$ denote the set of worker indices.
%The master and workers have computation and communication capabilities.
We assume that the master and each worker are connected by a fast communication link, and hence we omit the communication time, as in \cite{mat_mat_Draper_exploitation,mat_vec_Draper_hierarchical_ISIT,mat_mat_Lee_high_dimensional,mat_mat_Draper_hierarchical_TIT}.
We consider a general partial straggler model for the workers.
\TCOMb{Specifically,}
at any instant, the CPU cycle times of the $N$ workers, denoted by $T_n,n\in [N]$, are i.i.d. random variables. 
%whose common distribution is the same as that of a random variable $T$.
The values of $T_n,n\in [N]$ at each instant are not known to the master, but the common distribution is known to the master.
Let $F_T(\cdot)$ and $f_T(\cdot)$ denote the cumulative distribution function (CDF) and probability density function (PDF) of $T_n,n\in[N]$, respectively.
\TCOMb{We shall see that}
most theoretical results in this paper do not require any assumption on the distribution of $T_n, n\in[N]$.
\begin{Rem}[General Partial Straggler Model]
	The adopted straggler model is more general than those in~\cite{GC_Tandon_exact,GC_approx_Tandon_approx,approx_GC_Charles,Maity_approx,approx_Wang_fundamental,approx_GC_Bitar,Kadhe_approx,mat_mat_Yu_polynomial}.                                                                                                                                    
	Specifically, when $T_n,n\in[N]$ follow a Bernoulli distribution, the adopted straggler model degenerates to the full straggler model in \cite{GC_Tandon_exact,approx_GC_Charles,Maity_approx,approx_Wang_fundamental,approx_GC_Bitar,Kadhe_approx,mat_mat_Yu_polynomial};
	when $T_n\in[T_{\rm lb},T_{\rm ub}], n\in[N]$, with $T_{\rm lb},T_{\rm ub}>0$ and $T_{\rm ub}=\alpha T_{\rm lb}$ for some constant $\alpha>1$, the adopted straggler model degenerates to the $\alpha$-partial straggler model in \cite{GC_approx_Tandon_approx}.
\end{Rem}

%Denote $\mathbf{T}\triangleq \left(T_n\right)_{n\in [N]}$.

%Let $T_{(1)},T_{(2)},\cdots,T_{(N)}$ be $T_{n},n\in[N]$ arranged in increasing order, so that $T_{(n)}$ is the $n$-th smallest one.
As in \cite{GC_Tandon_exact,GC_approx_Tandon_approx,approx_GC_Charles,Maity_approx,Wang_approx_1,approx_Wang_fundamental,approx_GC_Bitar,Kadhe_approx,GC_Min_Ye_communication,GC_WH_improving_RS_Codes,mat_vec_Lee_speeding_up,mat_mat_Lee_high_dimensional,mat_mat_Yu_polynomial,mat_mat_Draper_exploitation,mat_vec_Draper_hierarchical_ISIT,mat_mat_Draper_hierarchical_TIT, mat_vec_Gunduz_speeding_up,GC_Gunduz_clustering}, we focus on the following distributed computation scenario. 
The master holds a data set of $M$ samples, denoted by $\mathcal{D}$ $\triangleq \{\mathbf{y}_i : i\in[M]\}$, and aims to train a model. 
The model is parameterized by an $L$-dimensional vector $\bm{\theta}\in\mathbb{R}^L$.
Notice that the model size $L$ is usually much larger than the number of workers $N$.
For a given $\bm{\theta}\in\mathbb{R}^L$, define the loss incurred by $\mathbf{y}_i$ as $\ell(\bm{\theta};\mathbf{y}_i)$.
We assume that $\ell(\cdot)$ is \REPLYb{differentiable} but not necessarily convex.
Then, the risk function $\hat{\ell}:\mathbb{R}^{L}\rightarrow \mathbb{R}$ of the model parameter $\bm{\theta}\in\mathbb{R}^L$ is defined as
\begin{equation*}
	\hat{\ell}(\mathbf{\bm{\theta}};\mathcal{D}) \triangleq \sum_{\mathbf{y}\in\mathcal{D}} \ell(\bm{\theta};\mathbf{y}).
\end{equation*}
The master aims to minimize the risk function with respect to the model parameter $\bm{\theta}$ using commonly used gradient descent methods.\footnote{We present the results based on gradient descent methods for ease of exposition. Note that our results also apply to mini-batch stochastic gradient descent methods~\cite{GC_Tandon_exact,GC_approx_Tandon_approx,approx_GC_Charles,Maity_approx,Wang_approx_1,approx_Wang_fundamental,approx_GC_Bitar,Kadhe_approx,GC_Min_Ye_communication,GC_WH_improving_RS_Codes,mat_vec_Lee_speeding_up,mat_mat_Lee_high_dimensional,mat_mat_Yu_polynomial,mat_mat_Draper_exploitation,mat_vec_Draper_hierarchical_ISIT,mat_mat_Draper_hierarchical_TIT, mat_vec_Gunduz_speeding_up,GC_Gunduz_clustering}, where the master randomly selects a mini-batch of samples and notifies the workers the index of the mini-batch.}

To handle a massive amount of training data, 
%Due to the explosion in the numbers of samples and features of modern datasets, \TCOMg{it may not be possible for a single node to solve large-scale machine learning problems.
%In this paper, 
the master implements a gradient descent method with the help of all workers.
Specifically,
the master partitions the whole data set and sends \TCOMb{to each worker some particular subsets}
so that the master and $N$ workers can collaboratively compute the gradient $\nabla_{\bm{\theta}} \hat{\ell}(\bm{\theta};\mathcal{D})\triangleq \sum_{\mathbf{y}\in\mathcal{D}}\nabla_{\bm{\theta}}\ell(\bm{\theta};\mathbf{y})$ in each iteration.
\TCOMb{In contrast with the case in \cite{GC_Tandon_exact,GC_Min_Ye_communication,GC_WH_improving_RS_Codes} where each worker starts to send the computation results after completing all $L$ subtasks,}
we \TCOMb{investigate} the case where each worker sequentially computes $L$ subtasks and sends the \TCOMb{computation} result of each subtask to the master once its computation is completed.
\TCOMb{Based on the computation results of the $l$-th subtasks of all workers, the master can compute the $l$-th partial derivative $\frac{\partial\hat{\ell}(\bm{\theta};\mathcal{D})}{\partial\theta_l}$, i.e., the $l$-th coordinate of the gradient $\nabla_{\bm{\theta}} \hat{\ell}(\bm{\theta};\mathcal{D})$.\footnote{We choose a coordinate as the basic computing and communication unit for ease of exposition. Note that our results also apply to the case where a block of coordinates 
(which can associate with one layer of a neural network) 
is viewed as the basic unit.}
We aim to design a coordinate gradient coding scheme for the considered case.}

\section{Coordinate Gradient Coding}
\label{sec:Coordinate_Gradient_Coding}

%To diversify the redundancy \TCOMr{in different partial derivatives across all workers} (for a partition of the whole data set), 
We propose a coordinate gradient coding scheme parameterized by the coding parameters $\mathbf{s}\triangleq (s_l)_{l\in [L]}$ for the $L$ coordinates, which satisfy
%\footnote{For ease of exposition, we present the gradient coding scheme for the case that the machine learning problem is directly solved by gradient descent.}
\begin{equation}
	s_l \in \{0,1,\cdots,N-1\},\ l\in [L].\label{equ:constraint_s_1}
\end{equation}
%---------------------------
%Here, $s_l$ represents the number of stragglers \TCOMc{that the master can tolerate} (only needs to receive computation results from arbitrary $N-s_l$ workers) when recovering the $l$-th partial derivative $\frac{\partial\hat{\ell}(\bm{\theta;\mathcal{D}})}{\partial\theta_l}$.
%---------------------------
%---------------------------
%\TCOMr{Note that this scheme generalizes the gradient coding scheme in \cite{GC_Tandon_exact} ($s_l,l\in[L]$ are identical).}
%Later, we investigate the optimization of the coding parameters $\mathbf{s}$ for the $L$ coordinates, 
%which eventually produces a
%block coordinate gradient coding scheme \TCOMb{with at most $N$ blocks, each with one coding parameter.
%We shall see that the resulting block coordinate gradient coding scheme can be more easily implemented in practice, as each worker can send the computing results of all $L$ subtasks with at most $N$ transmissions \TCOMc{rather than $L$ transmissions}.}
\TCOMb{Here, $s_l$ represents the redundancy in the computation of a coded local partial derivative corresponding to the $l$-th coordinate.
It also represents the number of stragglers
the master can tolerate when recovering the $l$-th partial derivative $\frac{\partial\hat{\ell}(\bm{\theta;\mathcal{D}})}{\partial\theta_l}$.
In other words, the master only needs to receive the coded partial derivatives corresponding to the $l$-th coordinate from arbitrary $N-s_l$ workers to recover $\frac{\partial\hat{\ell}(\bm{\theta};\mathcal{D})}{\partial\theta_l}$.}
The proposed scheme operates in two phases, as illustrated below.
\TCOMb{It generates the gradient coding scheme in \cite{GC_Tandon_exact,GC_WH_improving_RS_Codes} by allowing $s_l,l\in[L]$ to be different.
Later, we shall see that the proposed coordinate gradient coding scheme with optimal coding parameters $\mathbf{s}$ turns to a block coordinate gradient coding scheme, which can be implemented more easily.
}

\textbf{Sample Allocation Phase}:
First, the master partitions dataset $\mathcal{D}$ into $N$ subsets of size $\frac{M}{N}$, denoted by $\mathcal{D}_i,i\in [N]$~\cite{GC_Tandon_exact,GC_Min_Ye_communication,GC_approx_Tandon_approx,GC_WH_improving_RS_Codes}.
%over set $[N]$
Then, \REPLYb{the master allocates the same number of data subsets to each worker. Specifically,}
for all $n\in [N]$, the master allocates the $\max_{l\in [L]}s_l+1$ subsets, $\mathcal{D}_i,i\in \mathcal{I}_n\triangleq \{j\oplus(n-1):j\in[\max_{l\in [L]}s_l+1] \}$, to worker $n$, where the operator $\oplus$ is defined as: $a_1 \oplus a_2 \triangleq \left\{
	\begin{aligned}
		a_1+a_2,\quad {\rm if}\ a_1+a_2 \le N\\
		a_1+a_2-N,\quad {\rm if}\ a_1+a_2 > N
	\end{aligned}
	\right.$, for all $a_1,a_2\in[N]$.
\textbf{Collaborative Training Phase}:
%==================
In each iteration of a gradient descent method, the master first sends the latest $\bm{\theta}$ to all workers.
Then, each worker $n\in[N]$ sequentially computes and sends coded local partial derivatives corresponding to coordinates $1,2,\cdots,L$ to the master.
\REPLYb{Specifically, for all $l\in[L]$, worker $n\in[N]$ computes local partial derivatives $\frac{\partial\hat{\ell}(\bm{\theta};\mathcal{D}_i)}{\partial\theta_l},i \in \{j\oplus(n-1):j\in[s_l+1] \}$ \SecREPLYb{and} the coded local partial derivative $\mathbf{b}_{l,n} \mathbf{g}_l$ \SecREPLYb{and} sends $\mathbf{b}_{l,n} \mathbf{g}_l$ to the master, 
where $\mathbf{b}_{l,n}$ denotes the $n$-th row of encoding matrix $\mathbf{B}_l$ generated according to \cite{GC_Tandon_exact} with $s$ in [6] being $s_l$ (please refer to Alg.~\ref{alg:encoding_mat} in Appendix N for details), and
$\mathbf{g}_l \triangleq \left( \frac{\partial\hat{\ell}(\bm{\theta};\mathcal{D}_1)}{\partial\theta_l}, \frac{\partial\hat{\ell}(\bm{\theta};\mathcal{D}_2)}{\partial\theta_l}, \cdots, \frac{\partial\hat{\ell}(\bm{\theta};\mathcal{D}_N)}{\partial\theta_l} \right)^T$.}
Next, the master sequentially receives coded local partial derivatives \REPLYb{$\mathbf{b}_{l,1} \mathbf{g}_l, \mathbf{b}_{l,2} \mathbf{g}_l, \cdots, \mathbf{b}_{l,N} \mathbf{g}_l$} corresponding to coordinates $1,2,\cdots,L$ from each worker and recovers partial derivatives $\frac{\partial\hat{\ell}(\bm{\theta};\mathcal{D})}{\partial\theta_1},\frac{\partial\hat{\ell}(\bm{\theta};\mathcal{D})}{\partial\theta_2},\cdots,\frac{\partial\hat{\ell}(\bm{\theta};\mathcal{D})}{\partial\theta_L}$.
Let $T_{(1)},T_{(2)},\cdots,T_{(N)}$ be $T_{n},n\in[N]$ arranged in the increasing order, so that $T_{(n)}$ is the $n$-th smallest one.
Specifically, \REPLYb{for all $l\in[L]$, the master computes the $l$-th partial derivative $\frac{\partial\hat{\ell}(\bm{\theta};\mathcal{D})}{\partial\theta_l} = \text{supp}(\mathbf{a}_{l,j}) \mathbf{B}_{l,\mathcal{F}_l} \mathbf{g}_l$,} 
once it receives the coded local partial derivatives from the $N-s_l$ fastest workers with CPU cycle times $T_{(n)}, n\in[N-s_l]$, 
\REPLYb{where $\mathbf{a}_{l,j}$ denotes the $j$-th row of decoding matrix $\mathbf{A}_l$ generated according to [6] with $s$ in [6] being $s_l$ (please refer to Alg.~\ref{alg:decoding_mat} in Appendix N for details), 
supp($\mathbf{a}_{l,j}$) denotes the support of $\mathbf{a}_{l,j}$,  
$\mathcal{F}_l \triangleq \{i_1,i_2,\cdots,i_{N-s_l}\}$ denotes the indices of the $N-s_l$ fastest workers, 
$\mathbf{B}_{l,\mathcal{F}_l}$ denotes the sub-matrix of $\mathbf{B}_l$ with rows indexed by $\mathcal{F}_l$, and 
$j \in [\binom{N}{s_l}]$ is chosen such that $\text{supp}(\mathbf{a}_{l,j}) \mathbf{B}_{l,\mathcal{F}_l} = {\mathbf{1}_{N}}^T$.}
Note that the orders for computing, sending, and receiving the coded local partial derivatives are $1,\cdots,L$.
Once the master has recovered $\frac{\partial\hat{\ell}(\bm{\theta};\mathcal{D})}{\partial\theta_l}$, $l\in[L]$, it readily obtains the gradient $\nabla_{\bm{\theta}} \hat{\ell}(\bm{\theta};\mathcal{D})$ \cite{GC_Tandon_exact}.
%>>>>>>>>>>>>>>>>>>>>>>>>>>>>>>>>>>>>>>>>>>>>>>>>>>>>>>>>>>>>>>

Let $b$ denote the maximum of the numbers of CPU cycles for computing $\frac{\partial\hat{\ell}(\bm{\theta};\mathcal{D})}{\partial\theta_l}$, $l\in[L]$.\footnote{\TCOMb{For tractability,  we use the maximum, $b$, when optimizing the coding parameters in this paper.} The proposed optimization framework can be extended to consider the exact numbers of CPU cycles for computing $\frac{\partial \hat{\ell}(\bm{\theta};\mathcal{D})}{\partial \theta_l},l\in[L]$ in optimizing the coding parameters.}
We omit the computation loads for encoding at each worker and decoding at the master, as they are usually much smaller than the computation load for calculating the partial derivatives in practice.
Thus, for all $n\in[N]$ and $l\in[L]$, the completion time for computing the coded local partial derivative \TCOMb{corresponding to the $l$-th coordinate} at worker $n$ is $\frac{M}{N}bT_n\sum_{i=1}^l\left(s_i+1\right)$.
For all $l\in[L]$, the completion time for recovering $\frac{\partial\hat{\ell}(\bm{\theta};\mathcal{D})}{\partial\theta_l}$ at the master is $\frac{M}{N}bT_{(N-s_l)}\sum_{i=1}^l\left(s_i+1\right)$.
%\wqbr{Let $T_{(1)},T_{(2)},\cdots,T_{(N)}$ be $T_{n},n\in[N]$ arranged in increasing order, so that $T_{(n)}$ is the $n$-th smallest one.}
%+++++++++++++++++++++++++++++++++++++++++++++++++++++++++++++++++++++
Therefore, the overall runtime for the master and workers to collaboratively compute the gradient $\nabla_{\bm{\theta}} \hat{\ell}(\bm{\theta};\mathcal{D})$ is
\begin{equation}
	\tau(\mathbf{s,T}) = \frac{M}{N}b \max_{l\in [L]} T_{(N-s_l)} \sum_{i=1}^l\left(s_i+1\right),\label{equ:tau_calculate}	
\end{equation}
where $\mathbf{T}\triangleq \left(T_n\right)_{n\in [N]}$.
Note that $\tau(\mathbf{s,T})$ is a function of the parameters $\mathbf{s}$ and random vector $\mathbf{T}$ and hence is also random.
%In Fig.~\ref{subfig:motivating_example_proposed}, we provide an example to illustrate the idea of the proposed coordinate gradient coding scheme.
%Specifically, from the example in Fig.~\ref{fig:motivating_example},
%we can see that the proposed coordinate gradient coding scheme with coding parameters $\mathbf{s}=(1,1,2,2)$ has a shorter overall runtime than the gradient coding scheme in~\cite{GC_Tandon_exact} with coding parameter $s=1$ or $s=2$ at $\mathbf{T}=\left( \frac{1}{10},\frac{1}{10},\frac{1}{4},1 \right)$, as more computation results from partial stragglers are utilized.

\begin{figure*}[t]
\begin{center}
  \subfigure[\scriptsize{Gradient coding scheme in~\cite{GC_Tandon_exact} with $s=1$. The master only recovers partial derivatives \TCOMb{$\frac{\partial\hat{\ell}(\bm{\theta};\mathcal{D})}{\partial\theta_1}$ and $\frac{\partial\hat{\ell}(\bm{\theta};\mathcal{D})}{\partial\theta_2}$} at time $\frac{Mb}{4}T_0$. \TCOMbr{Here $ \tau(\mathbf{s,T}) = \frac{Mb}{2}T_0$.}}\label{subfig:system_Tandon_1}]
  {\resizebox{5cm}{!}{\includegraphics{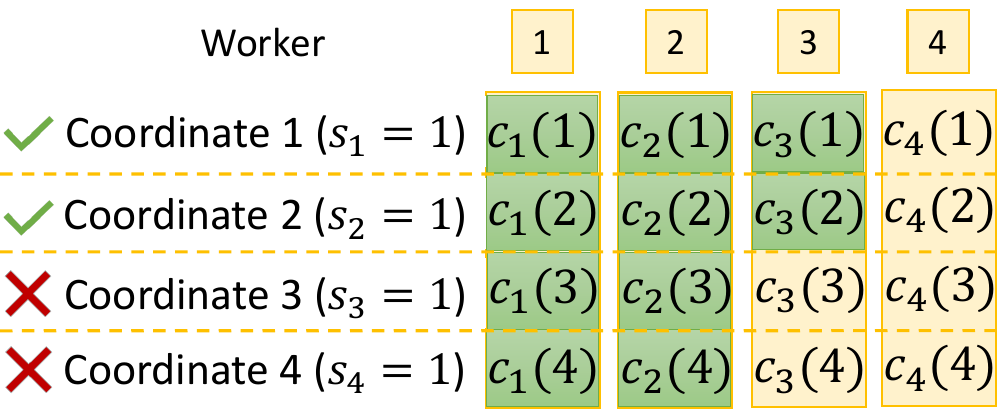}}}
  \ \
  \subfigure[\scriptsize{Gradient coding scheme in~\cite{GC_Tandon_exact} with $s=2$. The master only recovers partial derivatives \TCOMb{$\frac{\partial\hat{\ell}(\bm{\theta};\mathcal{D})}{\partial\theta_1}$, $\frac{\partial\hat{\ell}(\bm{\theta};\mathcal{D})}{\partial\theta_2}$, and $\frac{\partial\hat{\ell}(\bm{\theta};\mathcal{D})}{\partial\theta_3}$} at time $\frac{Mb}{4}T_0$. \TCOMbr{Here $\tau(\mathbf{s,T}) = \frac{3Mb}{10}T_0$.}}\label{subfig:system_Tandon_2}]
  {\resizebox{5cm}{!}{\includegraphics{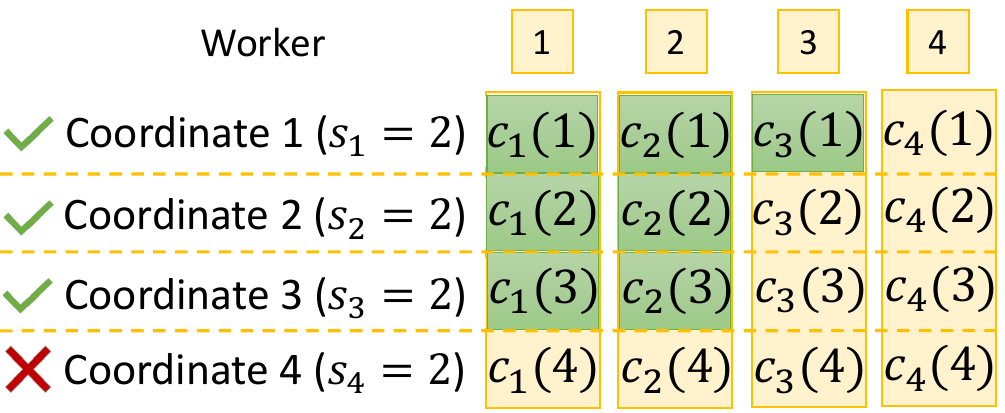}}}
  \ \
  \subfigure[\scriptsize{Proposed coordinate gradient coding scheme with $\mathbf{s}=(1,1,2,2)$. The master recovers partial derivatives \TCOMb{$\frac{\partial\hat{\ell}(\bm{\theta};\mathcal{D})}{\partial\theta_l},\ l=1,\cdots,4$,} at time $\frac{Mb}{4}T_0$. \TCOMbr{Here $\tau(\mathbf{s,T}) = \frac{Mb}{4}T_0$.}}\label{subfig:motivating_example_proposed}]
  {\resizebox{5cm}{!}{\includegraphics{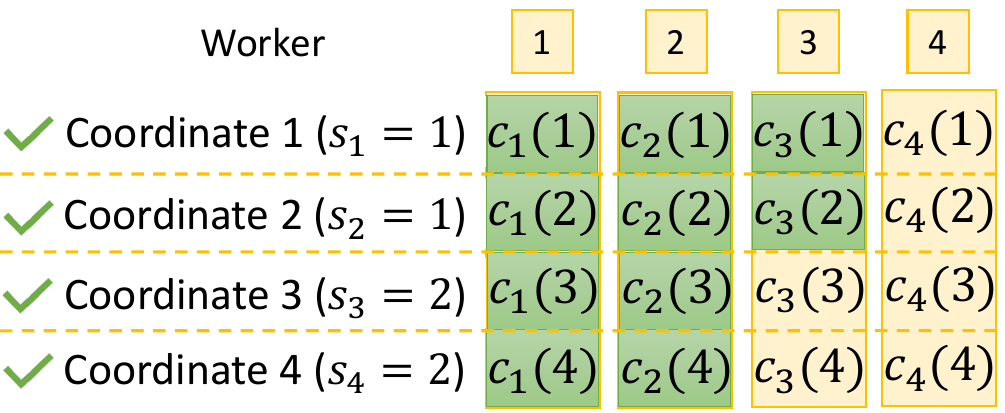}}}
  \end{center}
  \vspace{-0.4cm}
         \caption{
         Motivating examples at $N=4$, $L=4$, and $\mathbf{T}=\left( \frac{1}{10},\frac{1}{10},\frac{1}{4},1 \right)T_0$, with $T_0>0$.
%         \TCOMc{Here coding parameters $\mathbf{s} = (s_1,s_2,s_3,s_4)$.}
         Let $c_n(l)$ denote the coded local partial derivative \TCOMb{corresponding to the $l$-th coordinate} computed by worker $n$, for all $n\in[N]$ and $l\in[L]$.
         Here $c_1(l) = g_1(l) - g_2(l),\ c_2(l) = g_2(l) + g_3(l),\ c_3(l) = g_3(l) - g_4(l)$, and $c_4(l) = g_1(l) + g_4(l)$, for all $l\in\{l:s_l=1\}$; 
         and $c_1(l) = g_1(l) + \frac{1}{3}g_2(l) +\frac{2}{3}g_3(l),\ c_2(l) = g_2(l) + \frac{1}{2}g_3(l) + \frac{3}{2}g_4(l),\ c_3(l) = 2g_1(l) + g_3(l) - g_4(l)$, and $c_4(l) = -\frac{1}{2}g_1(l)+\frac{1}{2}g_2(l)+g_4(l)$, for all $l\in\{l:s_l=2\}$, where $g_n(l)\triangleq\frac{\partial \hat{\ell}(\bm\theta;\mathcal{D}_n)}{\partial \theta_l}$ for all $n\in[N]$ and $l\in[L]$ \cite{GC_Tandon_exact}.
         The green (\REPLYb{yellow}) part represents the coded local partial derivatives that have (\REPLYb{have not} \SecREPLYb{completely}) been computed by a worker \TCOMb{by} time $\frac{Mb}{4}T_0$.
	}\label{fig:motivating_example}
	\vspace{-0.5cm}
\end{figure*}

% ^^^^^^^^^^^^^^^^^^^^^^^^^^^^^^^^^^^^^^^^^^^^^^^^^^^^^^^^
\begin{Rem}[Motivation of Diversifying Redundancy]
\REPLYb{Each worker sequentially computes and sends coded local partial derivatives (corresponding to coordinates $1,2,\cdots,L$) to the master.
Note that for any $\mathbf{s}=(s_l)_{l\in[L]}$ and any $n\in[N]$,
the length of the interval between the times receiving the coded local partial derivatives from the fastest worker and from the $n$-th fastest worker, $\frac{M}{N}b (T_{(n)} - T_{(1)}) \sum_{i=1}^l\left(s_i+1\right)$, increases with $l$.
To reduce the waiting time for collecting additional coded local partial derivatives for decoding, we intended to reduce the  number of required coded local partial derivatives for recovering the $l$-th partial derivative $\frac{\partial\hat{\ell}(\bm{\theta};\mathcal{D})}{\partial\theta_l}$, $N-s_l$, when $l$ increases.
That is, intuitively, letting $s_l$ (representing the redundancy in the computation of a coded local partial derivative corresponding to the $l$-th coordinate) increase with $l$ helps reduce the completion time for computing the coded local partial derivative corresponding to the $l$-th coordinate, thereby reducing the overall runtime.
This reveals the ineffectiveness of posing identical redundancy 
%in the computation of coded local partial derivatives corresponding to all $L$ coordinates 
(i.e., identical $s_l,l\in[L]$) and motivates diverse (more specifically, increase) redundancies in the computation of coded local partial derivatives for $L$ coordinates (i.e., different $s_l,l\in[L]$).}
A motivating example is illustrated in Fig.~\ref{fig:motivating_example}.
%\REPLYb{Specifically, Fig. 1 (a) and Fig. 1 (b) illustrate the original gradient coding scheme with coding parameter $s=1$ and $s=2$, respectively, and Fig.1 (c) illustrates the proposed coordinate gradient coding scheme with coding parameters $\mathbf{s}=(1,1,2,2)$.}
Specifically, 
in Fig.~\ref{subfig:system_Tandon_1}, the computation results of the coded local partial derivatives \TCOMb{corresponding to the 3rd and 4th coordinates} from worker 1 and from worker 2 are not utilized;
in Fig.~\ref{subfig:system_Tandon_2}, the computation result of the coded local partial derivative \TCOMb{corresponding to the 1st coordinate} from worker 3 is not utilized;
in Fig.~\ref{subfig:motivating_example_proposed}, \TCOMb{the computation results of the coded local partial derivatives corresponding to all coordinates from all workers are utilized.}
\REPLYb{
%As shown in Fig. 1, their overall runtimes are $\frac{Mb}{2}T_0, \frac{3Mb}{10}T_0$, and $\frac{Mb}{4}T_0$, respectively.
Fig.~\ref{fig:motivating_example} shows that the overall runtime of the proposed coordinate gradient coding scheme with coding parameters $\mathbf{s}=(1,1,2,2)$ (shown in Fig.~\ref{subfig:motivating_example_proposed})  is shorter than those of the original gradient coding scheme with coding parameter $s=1$ (shown in Fig.~\ref{subfig:system_Tandon_1}) and $s=2$ (shown in Fig.~\ref{subfig:system_Tandon_2}).}
%\TCOMb{It is clear that} the proposed coordinate gradient coding scheme with coding parameters $\mathbf{s}=(1,1,2,2)$ \TCOMb{(shown in Fig.~\ref{subfig:motivating_example_proposed})} has a shorter overall runtime than the gradient coding scheme in~\cite{GC_Tandon_exact} with coding parameter $s=1$ \TCOMb{(shown in Fig.~\ref{subfig:system_Tandon_1})} or $s=2$ \TCOMb{(shown in Fig.~\ref{subfig:system_Tandon_2})}. 
\end{Rem}
% vvvvvvvvvvvvvvvvvvvvvvvvvvvvvvvvvvvvvvvvvvvvvvvvvvvvvvvv

%\cite{ mat_mat_Yu_polynomial, mat_mat_Dutta_PolyDot_Codes , GC_Tandon_exact, GC_approx_Tandon_approx, approx_GC_Bitar}. 
%\cite{mat_vec_Lee_speeding_up, mat_vec_LT_Codes, mat_vec_Dutta_Shot_Dot_Codes, mat_vec_Gunduz_speeding_up,   mat_vec_Draper_hierarchical_ISIT, mat_mat_Draper_hierarchical_TIT, mat_mat_Lee_high_dimensional, mat_mat_Yu_polynomial, mat_mat_Draper_exploitation, GC_WH_improving_RS_Codes, GC_Min_Ye_communication, GC_SA_near_optimal, GC_Gunduz_clustering, GC_Gunduz_dynamic_clustering_J, approx_mat_vec_Draper_anytime, approx_mat_vec_Coded_Sequential}

In this paper, we consider two performance metrics, namely, the expected overall runtime, $\mathbb{E}\left[\tau(\mathbf{s,T})\right]$ \DAIm{(as considered in \cite{GC_Tandon_exact,mat_vec_Lee_speeding_up,mat_vec_Dutta_Shot_Dot_Codes,mat_vec_Gunduz_speeding_up,mat_vec_Draper_hierarchical_ISIT, mat_mat_Draper_hierarchical_TIT, mat_mat_Lee_high_dimensional,mat_mat_Draper_exploitation, GC_WH_improving_RS_Codes, GC_Min_Ye_communication, GC_SA_near_optimal,GC_Gunduz_clustering, GC_Gunduz_dynamic_clustering_J,approx_mat_vec_Draper_anytime,approx_mat_vec_Coded_Sequential,Wang_approx_1,approx_Wang_fundamental})}, and the completion probability by time threshold $t$, $P(\mathbf{s},t)\triangleq \Pr[\tau(\mathbf{s},\mathbf{T})\le t]$ \DAIm{(as considered in \cite{mat_mat_Yu_polynomial,mat_vec_Lee_speeding_up,mat_vec_Gunduz_speeding_up,mat_vec_Draper_hierarchical_ISIT, mat_mat_Draper_hierarchical_TIT, mat_mat_Lee_high_dimensional,mat_mat_Draper_exploitation, GC_WH_improving_RS_Codes, GC_Min_Ye_communication, GC_SA_near_optimal,GC_Gunduz_clustering, GC_Gunduz_dynamic_clustering_J,approx_mat_vec_Draper_anytime})}. 
%referred to as the completion probability.
By \cite{mat_vec_Draper_hierarchical_ISIT}, we have
\begin{small}
\begin{equation}
	P(\mathbf{s},t) 
%	\triangleq \Pr[\tau(\mathbf{s},\mathbf{T})\le t] 
%	= \Pr \Bigg[ \bigcup_{\mathbf{k} \in \mathcal{K}(\mathbf{s},N,L)} A_{\mathbf{k}} \Bigg]
	= \sum_{\mathbf{k} \in \mathcal{K}(\mathbf{s},N,L)} \binom{N}{k_0,k_1,\cdots,k_L} \prod_{l=0}^L\big( F_l(\mathbf{s},t)-F_{l+1}(\mathbf{s},t) \big)^{k_l},\label{equ:CDF_fobj_s}
\end{equation}
\end{small}where $\mathbf{k} \triangleq (k_l)_{l=0,\cdots,L}$, 
%with $k_i$ representing the number of workers completing $i$ coded local partial derivatives but not completing $i+1$ ones, 
$\mathcal{K}(\mathbf{s},N,L)\triangleq \left\{ \mathbf{k}\in\mathbb{N}^{L+1} : \sum_{i=0}^{l-1} k_i\le s_{l}, l\in[L], \sum_{i=0}^{L}k_i=N \right\}$, 
%\TCOMg{$\mathbb{N}$ denotes the set of natural numbers,} 
and $F_l(\mathbf{s},t) \triangleq \Pr\big[\frac{M}{N}bT\sum_{i=1}^l\left(s_i+1\right) < t\big] = F_T\left(\frac{t}{\frac{M}{N}b\sum_{j=1}^l(s_j+1)}\right)$, $l\in[L]$.
For ease of exposition, we let $F_0(\mathbf{s},t)= 1$ and $F_{L+1}(\mathbf{s},t) = 0$.
In Sec.~\ref{sec:Expected_Overall_Runtime_Minimization} and Sec.~\ref{sec:Completion_Probability_Maximization}, we will minimize $\mathbb{E}\left[\tau(\mathbf{s,T})\right]$ and maximize $P(\mathbf{s},t)$, respectively, by optimizing the coding parameters $\mathbf{s}$ for the $L$ coordinates under the constraints in~\eqref{equ:constraint_s_1}.

%\begin{figure}
%\begin{center}
%  {\resizebox{5cm}{!}{\includegraphics{motivating_exam_1.pdf}}}
%  \vspace{-2mm}
%         \caption{Proposed coordinate gradient coding scheme with $\mathbf{s}   =(1,1,2,2)$ at $N=4$, $L=4$, and $\mathbf{T}=\left( \frac{1}{10},\frac{1}{10},\frac{1}{4},1 \right)T_0$.
%         The overall runtime is $\frac{Mb}{4}T_0$.
%         \vspace{-6mm}
%         }\label{fig:motivating_example_proposed}
%\end{center}
%\end{figure}

\section{Expected Overall Runtime Minimization}
\label{sec:Expected_Overall_Runtime_Minimization}

In this section, we first formulate the minimization of the expected overall runtime with respect to the coding parameters for coordinates.
Then, we obtain an optimal solution of the continuous relaxation of the original problem.
Finally, we obtain two low-complexity closed-form approximate solutions of the relaxed problem.

\subsection{Problem Formulation}
\label{ssec:Problem_Formulation_E} 

We would like to minimize $\mathbb{E}\left[\tau(\mathbf{s,T})\right]$ by optimizing the coding parameters $\mathbf{s}$ for the $L$ coordinates under the constraints in~\eqref{equ:constraint_s_1}.

\begin{Prob}[Expected Overall Runtime Minimization]\label{prob:E_original}
	\begin{align}
		\tau_{\rm avg}^* \triangleq \min_{\mathbf{s}} \ \  &\mathbb{E}\left[ \tau(\mathbf{s,T}) \right] \nonumber\\
		  \rm{s.t.}\ \  &\eqref{equ:constraint_s_1}.\nonumber
	\end{align}
\end{Prob}
%\vspace{-0.1cm}
%where $\tau(\mathbf{s,T})$ is given by~\eqref{equ:tau_calculate}.

In general, the objective function $\mathbb{E}\left[\tau(\mathbf{s,T})\right]$ does not have an analytical expression, and the number of variables (model size $L$) is usually quite large.\footnote{\REPLYb{For example, AlexNet has 60 million parameters \cite{AlexNet}; VGG-16 has 138 million parameters \cite{VGG}; GoogleNet has 13 million parameters \cite{GoogleNet}; ResNet-152 has 60 million parameters \cite{ResNet}.}}
Thus, Problem~\ref{prob:E_original} is a challenging stochastic optimization problem.
First, we characterize the monotonicity of an optimal solution of Problem~\ref{prob:E_original}.
\begin{Lem}[Monotonicity of Optimal Solution of Problem~\ref{prob:E_original}]\label{lem:E_monotonicity}
	An optimal solution $\mathbf{s^*}\triangleq (s^*_l)_{l\in [L]}$ of Problem~\ref{prob:E_original} satisfies $ s^*_1 \le s^*_2 \le \cdots \le s^*_L $.
\end{Lem}

%\begin{Proof}[Sketch]
%	We prove Lemma~\ref{lem:E_monotonicity} by contradiction.
%	Suppose that for any optimal solution $\mathbf{s^*}$, there exists $k\in [L]$ such that $s^*_k > s^*_{k+1}$.
%	Construct a feasible solution $\tilde{\mathbf{s}} \triangleq (\tilde{s}_l)_{l\in [L]}$, where $\tilde{s}_l = s^*_l, l\neq k$ and $\tilde{s}_l = s^*_{k+1}, l=k$.
%	By~\eqref{equ:tau_calculate} and the definition of $\tilde{\mathbf{s}}$, we can prove that $\tau(\tilde{\mathbf{s}},\mathbf{T}) < \tau(\mathbf{s^*,T})$.
%	This indicates that $\mathbf{s}^*$ is not an optimal solution, which contradicts with the assumption.
%	$\hfill\IEEEQED$ 
%\end{Proof}

\begin{IEEEproof}
Please refer to Appendix A.	
\end{IEEEproof}

\begin{figure}[t]
\begin{center}
  {\resizebox{3.3cm}{!}{\includegraphics{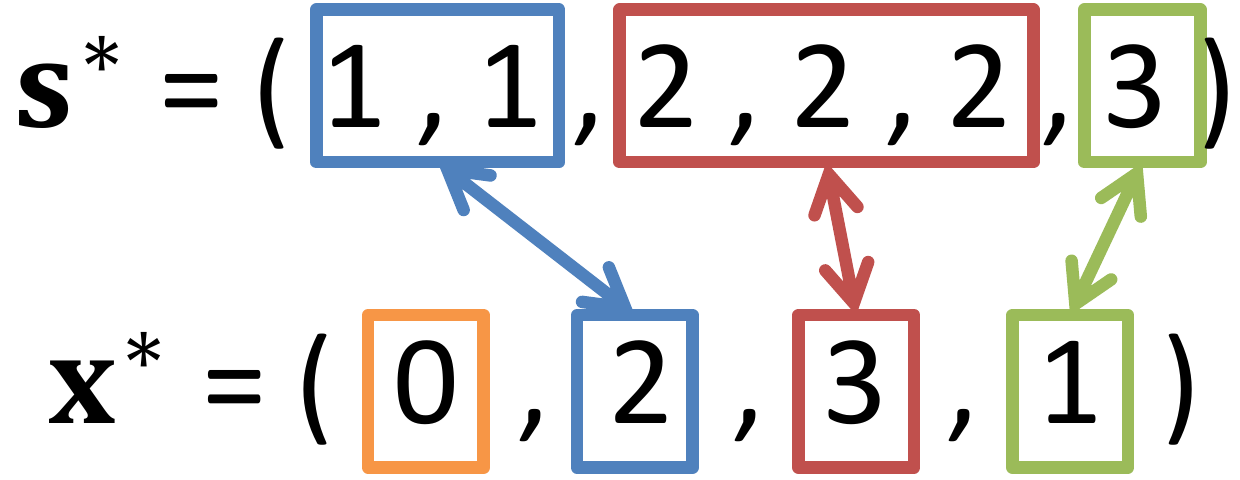}}}
  \quad\quad 
  {\resizebox{3.3cm}{!}{\includegraphics{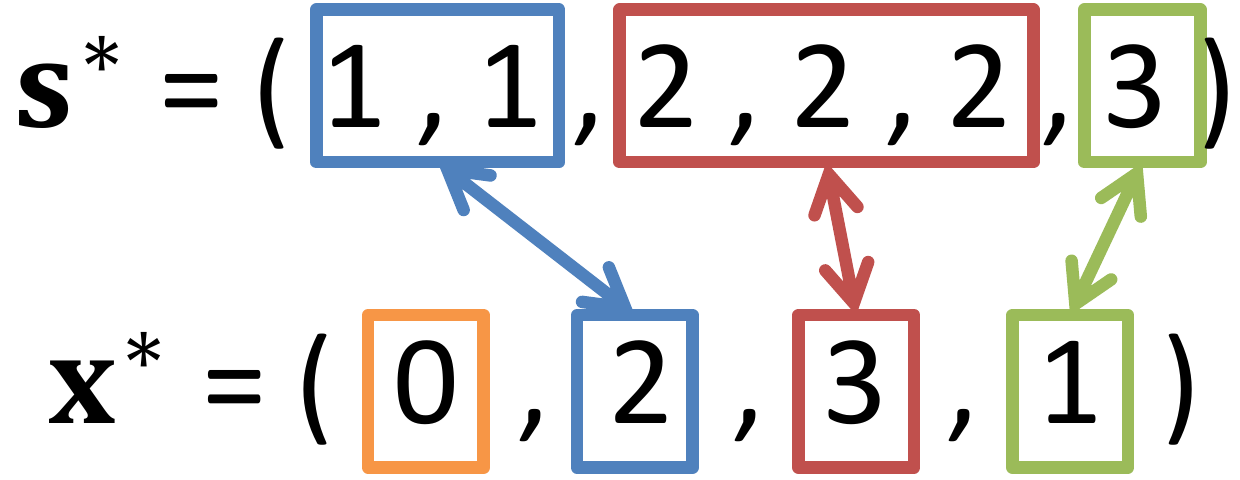}}}
  \end{center}
  \vspace{-0.4cm}
         \caption{Illustration of relation between $\mathbf{s}^*$ and $\mathbf{x}^*$ at $N=4$ and $L=6$.}\label{fig:example_illustration}
         \vspace{-0.4cm}
\end{figure} 

\REPLYb{Lemma~\ref{lem:E_monotonicity} verifies the intuition in Remark 2.}
Lemma~\ref{lem:E_monotonicity}, \TCOMb{together with the constraints in \eqref{equ:constraint_s_1}}, indicates that we can optimally partition the $L$ coordinates into $N$ blocks, each with identical redundancy, for tolerating $0, 1,\cdots, N-1$ stragglers, respectively.
That is, the proposed coordinate gradient coding scheme with $\mathbf{s}^*$ becomes a block coordinate gradient coding scheme \TCOMb{with at most $N$ blocks, each with one coding parameter.
It} \TCOMb{can be implemented with lighter overhead. Specifically, each worker can send multiple coded local partial derivatives corresponding to one block together in one transmission once their computations are completed without influencing the overall runtime.}

Next, based on Lemma~\ref{lem:E_monotonicity} \TCOMb{and the constraints in \eqref{equ:constraint_s_1}}, we transform Problem~\ref{prob:E_original} to an equivalent problem with $N$ variables, which optimizes the partition of the $L$ coordinates into $N$ blocks.
\begin{Prob}[Equivalent Problem of Problem~\ref{prob:E_original}]\label{prob:equivalent_1}
		\begin{align}
			\hat{\tau}_{\rm avg}^* \triangleq \min_{\mathbf{ x}} \ \  &\mathbb{E}\left[ \hat{\tau}(\mathbf{ x},\mathbf{T}) \right] \nonumber\\
		  \rm{s.t.}\ \  &\sum_{n=0}^{N-1}  x_n=L,\label{equ:constraint_x_1}\\
		  				& x_n \in \mathbb{N},\quad n=0,1,\cdots,N-1,\label{equ:constraint_x_2}
		\end{align}
		where $\mathbf{ x} \hspace{-0cm} \triangleq \hspace{-0cm} ( x_n)_{n=0,1,\cdots,N-1}$\hspace{-0cm} 
%		$\mathbb{N}$ denotes the set of  natural numbers, 
		and 
		\begin{align}
			\hat{\tau}(\mathbf{ x},\mathbf{T})\ \triangleq \frac{M}{N}b\underset{n=0,1,\cdots,N-1}{\max}  T_{(N-n)}\sum_{i=0}^n(i+1) x_i.\label{equ:tau_hat}
		\end{align}
\end{Prob}
\begin{Thm}[Equivalence between Problem~\ref{prob:E_original} and Problem~\ref{prob:equivalent_1}]\label{thm:E_prob_equvalence}
	An optimal solution of Problem~\ref{prob:E_original}, denoted by $\mathbf{s}^*=(s^*_l)_{l\in [L]}$, and an optimal solution of Problem~\ref{prob:equivalent_1}, denoted by $\mathbf{ x}^*=( x^*_n)_{n=0,1,\cdots,N-1}$, satisfy
	\begin{align}
		x_n^* &= \sum_{l\in[L]} I(s_l^*=n),\ n=0,1,\cdots,N-1,\label{equ:change_of_var_x}\\
		s_l^* &= \min\left\{ i : \sum_{n=0}^{i} x_n^* \ge l \right\},\ l\in[L].\label{equ:change_of_var_s}
	\end{align}
%	\TCOMg{where $I(\cdot)$ denotes the indicator function.}
	Furthermore, their optimal values, $\tau_{\rm avg}^*$ and $\hat{\tau}_{\rm avg}^*$, satisfy $\tau_{\rm avg}^*=\hat{\tau}_{\rm avg}^*$.
\end{Thm}
%\begin{Proof}[Sketch]
%	Consider $\mathbf{s}^*$ satisfying $ s^*_1 \le s^*_2 \le \cdots \le s^*_L $.
%	We show that the change of variables given by \eqref{equ:change_of_var_x} and \eqref{equ:change_of_var_s} is one to one.
%	Then, we show $\tau(\mathbf{s}^*,\mathbf{T}) = \hat{\tau}(\mathbf{x}^*,\mathbf{T})$.
%	$\hfill\IEEEQED$
%\end{Proof}
\begin{IEEEproof}
Please refer to Appendix B.	
\end{IEEEproof}

%\begin{figure*}[ht]
%\normalsize{
%\begin{small}
%	\begin{align}
%		t_{n}'\hspace{-0.0cm} 
%		=\hspace{-0.0cm}  -1 \Bigg/ \Bigg(\mu n\binom{N}{n-1}\hspace{-0.0cm} \sum_{i=0}^{n-1}(-1)^i\binom{n-1}{i}e^{\mu t_0 (N-n+i+1)} E_i(-\mu t_0 (N-n+i+1)) \Bigg).\label{equ:t'}
%	\end{align}
%\end{small}
%%\vspace{-1mm}
%}\hrulefill
%%\vspace{-5mm}
%\end{figure*}

Fig.~\ref{fig:example_illustration} illustrates the relationship between $\mathbf{x}^*$ and $\mathbf{s}^*$. 
\TCOMb{Based on Theorem \ref{thm:E_prob_equvalence}, we can optimally design a block coordinate gradient coding scheme instead.
Specifically, $x_n$ represents} the number of coordinates with identical redundancy for tolerating $n$ stragglers, \TCOMb{where $n\in[N]$}, \TCOMb{and $\mathbf{x}$ reflects the coding parameters for a block coordinate gradient coding scheme with} 
$N$ blocks, each with identical redundancy.
%Thus, $\mathbf{x}^*$ specifies the optimal block coordinate gradient coding scheme.
\TCOMb{The optimal block coordinate gradient coding scheme can be obtained by solving Problem \ref{prob:equivalent_1}.}
As the number of model parameters $L$ is usually much larger than the number of workers $N$, the computational complexity can be greatly reduced if we solve Problem~\ref{prob:equivalent_1} rather than Problem~\ref{prob:E_original}.
By relaxing the integer constraints in~\eqref{equ:constraint_x_2}, we have the following continuous relaxation of  Problem~\ref{prob:equivalent_1}, which is more tractable.

\begin{Prob}[Relaxed Continuous Problem of Problem~\ref{prob:equivalent_1}]\label{prob:relaxed_prob}
	\begin{align}
		\hat{\tau}^*_{\rm avg-ct} \triangleq \min_{\mathbf{ x}} \ \  &\mathbb{E}\left[ \hat{\tau}(\mathbf{ x},\mathbf{T}) \right] \nonumber\\
		  \rm{s.t.}\ \  &\eqref{equ:constraint_x_1}, \nonumber\\
		  				& x_n \ge 0,\quad n=0,1,\cdots,N-1. \label{equ:constraint_x_3}
	\end{align}
\end{Prob}

One can apply the rounding method in \cite[pp. 386]{Boyd_cvxbook} to round an optimal solution (or a suboptimal solution) of Problem~\ref{prob:relaxed_prob} to an integer-valued feasible point of Problem~\ref{prob:equivalent_1}, which is a good approximate solution when $N\ll L$ (usually satisfied in most machine learning problems).
\TCOMb{Thus}, in the following subsections, we focus on solving the relaxed problem in Problem~\ref{prob:relaxed_prob}.

\subsection{Optimal Solution}
\label{ssec:Optimal_Solution_E}

Problem~\ref{prob:relaxed_prob} is a stochastic convex problem whose objective function is the expected value of a (non-differentiable) piecewise-linear function.
An optimal solution of Problem~\ref{prob:relaxed_prob}, denoted by $\mathbf{x}^{\rm (E,opt)}$, can be obtained by the stochastic projected subgradient method~\cite{Boyd_stochastic_subgradient_methods}.
The main idea is to compute a noisy unbiased subgradient of the objective function and carry out a projected subgradient update based on it at each iteration.
Specifically, in the $i$-th iteration, we generate $S$ random samples of $\mathbf{T}$, denoted by $\mathbf{t}^{1,(i)},\cdots,\mathbf{t}^{S,(i)}$,
%By applying the rules for the subgradient of the maximum, we obtain 
and calculate the noisy unbiased subgradient of $\mathbb{E}\left[ \hat{\tau}(\mathbf{x,T}) \right]$ at $\mathbf{x}^{(i-1)}$ for the $S$ random samples, i.e., the subgradient of $\frac{1}{S}\sum_{j=1}^S \hat{\tau}(\mathbf{x}^{(i-1)},\mathbf{t}^{j,(i)})$, by applying the rules for the subgradient of the maximum~\cite[Sec. 3.4]{Boyd_subgradients}.
\begin{small}
\begin{align}
			\tilde{g}\left( \mathbf{x}^{(i-1)} \right ) \triangleq \frac{1}{S}\sum_{j=1}^{S} \frac{Mb}{N} t_{(N-n^{j,(i)})}^{j,(i)} \left(1,2,3,\cdots,n^{j,(i)} +1,0,\cdots,0\right),\label{equ:subgrad_1}
\end{align}
\end{small}where $ n^{j,(i)} \triangleq \underset{{n=0,1,\cdots,N-1}}{\argmax}  t^{j,(i)}_{(N-n)} \sum_{k=0}^n (k+1) x_k $.
Then, we update $\mathbf{x}^{(i)}$ by
\begin{align}
	\hat{\mathbf{x}}^{(i)} = \mathbf{x}^{(i-1)}-\sigma^{(i)} \tilde{g} \left( \mathbf{x}^{(i-1)} \right),\label{equ:subgrad_2}
\end{align}
where $\{\sigma^{(i)}\}$ is a positive diminishing stepsize \TCOMb{sequence} satisfying
\begin{equation}
	\sigma^{(i)}>0,\quad \sigma^{(i)} \rightarrow 0,\quad \sum_{i=1}^{\infty}\sigma^{(i)}=\infty,\quad  \sum_{i=1}^{\infty}\left(\sigma^{(i)}\right)^2<\infty.\label{equ:stepsize_2}
\end{equation}
Next, we project $\hat{\mathbf{x}}^{(i)}$ on the feasible set of Problem~\ref{prob:relaxed_prob} which is convex.
The projection is given by the solution of the following problem.
\begin{Prob}[Projection \TCOMb{of $\hat{\mathbf{x}}^{(i)}$}]
\label{prob:projection}
\begin{align}
\setlength{\abovedisplayskip}{0pt}
\setlength{\belowdisplayskip}{0pt}
			\mathbf{x}^{(i)}=\argmin_{\mathbf{x}} &\left\|\hat{\mathbf{x}}^{(i)} - \mathbf{x} \right\|_2^2\nonumber\\
			\rm{s.t.} \ \  &\eqref{equ:constraint_x_1},\eqref{equ:constraint_x_3}.\nonumber
\end{align}
\end{Prob}

Problem~\ref{prob:projection} is a convex \TCOMb{quadratic programming}.
By the Karush-Kuhn-Tucker (KKT) conditions \cite[Sec. 5.5.3]{Boyd_cvxbook}, we obtain the optimal solution of it.
\begin{Lem}[Optimal Solution of Problem~\ref{prob:projection}]
\label{lem:opt_projection}
	The optimal solution $\mathbf{x}^{(i)}\triangleq \left(x^{(i)}_n\right)_{n=0,1,\cdots,N-1}$ of Problem~\ref{prob:projection} is given by	
	\begin{align}
		x_n^{(i)} = \max\left\{\hat{x}_n^{(i)} + \lambda,0\right\},\ n=0,1,\cdots,N-1,\label{equ:E_projection_closed_form_1} 
	\end{align}
	where $\lambda$ satisfies
	\begin{align}
		\sum_{n=0}^{N-1} \max\left\{ \hat{x}_n^{(i)} + \lambda, 0\right\}=L.\label{equ:E_projection_closed_form_2}
	\end{align}
\end{Lem}
\begin{IEEEproof}
%The proof is omitted due to page limitation.
\REPLYb{Please refer to Appendix C.}
\end{IEEEproof}

Note that $\lambda$ can be obtained by bisection search.
The details of the algorithm are summarized in Alg.~\ref{alg:optimal_solution}.
We can easily verify that the computational complexities of Steps 3, 4, 5, 6, and 7 in Alg.~\ref{alg:optimal_solution} are $\mathcal{O}(N)$, $\mathcal{O}(N^2)$, $\mathcal{O}(N)$, $\mathcal{O}(N)$, and $\mathcal{O}(N)$, respectively.
Thus, the overall computational complexity of Alg.~\ref{alg:optimal_solution} is $\mathcal{O}(N^2)$.
%\TCOMr{The algorithm is illustrated in Alg.~\ref{alg:optimal_solution}}

%\setlength{\intextsep}{-5pt}
%\setlength{\textfloatsep}{5pt}
\begin{algorithm}[t]
	\caption{Algorithm for Obtaining An Optimal Solution of Problem~\ref{prob:relaxed_prob}}
\begin{small}
	\begin{algorithmic}[1]
		\STATE \textbf{initialization}: Set $i=1$ and choose any feasible point $\mathbf{x}^{(0)}$ of Problem~\ref{prob:relaxed_prob}.
		\STATE \textbf{repeat} 
		\STATE \quad Generate $S$ random samples of $\mathbf{T}$, i.e.,  $\mathbf{t}^{1,(i)},\cdots,\mathbf{t}^{S,(i)}$, according to the distribution of $T_n,n\in[N]$.
		\STATE \quad Obtain a noisy unbiased subgradient $\tilde{g}\left( \mathbf{x}^{(i-1)} \right )$ according to \eqref{equ:subgrad_1}.
		\STATE \quad Update $\hat{\mathbf{x}}^{(i)}$ according to \eqref{equ:subgrad_2}.
		\STATE \quad Compute $\lambda$ by solving the equation in \eqref{equ:E_projection_closed_form_2} with bisection search, and compute $\mathbf{x}^{(i)}$ according to \eqref{equ:E_projection_closed_form_1}. 
		\STATE \quad Set $i=i+1$.
		\STATE \textbf{until} Some convergence criterion is met.
	\end{algorithmic}\label{alg:optimal_solution}
\end{small} 
\end{algorithm}

%\vspace{-0.1cm}
\subsection{Approximate Solutions}
\label{ssec:approx_solution}
%\vspace{-0.1cm}

%\subsubsection{\wqm{Approximate Solution Derived from Mean Runtime}}
In the following, we obtain two closed-form approximate solutions of Problem~\ref{prob:relaxed_prob} which are more computationally efficient than the optimal solution \TCOMb{given by Lemma~\ref{lem:opt_projection}}.
%\wqm{First, we propose an approximate solution derived from mean runtime.}
\subsubsection{Approximate Solution Based On Deterministic CPU Cycle Times}
\label{sssec:Senario_1}
We approximate the objective function of Problem~\ref{prob:relaxed_prob} by replacing the random vector $\mathbf{T}$ with the deterministic vector $\mathbf{t}\triangleq (t_n)_{n\in[N]}$, where $t_n \triangleq \mathbb{E}\left[T_{(n)}\right]$ (which can be numerically computed for an arbitrary \TCOMb{form of} $F_T(\cdot)$ and analytically computed for some special forms of $F_T(\cdot)$).
%Consider the following relaxed problem of Problem \ref{prob:relaxed_prob}.
\begin{Prob}[Approximation of Problem~\ref{prob:relaxed_prob} at $\mathbf{t}$]\label{prob:approx_solution_time}
	\begin{align}
		\mathbf{x}^{\rm (E,t)} \triangleq \argmin_{\mathbf{x}} \ \  &\hat{\tau}(\mathbf{x},\mathbf{t}) \nonumber\\
		  \rm{s.t.}\ \  &\eqref{equ:constraint_x_1},\eqref{equ:constraint_x_3}.\nonumber
	\end{align}
\end{Prob}

Note that Problem~\ref{prob:approx_solution_time} is a challenging convex problem with a non-differentiable objective function due to the point-wise maximum in $\hat{\tau}(\mathbf{x},\mathbf{T})$.
By contradiction and construction, we obtain the optimal solution of Problem~\ref{prob:approx_solution_time} given as follows.
%++++++++++++++++++++++++++++++++++++++++++++++++++++++++++
\begin{Thm} [Closed-form Optimal Solution of Problem~\ref{prob:approx_solution_time}]\label{thm:E_approx_solution_time}
$x^{\rm (E,t)}_0 \hspace{-0.00cm} = \hspace{-0.00cm} \frac{1} {t_N}z^{\rm (E,t)} \hspace{-0.0cm},\ x^{\rm (E,t)}_n \hspace{-0.00cm} = \hspace{-0.00cm} \frac{1}{n+1}\hspace{-0cm}\Big(\hspace{-0cm}\frac{1} {t_{N-n}} \hspace{-0.00cm} - \hspace{-0cm} \frac{1} {t_{N+1-n}}\hspace{-0cm}\Big)\hspace{-0.00cm}z^{\rm (E,t)} \hspace{-0.0cm},\ n\in[N-1],$
where $z^{\rm (E,t)} \triangleq \frac{L} {\sum_{n=1}^{N-1} \frac{1}{n(n+1)t_{N+1-n}} + \frac{1}{N t_{1}}} $.
\end{Thm}
\begin{IEEEproof}
Please refer to Appendix \REPLYb{D}.	
\end{IEEEproof}

${\mathbf{x}}^{\rm (E,t)}$ can be interpreted as an optimal solution of Problem~\ref{prob:relaxed_prob} for a \TCOMb{master-worker distributed computation system} where $N$ workers have deterministic CPU cycle times $\mathbf{t}$, and can be treated as an approximate solution of Problem~\ref{prob:relaxed_prob}.
%Given $\mathbf{t}$, the computational complexity for calculating $\mathbf{x}^{\rm {(t)}}$ is $\mathcal{O}(N)$.

Finally, we characterize the computational complexity and suboptimality of ${\mathbf{x}}^{\rm (E,t)}$ under the assumption that the distribution of $T_n,n\in[N]$ is a shifted-exponential distribution, i.e., 
\begin{equation}
	F_T(t)=1-e^{-\mu(t-t_0)},\ t\ge t_0,\label{equ:E_shifted-exponential_distribution}
\end{equation}
where $\mu$ is the rate parameter and $t_0 \TCOMb{> 0}$ is the shift parameter.
Note that shifted-exponential distributions are widely considered in modeling stragglers in distributed computation systems~\cite{mat_vec_Lee_speeding_up,mat_vec_Draper_hierarchical_ISIT,mat_mat_Draper_hierarchical_TIT,GC_Min_Ye_communication,GC_WH_improving_RS_Codes}.
By R\'enyi representation theorem of exponential order statistics~\cite[Corrollary 2.26]{Order_Statistics}, we have
%{\setlength{\abovedisplayskip}{5pt}
%\setlength{\belowdisplayskip}{5pt}
\begin{align}
	t_n = \mathbb{E}[T_{(n)}] = \frac{1}{\mu}(H_N-H_{N-n})+t_0,\ n\in[N],\label{equ:Renyi}
\end{align}
\begin{figure*}[ht]
\normalsize{
\begin{small}
	\begin{align}
		t_{n}'\hspace{-0.0cm} 
		&=\hspace{-0.0cm}  -1 \Bigg/ \Bigg(\mu (N+1-n)\binom{N}{n-1}\hspace{-0.0cm} \sum_{i=0}^{n-1}(-1)^i\binom{n-1}{i}e^{\mu t_0 (N-n+i+1)} E_i(-\mu t_0 (N-n+i+1)) \Bigg),
		 n\in[N],\label{equ:t'}\\
		g(\mathbf{x},t,\hat{\mathbf{k}}) 
		&\triangleq  \binom{N} {\hat{k}_0,\hat{k}_1,\cdots,\hat{k}_{N-1}} \prod_{n=0}^{N-2}\left( F_T\left( \frac{t} {\frac{M}{N}b\sum_{i=0}^n (i+1)x_i} \right) \hspace{-0.1cm} - \hspace{-0.1cm} F_T\left( \frac{t} {\frac{M}{N}b\sum_{i=0}^{n+1} (i+1)x_i} \right) \right)^{\hspace{-0.1cm}\hat{k}_n} 
		\hspace{-0.3cm}F_T\left( \frac{t} {\frac{M}{N}b\sum_{i=0}^{N-1} (i+1)x_i} \right)^{\hspace{-0.1cm}\hat{k}_{N-1}}\hspace{-0.4cm}.\label{equ:g}
	\end{align}
\end{small}
\vspace{-2mm}
}\hrulefill
\vspace{-2mm}
\end{figure*}where $H_n\triangleq\sum_{i=1}^n \frac{1}{i}$ is the $n$-th harmonic number.
The computational complexity for calculating $\mathbf{t}$ \TCOMb{according to \eqref{equ:Renyi}} is $\mathcal{O}(N)$.
Given $\mathbf{t}$, the computational complexity for calculating $z^{\rm (E,t)}$ is $\mathcal{O}(N)$.
Thus, the computational complexity for calculating ${\mathbf{x}}^{\rm (E,t)}$ is $\mathcal{O}(N)$.
The suboptimality of ${\mathbf{x}}^{\rm (E,t)}$ is given as follows.
\begin{Thm}[Sub-optimality of $\mathbf{x}^{\rm (E,t)}$]
\label{thm:multiplicative_gap_1}
%\begin{small}
%\begin{align*}
	$\frac{\mathbb{E}\left[\hat{\tau}({\mathbf{x}}^{\rm (E,t)},\mathbf{T})\right]} {\hat{\tau}^*_{\rm avg-ct}} = \mathcal{O}\left(\log^2(N)\right)$, where $\hat{\tau}^*_{\rm avg-ct}$ denotes the optimal value of Problem~\ref{prob:relaxed_prob}.
%\end{align*}
%\begin{align*}
%	\frac{\mathbb{E}\left[\hat{\tau}({\mathbf{x}}^{\rm (E,f)},\mathbf{T})\right]} {\hat{\tau}^*_{\rm avg-ct}} = \mathcal{O}\left(\log(N)\right).
%\end{align*}
%\end{small}
\end{Thm}
\begin{IEEEproof}
Please refer to Appendix \REPLYb{E}.	
\end{IEEEproof} 

Theorem~\ref{thm:multiplicative_gap_1} indicates that for any model size $L$, the expected overall runtime of ${\mathbf{x}}^{\rm (E,t)}$ and the minimum expected overall runtime have a sub-linear multiplicative gap in $N$.
As \TCOMb{$\log^2(N)=5.3$ at $N=10$}, the analytical upper bound on the gap is not large at a small or moderate $N$. 
Later in Sec.~\ref{sec:Numerical_Results}, we shall see that the actual gap is very small even at $N=50$.

\subsubsection{Approximate Solution Based on Deterministic CPU Frequencies}
We approximate the objective function of Problem~\ref{prob:relaxed_prob} by replacing random vector $\mathbf{T}$ with deterministic vector $\mathbf{t}'\triangleq (t'_n)_{n\in[N]}$, where $t'_n \triangleq 1 \Big/ \mathbb{E}\left[\frac{1}{T_{(n)}}\right],n\in[N]$ (which can be numerically computed for an arbitrary \TCOMb{form of} $F_T(\cdot)$ and analytically computed for some special forms of $F_T(\cdot)$).

\begin{Prob}[Approximation of Problem~\ref{prob:relaxed_prob} at $\mathbf{t}'$]\label{prob:approx_solution_speed}
	\begin{align}
		{\mathbf{x}}^{\rm (E,f)} \triangleq \argmin_{\mathbf{x}} \ \  &\hat{\tau}(\mathbf{x},\mathbf{t}') \nonumber\\
		  \rm{s.t.}\ \  &\eqref{equ:constraint_x_1},\eqref{equ:constraint_x_3}.\nonumber
	\end{align}
\end{Prob}

%\TCOMc{By similar methods in the proof of Theorem~\ref{thm:E_approx_solution_time}, we obtain}
%the optimal solution of Problem~\ref{prob:approx_solution_speed} given as follows.
Problem~\ref{prob:approx_solution_speed} has the same structure as Problem~\ref{prob:approx_solution_time} and can be solved with the same method.
\begin{Thm} [Closed-form Optimal Solution of Problem~\ref{prob:approx_solution_speed}]\label{thm:approx_solution_speed}
$x^{\rm (E,f)}_0 \hspace{-0.00cm} = \hspace{-0.00cm} \frac{1} {t_N'}z^{\rm (E,f)} \hspace{-0.0cm},\ x^{\rm (E,f)}_n \hspace{-0.00cm} = \hspace{-0.00cm} \frac{1}{n+1}\hspace{-0cm}\Big(\hspace{-0cm}\frac{1} {t_{N-n}'} \hspace{-0.00cm} - \hspace{-0cm} \frac{1} {t_{N+1-n}'}\hspace{-0cm}\Big)\hspace{-0.00cm}z^{\rm (E,f)} \hspace{-0.0cm},\ n\in[N-1]$,
where $z^{\rm (E,f)} \triangleq \frac{L} {\sum_{n=1}^{N-1} \frac{1}{n(n+1)t_{N+1-n}'} + \frac{1}{N t_{1}'}}$.
\end{Thm}
%\vspace{0.1cm}
\begin{IEEEproof}
%	The\hspace{-0.05cm} proof\hspace{-0.05cm} is\hspace{-0.05cm} similar\hspace{-0.05cm} to\hspace{-0.05cm} that\hspace{-0.05cm} of\hspace{-0.05cm} Theorem\hspace{-0.05cm} \ref{thm:E_approx_solution_time}.
	The proof is similar to that of Theorem~\ref{thm:E_approx_solution_time} and is omitted due to page limitation.
%	$\hfill\IEEEQED$
\end{IEEEproof}

Let $F_n \triangleq \frac{1}{T_n}, n\in[N]$ denote the CPU frequencies of the $N$ workers.
Thus, ${\mathbf{x}}^{\rm (E,f)}$ can be interpreted as an optimal solution of Problem~\ref{prob:relaxed_prob} for a distributed computation system where $N$ workers have deterministic CPU frequencies \TCOMb{$\mathbf{t}^{\prime}$}, and can be treated as an approximate solution of Problem~\ref{prob:relaxed_prob}. 
Given $\mathbf{t}'$, the computational complexity of calculating $\mathbf{x}^{\rm {(E,f)}}$ is $\mathcal{O}(N)$.

Finally, we characterize the computational complexity and suboptimality of ${\mathbf{x}}^{\rm (E,f)}$ under \TCOMb{a shifted-exponential distribution as in Subsec.~\ref{sssec:Senario_1}}. 
We first derive the expression of $\mathbf{t}'$ in the following Lemma.
\begin{Lem}[\TCOMb{Parameters of Problem \ref{prob:approx_solution_speed} under a shifted-exponential distribution}]
\label{lem:t'}
%If $t_0>0$,\footnote{When $t_0=0$, $E_i(0)$ does not exist.} for all $n\in[N]$, 
We have \TCOMbr{$t_{n}'$ given in \eqref{equ:t'} as shown at the top of this page,}
%\begin{align}
%		t_{n}'\hspace{-0.0cm} 
%		=\hspace{-0.0cm}  -1 \Bigg/ \Bigg(\mu (N+1-n)\binom{N}{n-1}\hspace{-0.0cm} \sum_{i=0}^{n-1}(-1)^i\binom{n-1}{i}e^{\mu t_0 (N-n+i+1)} E_i(-\mu t_0 (N&-n+i+1)) \Bigg),\nonumber\\
%		 &n\in[N],\label{equ:t'}
%\end{align}
where $E_i(x)\triangleq \int_{-\infty}^x \frac{e^t}{t}dt$ is the exponential integral.
\end{Lem}
\begin{IEEEproof}
Please refer to Appendix \REPLYb{F}.	
\end{IEEEproof} 

%\TCOMr{The computational complexity for calculating ${\mathbf{x}}^{\rm (E,f)}$ is $\mathcal{O}(N^2)$ ($\mathcal{O}(N^2)$ for calculating $\mathbf{t}'$, $\mathcal{O}(N)$ for calculating $z^{\rm (E,f)}$).}
The computational complexity for calculating $\mathbf{t}'$ \TCOMb{according to \eqref{equ:t'}} is $\mathcal{O}(N^2)$.
Given $\mathbf{t}'$, the computational complexity for calculating $z^{\rm (E,f)}$ is $\mathcal{O}(N)$.
Thus, the computational complexity for calculating ${\mathbf{x}}^{\rm (E,f)}$ is $\mathcal{O}(N^2)$.
%Note that the computational complexity for calculating the parameters for ${\mathbf{x}}^{\rm (E,f)}$, i.e., $\mathbf{t}'$, is $\mathcal{O}(N^2)$.
%Then, we characterize the sub-optimalitiy of   ${\mathbf{x}}^{\rm (E,f)}$.
%Recall that $\hat{\tau}^*_{\rm avg-ct}$ denotes the optimal value of Problem~\ref{prob:relaxed_prob}.
The suboptimality of ${\mathbf{x}}^{\rm (E,f)}$ is given as follows.
\begin{Thm}[Sub-optimality of $\mathbf{x}^{\rm (E,f)}$]
\label{thm:multiplicative_gap_2}
%\begin{small}
%\begin{align*}
%	\frac{\mathbb{E}\left[\hat{\tau}({\mathbf{x}}^{\rm (E,t)},\mathbf{T})\right]} {\hat{\tau}^*_{\rm avg-ct}} = \mathcal{O}\left(\log^2(N)\right),
%\end{align*}
%\begin{align*}
	$\frac{\mathbb{E}\left[\hat{\tau}({\mathbf{x}}^{\rm (E,f)},\mathbf{T})\right]} {\hat{\tau}^*_{\rm avg-ct}} = \mathcal{O}\left(\log(N)\right)$, where $\hat{\tau}^*_{\rm avg-ct}$ denotes the optimal value of Problem~\ref{prob:relaxed_prob}.
%\end{align*}
%\end{small}
\end{Thm}
\begin{IEEEproof}
Please refer to Appendix \REPLYb{G}.	
\end{IEEEproof} 

Similarly, Theorem~\ref{thm:multiplicative_gap_2} indicates that for any model size $L$, the expected overall runtime of ${\mathbf{x}}^{\rm (E,f)}$ and the minimum expected overall runtime have a sub-linear multiplicative gap in $N$.
\TCOMb{In summary}, the multiplicative gap for ${\mathbf{x}}^{\rm (E,f)}$ is smaller than that for ${\mathbf{x}}^{\rm (E,t)}$, but the computational complexity for calculating ${\mathbf{x}}^{\rm (E,f)}$ is higher than that for ${\mathbf{x}}^{\rm (E,t)}$.

\section{Completion Probability Maximization}
\label{sec:Completion_Probability_Maximization}

In this section, we first formulate the maximization of the completion probability with respect to the coding parameters for coordinates.
Then, we obtain a stationary point of the continuous relaxation of the original problem.
Finally, we obtain a low-complexity closed-form approximate solution of the relaxed problem at large threshold $t$ which is close to the optimal one.   

%\subsection{Optimization of Coding Parameters and Block Coordinate Gradient Coding}
\subsection{Problem Formulation}
\label{ssec:Problem_Formulation_CDF}

We would like to maximize $P(\mathbf{s},t)$ by optimizing the coding parameters $\mathbf{s}$ for the $L$ coordinates under the constraints in \eqref{equ:constraint_s_1}.
\begin{Prob}[Completion Probability Maximization]\label{prob:CDF_original}
	\begin{align}
		P^*(t) \triangleq \max_{\mathbf{s}} \ \  &P(\mathbf{s},t) \nonumber\\
		 {\rm s.t.}\ \  &\eqref{equ:constraint_s_1},\nonumber
	\end{align}
\end{Prob}
where $P(\mathbf{s},t)$ is given by~\eqref{equ:CDF_fobj_s}.

The objective function $P(\mathbf{s},t)$ is non-convex 
and has a large number of summands for large $N$ or $L$.
In addition, the number of variables (model size $L$) is usually quite large.
Hence, Problem~\ref{prob:CDF_original} is a challenging non-convex problem.
First, we characterize the monotonicity of an optimal solution of Problem~\ref{prob:CDF_original}.

\begin{Lem}[Monotonicity of Optimal Solution of Problem~\ref{prob:CDF_original}]\label{lem:CDF_monotonicity}
	\Jwqm{The optimal solution} $\mathbf{s^*}\triangleq (s^*_l)_{l\in [L]}$ of Problem~\ref{prob:CDF_original} satisfies $ s^*_1 \le s^*_2 \le \cdots \le s^*_L $.
\end{Lem}

\begin{IEEEproof}
	Please refer to Appendix \REPLYb{H}.
\end{IEEEproof}

Notice that Lemma \ref{lem:CDF_monotonicity} resembles Lemma \ref{lem:E_monotonicity}.
Thus, similar conclusions can be drawn.
Next, based on Lemma \ref{lem:CDF_monotonicity}, we transform Problem~\ref{prob:CDF_original} to an equivalent problem with $N$ variables, which optimizes the partition of the $L$ coordinates into $N$ blocks.

\begin{Prob}[Equivalent Problem of Problem~\ref{prob:CDF_original}]\label{prob:CDF_equivalent}
	\begin{align}
		\hat{P}^*(t) \triangleq \max_{\mathbf{x}} \ \  &\hat{P}(\mathbf{x},t) \nonumber\\
		 {\rm s.t.}\ \  &\eqref{equ:constraint_x_1}, \eqref{equ:constraint_x_2},\nonumber
	\end{align}
\end{Prob}
where $\hat{P}(\mathbf{x},t)$ is given by
%\begin{equation}
%	\hat{P}(\mathbf{x},t) \triangleq \sum_{\hat{\mathbf{k}} \in \hat{\mathcal{K}}} \binom{N} {\hat{k}_0,\hat{k}_1,\cdots,\hat{k}_{N-1}} \prod_{n=0}^{N-1}\left( \hat{F}_n(\mathbf{x},t)-\hat{F}_{n+1}(\mathbf{x},t) \right)^{\hat{k}_n},\label{equ:CDF_fobj_x}
%\end{equation}
\begin{equation}
	\hat{P}(\mathbf{x},t) \triangleq \sum_{\hat{\mathbf{k}} \in \hat{\mathcal{K}}(N)} g(\mathbf{x},t,\hat{\mathbf{k}}),\label{equ:CDF_fobj_x}
\end{equation}
with  
$\hat{\mathbf{k}} \triangleq (\hat{k}_i)_{i\in\{0\}\cup[N-1]}$,
\begin{small}
\begin{equation}
	\hat{\mathcal{K}}(N) \triangleq \Bigg\{ \hat{\mathbf{k}}\in \mathbb{N}^{N} : \sum_{i=0}^{n-1} \hat{k}_i \le n, n\in[N], \sum_{i=0}^{N-1} \hat{k}_i=N \Bigg\},\label{equ:hat_K}
\end{equation}
\end{small}and \TCOMbr{$g(\mathbf{x},t,\hat{\mathbf{k}})$ is given in \eqref{equ:g} as shown at the top of this page.}
%\begin{align*}
%	g(\mathbf{x},t,\hat{\mathbf{k}}) 
%	&\triangleq  \binom{N} {\hat{k}_0,\hat{k}_1,\cdots,\hat{k}_{N-1}} \prod_{n=0}^{N-2}\left( F_T\left( \frac{t} {\frac{M}{N}b\sum_{i=0}^n (i+1)x_i} \right) - F_T\left( \frac{t} {\frac{M}{N}b\sum_{i=0}^{n+1} (i+1)x_i} \right) \right)^{\hat{k}_n} \\
%	&\quad\times F_T\left( \frac{t} {\frac{M}{N}b\sum_{i=0}^{N-1} (i+1)x_i} \right)^{\hat{k}_{N-1}}.
%\end{align*}
%\TCOMc{Here $\binom{n}{k_1,k_2,\cdots,k_r} \triangleq \frac{n!}{k_1! k_2! \cdots k_r!}$ denotes the multinomial coefficient, where $\sum_{i=1}^r k_i = n$.} 
%For ease of exposition, we let $F_T\left( \frac{t} {\frac{M}{N}b\sum_{i=0}^{N+1} (i+1)x_i} \right) = 0$ and omit $t$ in function $g(\cdot)$.

\begin{Thm} [Equivalence between Problem~\ref{prob:CDF_original} and Problem~\ref{prob:CDF_equivalent}]\label{thm:CDF_prob_equvalence}
	An optimal solution of Problem~\ref{prob:CDF_original}, denoted by $\mathbf{s}^*=(s^*_l)_{l\in [L]}$, and an optimal solution of Problem~\ref{prob:CDF_equivalent}, denoted by $\mathbf{ x}^*=( x^*_n)_{n\in [N-1]}$, satisfy \eqref{equ:change_of_var_x} and \eqref{equ:change_of_var_s},
	and their optimal values, $P^*(t)$ and $\hat{P}^*(t)$, satisfy $P^*(t) = \hat{P}^*(t)$.
%	\footnote{By a slight abuse of notation, here we use the same letters $\mathbf{s}^*$ and $\mathbf{x}^*$ as the ones in Sec.~\ref{ssec:Problem_Formulation_E}.}
\end{Thm}

\begin{IEEEproof}
	Please refer to Appendix \REPLYb{I}.
\end{IEEEproof}

Theorem \ref{thm:CDF_prob_equvalence} resembles Theorem \ref{thm:E_prob_equvalence}, and hence similar conclusions can be drawn. 
We can solve Problem \ref{prob:CDF_equivalent} whose number of variables, $N$, is usually much smaller than the number of variables of Problem \ref{prob:CDF_original}, $L$.
By relaxing the integer constraints in~\eqref{equ:constraint_x_2}, we obtain a more tractable problem compared with Problem~\ref{prob:CDF_original}.

\begin{Prob}[Relaxed Continuous Problem of Problem~\ref{prob:CDF_equivalent}]\label{prob:CDF_relaxed}
	\begin{align}
		\hat{P}^*_{\rm ct}(t) \triangleq \max_{\mathbf{x}} \ \  & \hat{P}(\mathbf{x},t)\nonumber\\
%		\hat{\tau}^*_{\rm avg-ct} \triangleq \min_{\mathbf{ x}} \ \  &\mathbb{E}\left[ \hat{\tau}(\mathbf{ x},\mathbf{T}) \right] \nonumber\\
		  \rm{s.t.}\ \  &\eqref{equ:constraint_x_1},\eqref{equ:constraint_x_3}. \nonumber		
	\end{align}
\end{Prob}
%Let $\mathbf{x}^{\rm (E,opt)}$ denote an optimal solution of Problem~\ref{prob:relaxed_prob}.

As illustrated in Sec.~\ref{ssec:Problem_Formulation_E}, the rounding method \cite[pp. 386]{Boyd_cvxbook} can be applied to obtain a good approximate solution of Problem~\ref{prob:CDF_original} from an optimal solution of Problem~\ref{prob:CDF_relaxed}.
Thus, in the following section, we focus on solving the relaxed problem in Problem~\ref{prob:CDF_relaxed}.

\subsection{Stationary Point}
\label{ssec:CDF_Stationary_Point}

First, we calculate the number of summands in $\hat{P}(\mathbf{x},t)$. 
%is prohibitively large when $N$ is large.
\begin{Lem}\label{lem:CDF_equi_stoch_prob}
	The number of summands in $\hat{P}(\mathbf{x},t)$ is given by
	\begin{equation}
		\left| \hat{\mathcal{K}}(N) \right| 
		= \frac{2}{N+1}\binom{2N-1}{N-1}
		\ge 2^{N-1},\label{equ:num_summands}
	\end{equation}
	where the the equality holds if and only if $N=1$.
\end{Lem}

%$(a)$ is due to $\left| \hat{\mathcal{K}} \right| = \frac{2}{N+1}\binom{2N-1}{N-1} = \prod_{i=2}^N\frac{i+N}{i} \ge 2^{N-1}$ and

\begin{IEEEproof}
	Please refer to Appendix \REPLYb{J}.
\end{IEEEproof}

As the number of summands in $\hat{P}(\mathbf{x},t)$ is prohibitively large when $N$ is large, directly tackling Problem~\ref{prob:CDF_equivalent} is not computationally efficient.
We hence solve its equivalent stochastic version to reduce the computation time.
%We consider the following stochastic problem.
\begin{Prob} [Equivalent Stochastic Problem of Problem~\ref{prob:CDF_relaxed}]\label{prob:CDF_stoch}
\setlength{\abovedisplayskip}{-2pt}
\setlength{\belowdisplayskip}{-0pt}
	\begin{align}
		\max_{\mathbf{x}} \ \  
%		&f(\mathbf{x},t) \triangleq
		&\mathbb{E}\left[ g(\mathbf{x},t,\bm\xi) \right] \nonumber\\
		{\rm s.t.}\ \  	&\eqref{equ:constraint_x_1},\eqref{equ:constraint_x_3},\nonumber
	\end{align}
\end{Prob}
where random vector $\bm\xi\triangleq (\xi_n)_{n=0,1,\cdots,N-1}$ follows the uniform probability distribution on \TCOMb{$\hat{\mathcal{K}}(N)$}.
It is obvious that $\hat{P}(\mathbf{x},t)=\big| \hat{\mathcal{K}}(N) \big| \mathbb{E}\left[ g(\mathbf{x},t,\bm\xi) \right]$, implying that Problem~\ref{prob:CDF_relaxed} and Problem~\ref{prob:CDF_stoch} are equivalent.
Problem~\ref{prob:CDF_stoch} is a stochastic non-convex problem 
with deterministic convex constraints.
We obtain a stationary point of Problem~\ref{prob:CDF_stoch} by using SSCA~\cite{yang2016parallel}.
The idea is to iteratively solve a sequence of convex approximations of Problem~\ref{prob:CDF_stoch}, each of which is obtained by replacing the objective function with a convex surrogate function \TCOMb{based on a randomly generated realization of $\bm \xi$}.
Specifically, in the $i$-th iteration, generate a realization of  $\bm\xi$, denoted by $\bm{\xi}^{(i)}$, and approximate Problem~\ref{prob:CDF_stoch} with the following convex problem.
\begin{Prob} [Approximate Convex Problem of Problem~\ref{prob:CDF_stoch} in the $i$-th Iteration]\label{prob:CDF_SSCA_ith_1}
	\begin{align}
		\hat{\mathbf{x}}^{(i)}\triangleq \argmax_{\mathbf{x}} \ \  &f^{(i)}(\mathbf{x},\mathbf{x}^{(i-1)},t,\bm{\xi}^{(i)}) \nonumber\\
		{\rm s.t.}\ \  	&\eqref{equ:constraint_x_1},\eqref{equ:constraint_x_3},\nonumber
	\end{align}
\end{Prob}
where \TCOMb{$f^{(i)} (\mathbf{x},\mathbf{x}^{(i-1)},t,\bm{\xi}^{(i)}) \hspace{-0cm} \triangleq 
\hspace{-0cm} {\mathbf{h}^{(i)}}^{\rm T} \hspace{-0cm} (\mathbf{x}-\mathbf{x}^{(i-1)}) \hspace{-0cm}
- \hspace{-0cm} \tau \|\mathbf{x}-\mathbf{x}^{(i-1)}\|_2^2$} is a concave surrogate function of $\mathbb{E}\left[ g(\mathbf{x},t,\bm\xi) \right]$ around $\mathbf{x}^{(i-1)}$,
with $\mathbf{h}^{(i)} \triangleq (1-\sigma^{(i)}) \mathbf{h}^{(i-1)} + \sigma^{(i)} \nabla g(\mathbf{x}^{(i)},t,\bm{\xi}^{(i)})\in\mathbb{R}^N $ as an approximation of $\nabla \mathbb{E}[g(\mathbf{x},\mathbf{\bm\xi})]$.
%\TCOMc{$f^{(i)} (\mathbf{x},\mathbf{x}^{(i-1)},t,\bm{\xi}^{(i)}) \hspace{-0cm} \triangleq 
% \sigma^{(i)} \nabla g(\mathbf{x}^{(i-1)},t,\bm{\xi}^{(i)})^{\rm T} \hspace{-0cm} (\mathbf{x}-\mathbf{x}^{(i-1)}) \hspace{-0cm}
%+ \hspace{-0cm} (1-\sigma^{(i)}) {\mathbf{h}^{(i-1)}}^{\rm T} \hspace{-0cm} (\mathbf{x}-\mathbf{x}^{(i-1)}) \hspace{-0cm}
%- \hspace{-0cm} \tau \|\mathbf{x}-\mathbf{x}^{(i-1)}\|_2^2$}
Here, $\{\sigma^{(i)}\}$ is a positive diminishing stepsize sequence satisfying \eqref{equ:stepsize}, and $\mathbf{h}^{(0)} = \mathbf{0}$.
\TCOMb{Problem~\ref{prob:CDF_SSCA_ith_1} is a convex quadratic programming.}
By the KKT conditions \cite[Sec. 5.5.3]{Boyd_cvxbook}, we obtain the optimal solution of Problem~\ref{prob:CDF_SSCA_ith_1}.

\begin{Lem}[Optimal Solution of Problem~\ref{prob:CDF_SSCA_ith_1}]
\label{lem:CDF_semi_closed_form}
	The optimal solution $\hat{\mathbf{x}}^{(i)}\triangleq \left(\hat{x}^{(i)}_n\right)_{n=0,1,\cdots,N-1}$ of Problem~\ref{prob:CDF_SSCA_ith_1} is given by
	\begin{align}
		x_n^{(i)} = \max\bigg\{x^{(i-1)}_n + \frac{1}{2\tau}\bigg( \sigma^{(i)} \frac{\partial}{\partial x_n} g(\mathbf{x}^{(i-1)},{\bm\xi}^{(i)}) \nonumber\\
		- (1-\sigma^{(i)})h_n^{(i-1)}  \bigg) + \frac{1}{\tau}\lambda,0\bigg\},\ n=0,1,\cdots,N-1,
		\label{equ:CDF_semi_closed_form_1}
	\end{align}
	where 
	$\lambda$ satisfies
	\begin{align}
		\sum_{n=0}^{N-1} \max\bigg\{x^{(i-1)}_n + \frac{1}{2\tau}\bigg( \sigma^{(i)} \frac{\partial}{\partial x_n} g(\mathbf{x}^{(i-1)},{\bm\xi}^{(i)}) \nonumber\\
		- (1-\sigma^{(i)})h_n^{(i-1)}  \bigg) + \frac{1}{\tau}\lambda,0\bigg\} = L.
		\label{equ:CDF_semi_closed_form_2}
	\end{align}
\end{Lem}
%\begin{IEEEproof}
%\TCOMbr{The proof is omitted due to page limitation.}
%\end{IEEEproof}

\TCOMb{Note that} $\lambda$ can be obtained by bisection search.
Given $\hat{\mathbf{x}}^{(i)}$, we update $\mathbf{x}^{(i)}$ according to
\begin{align}
	\mathbf{x}^{(i)} = \gamma^{(i)} \hat{\mathbf{x}}^{(i)} + (1-\gamma^{(i)}) \mathbf{x}^{(i-1)},\label{equ:SSCA_smooth}
\end{align}
where $\{\gamma^{(i)}\}$ is a positive diminishing stepsize sequence satisfying
\begin{equation}
	\gamma^{(i)}>0, 
	\gamma^{(i)} \rightarrow 0, 
	\sum_{i=1}^{\infty}\gamma^{(i)}=\infty, 
	\sum_{i=1}^{\infty}\left(\gamma^{(i)}\right)^2<\infty, 
	\lim_{i\rightarrow\infty}\frac{\gamma^{(i)}}{\sigma^{(i)}} = 0. \nonumber\label{equ:stepsize}
\end{equation}

We discuss how to generate ${\bm\xi}^{(i)}$ according to the uniform distribution on $\hat{\mathcal{K}}(N)$.
A naive way is to first determine and store all the elements in $\hat{\mathcal{K}}(N)$ and \TCOMb{then} uniformly select an element in $\hat{\mathcal{K}}(N)$.
By \eqref{equ:num_summands} and \TCOMb{the uniform distribution of $\bm\xi^{(i)}$} on $\hat{\mathcal{K}}(N)$, we know that the time complexity and space complexity are both $\mathcal{O}(N2^N)$.\footnote{The time complexity of generating a discrete random variable with $N$ outcomes in the sample space is $\mathcal{O}(\log{N})$ \cite{wiki_random_number_sampling}.}
To handle this issue, we propose a more efficient method \TCOMb{which generates} ${\bm\xi}^{(i)}$ with much lower time complexity and space complexity.
\TCOMb{First, we characterize the \REPLYb{probability mass function} (PMF) of $\bm\xi$.}

\begin{Thm}[\TCOMb{Joint PMF and conditional PMF of $\xi_n,n=0,\cdots,N-1$}]
\label{thm:CDF_conditional_probability}	
	$\Pr\left({\bm\xi}=\hat{\mathbf{k}}\right)=\frac{1}{|\hat{\mathcal{K}}(N)|},\ \hat{\mathbf{k}}\in\hat{\mathcal{K}}(N)$
	if and only if
	\vspace{-0.1cm}
	\begin{small}
	\begin{align}
		\Pr\left(\xi_0=\hat{k}_0\right) = \frac{\left(3-\hat{k}_0\right)(N+1)\binom{2N-\hat{k}_0}{N-1}}{2\left(N+2-\hat{k}_0\right)\binom{2N-1}{N-1}},\ \hat{k}_0\in\{0,1\},\label{equ:CDF_conditional_probability_1}
	\end{align}
\begin{figure*}[ht]
\normalsize{
\begin{small}
\begin{align}
	E_{\rm lg}^{\rm ub} (\mathbf{x},t) 
	\triangleq \left(\sum_{n=0,1,\cdots,N-1}\binom{N} {n+2} e^{\mu t_0(n+2)}\right) e^{-\frac{\mu N}{Mb}\underset{n=0,1,\cdots,N-1}{\min} \frac{n+1}{\sum_{i=0}^n (i+1)x_i} t}.\label{equ:CDF_ub}
\end{align}
\end{small}
\vspace{-2mm}
}\hrulefill
\vspace{-2mm}
\end{figure*}
	\begin{align}
		&\Pr\left(\xi_n=\hat{k}_n | \xi_0=\hat{k}_0,\cdots,\xi_{n-1}=\hat{k}_{n-1}\right)
		\hspace{-0.1cm}\nonumber\\
		&=\hspace{-0.1cm}\frac{\left(n+3-\sum_{i=0}^n\hat{k}_i\right) \hspace{-0.1cm} \left(N+n+1-\sum_{i=0}^{n-1}\hat{k}_i\right) \hspace{-0.1cm} \binom{2N+n-\sum_{i=0}^n\hat{k}_i}{N-1}} {\left(n+2-\sum_{i=0}^{n-1}\hat{k}_i\right) \hspace{-0.1cm} \left(N+n+2-\sum_{i=0}^n\hat{k}_i\right) \hspace{-0.1cm} \binom{2N+n-1-\sum_{i=0}^{n-1}\hat{k}_i}{N-1}},\nonumber\\
		&\hat{k}_n \in \Bigg\{0,1,\cdots,n+1-\sum_{i=0}^{n-1}\hat{k}_i\Bigg\},\
		\Big(\hat{k}_0,\cdots,\hat{k}_{n-1}\Big) \in \Bigg\{ \hat{\mathbf{k}}\in \mathbb{N}^{n} : \nonumber\\
		&\sum_{i=0}^{j-1} \hat{k}_i \le n, j\in[n] \Bigg\},\
		n \in [N-1].
		\label{equ:CDF_conditional_probability_2}
	\end{align}
	\end{small}
\end{Thm}

\begin{IEEEproof}
	Please refer to Appendix \REPLYb{K}.
\end{IEEEproof}

\TCOMb{Then}, based on Theorem~\ref{thm:CDF_conditional_probability}, we generate ${\bm\xi}^{(i)}$ by generating $\xi_0,\cdots,\xi_{N-1}$ \TCOMb{successively according to \eqref{equ:CDF_conditional_probability_1} and \eqref{equ:CDF_conditional_probability_2}}.
From \eqref{equ:CDF_conditional_probability_1} and \eqref{equ:CDF_conditional_probability_2}, we know that the time complexity and space complexity of the proposed method for generating ${\bm\xi}^{(i)}$ are $\mathcal{O}(N^2)$ and $\mathcal{O}(N)$, respectively.
Obviously, both the time complexity and space complexity of the new method are much lower than those of the naive method mentioned above.
%($\mathcal{O}(N2^N)$ and $\mathcal{O}(N2^N)$)

Finally, we summarize the details of the algorithm for obtaining a stationary point of Problem~\ref{prob:CDF_stoch} in Alg.~\ref{alg:CDF_SSCA}.
Here, $I$ denotes the total number of iterations, irrelevant to the problem size~\cite{yang2016parallel}.
%The computational complexity of each iteration of Alg.~\ref{alg:CDF_SSCA} is analyzed as follows.
The computational complexity of Steps 4 in Alg.~\ref{alg:CDF_SSCA} is $\mathcal{O}(N)$,
and the computational complexities of Steps 6 and 7 in Alg.~\ref{alg:CDF_SSCA} are $\mathcal{O}(N)$.
Thus, the computational complexity of Alg.~\ref{alg:CDF_SSCA} is $\mathcal{O}(N^2)$.
It has been shown in \cite{yang2016parallel} that every limit point of the sequence $\left\{x^{(i)}\right\}$ generated by Alg.~\ref{alg:CDF_SSCA} is a stationary point, denoted by $\mathbf{x}^{\rm (CDF,st)}$, of Problem~\ref{prob:CDF_stoch} almost surely.

\begin{algorithm}[t]
	\caption{Algorithm for Obtaining a Stationary Point of Problem~\ref{prob:CDF_stoch}}
\begin{small}
	\begin{algorithmic}[1]
		\STATE \textbf{initialization}: 
		Choose any feasible point $\mathbf{x}^{(0)}$ of Problem~\ref{prob:CDF_stoch}.
		\STATE \textbf{for} $i=1,2,\cdots,I$ \textbf{do}
		\STATE \quad \textbf{for} $n=0,1,\cdots,N-1$ \textbf{do}
		\STATE \quad \quad Generate $\xi^{(i)}_n$ \TCOMb{for given} $\xi_0,\cdots,\xi_{n-1}$, according to \eqref{equ:CDF_conditional_probability_2}.
		\STATE \quad \textbf{end for}
		\STATE \quad \TCOMb{Compute $\lambda$ by solving the equation in \eqref{equ:CDF_semi_closed_form_2} with bisection search,}
		and compute $\hat{\mathbf{x}}^{(i)}$ according to \eqref{equ:CDF_semi_closed_form_1}.  
		\STATE \quad Update $\mathbf{x}^{(i)}$ according to \eqref{equ:SSCA_smooth}.
		\STATE \textbf{end for}
	\end{algorithmic}\label{alg:CDF_SSCA}
\end{small}	
\end{algorithm}

%\subsection{Asumptotically Optimal Solution at Large $t$}
%\label{ssec:Asympototically_Optimal_Solution_at_Large_t}

\subsection{Approximate Solution at Large \TCOMb{threshold}}
\label{ssec:CDF_Approximate_Solutions}

In this part, assuming that the distribution of \TCOMb{$T_n,n\in[N]$} follows the shifted-exponential distribution given in \eqref{equ:E_shifted-exponential_distribution}, we obtain a low-complexity approximate solution of Problem \ref{prob:CDF_relaxed} at large $t$.
First, we obtain an asymptotic approximation of $1-\hat{P}(\mathbf{x},t)$ at large $t$.\footnote{We obtain an asymptotic approximation of $1-\hat{P}(\mathbf{x},t)$ which goes to $0$ as $t\rightarrow\infty$.}
\begin{Lem} [Asymptotic Failure Probability at Large $t$]\label{lem:CDF_large_t_asymp_completion_probability}
	$1-\hat{P}(\mathbf{x},t) \overset{t\rightarrow\infty}{\sim} E_{\rm lg}(\mathbf{x},t)$,\footnote{$f(x) \overset{t\rightarrow\infty}{\sim} g(x)$ means $\underset{t\rightarrow\infty}{\lim} \frac{f(x)}{g(x)} = 1$.} where
	\begin{small} 
	\begin{equation}
		E_{\rm lg} (\mathbf{x},t) \hspace{-0.1cm}
		\triangleq \hspace{-0.1cm} \left(\sum_{n\in\mathcal{N}_2(\mathbf{x})} \hspace{-0.2cm} \binom{N} {n+2} e^{\mu t_0(n+2)} \hspace{-0.1cm} \right) \hspace{-0.1cm} e^{-\frac{\mu N}{Mb}\underset{n\in\mathcal{N}_1(\mathbf{x})}{\min}\frac{n+1}{\sum_{i=0}^n (i+1)x_i}t} \hspace{-0.1cm}.\label{equ:CDF_large_completion_probability}
	\end{equation}
    \end{small}Here, $\mathcal{N}_1(\mathbf{x}) \triangleq \left\{n\in[N-1] : x_n\neq 0\right\}$ and $\mathcal{N}_2(\mathbf{x}) \triangleq \underset{n\in\mathcal{N}_1(\mathbf{x})}{\argmin} \frac{n+1}{\sum_{i=0}^{n} (i+1)x_i}$ (i.e., $\mathcal{N}_2(\mathbf{x}) \triangleq \Big\{ n\in\mathcal{N}_1(\mathbf{x}) : \frac{n+1}{\sum_{i=0}^n(i+1)x_i}=\underset{n^{\prime}\in\mathcal{N}_1(\mathbf{x})}{\min} \frac{n^{\prime}}{\sum_{i=0}^{n^{\prime}}(i+1)x_i} \Big\}$).
\end{Lem}
\begin{IEEEproof}
	Please refer to Appendix \REPLYb{L}.
\end{IEEEproof}

By Lemma~\ref{lem:CDF_large_t_asymp_completion_probability}, Problem~\ref{prob:CDF_relaxed} is asymptotically equivalent to the following problem, as $t\rightarrow\infty$.

\begin{Prob}[Asymptotically Equivalent Problem of Problem~\ref{prob:CDF_relaxed} as $t\rightarrow\infty$]\label{prob:CDF_large_t}
	\begin{align}
			\min_{\mathbf{x}} \ \  &E_{\rm lg} (\mathbf{x},t) \nonumber\\
		  	{\rm s.t.}\ \  &\eqref{equ:constraint_x_1}, \eqref{equ:constraint_x_3}. \nonumber
	\end{align}
\end{Prob}

Problem~\ref{prob:CDF_large_t} is still a quite challenging non-convex problem. 
For tractability, we approximate $E_{\rm lg} (\mathbf{x},t)$ with an upper bound, denoted as $E_{\rm lg}^{\rm ub} (\mathbf{x},t)$, which is given by \TCOMbr{\eqref{equ:CDF_ub} as shown at the top of this page.}
%\begin{small}
%\begin{align}
%	E_{\rm lg}^{\rm ub} (\mathbf{x},t) 
%	\triangleq \left(\sum_{n=0,1,\cdots,N-1}\binom{N} {n+2} e^{\mu t_0(n+2)}\right) \nonumber\\
%	\times e^{-\frac{\mu N}{Mb}\underset{n=0,1,\cdots,N-1}{\min} \frac{n+1}{\sum_{i=0}^n (i+1)x_i} t}.\label{equ:CDF_ub}
%\end{align}
%\end{small}
It is clear that $\hat{P}_{\rm lg} (\mathbf{x},t) \le E_{\rm lg}^{\rm ub} (\mathbf{x},t)$, as $\mathcal{N}_2(\mathbf{x}) \subseteq \mathcal{N}_1(\mathbf{x}) \subseteq \{0,1,\cdots,N-1\}$.
Therefore, we can approximate Problem  \ref{prob:CDF_large_t} with the following problem.
\begin{Prob}[Approximate Problem of Problem \ref{prob:CDF_large_t}]\label{prob:CDF_large_t_approx}
	\begin{align}
			\mathbf{x}^{\rm (CDF,lg)} 
			\triangleq \argmin_{\mathbf{x}} \ \  &E_{\rm lg}^{\rm ub} (\mathbf{x},t) \nonumber\\
		  	{\rm s.t.}\ \  &\eqref{equ:constraint_x_1}, \eqref{equ:constraint_x_3}. \nonumber
	\end{align}
%	where $\mathbf{x}^{\rm (CDF,lg)} \triangleq \left( x^{\rm (CDF,lg)}_n \right)_{n\in{\{0\}}\cup[N-1]}$.
\end{Prob}

Then, by contradiction and construction, we obtain an optimal solution of Problem \ref{prob:CDF_large_t_approx}.
\begin{Thm}[Closed-form Optimal Solution of Problem \ref{prob:CDF_large_t_approx}]\label{thm:CDF_large_t_approx_closedform}
	An optimal solution of Problem \ref{prob:CDF_large_t_approx} is given by
	\begin{equation}
		x_0^{\rm (CDF,lg)} = \frac{L}{H_N},\ x_n^{\rm (CDF,lg)} = \frac{1}{n+1}x_0^{\rm (CDF,lg)},\ n\in[N-1],\label{equ:CDF_large_t_approx_closedform}
	\end{equation}
	where $H_n\triangleq\sum_{i=1}^n \frac{1}{i}$ is the $n$-th harmonic number.
\end{Thm}
\begin{IEEEproof}
	Please refer to Appendix \REPLYb{M}.
%\TCOMp{The proof is omitted due to page limitation.}
\end{IEEEproof}

From Theorem~\ref{thm:CDF_large_t_approx_closedform}, we can see that $x^{\rm (CDF,lg)}_n$ decreases with $n$,
i.e., blocks with less redundancy include more coordinates.
Obviously, the computational complexity for calculating $\mathbf{x}^{\rm (CDF,lg)}$ \TCOMb{according to} \eqref{equ:CDF_large_t_approx_closedform} is $\mathcal{O}(N)$.

\section{Numerical Results}
\label{sec:Numerical_Results}

\begin{figure*}[t]
\begin{center}
  \subfigure[\scriptsize{$L=2\times 10^4$ and $\mu=10^{-3}$.}\label{subfig:E_vs_N}]
%  {\resizebox{5.9cm}{!}{\includegraphics{J_fig/fig_E_vs_N.eps}}}
  {\resizebox{4.35cm}{!}{\includegraphics{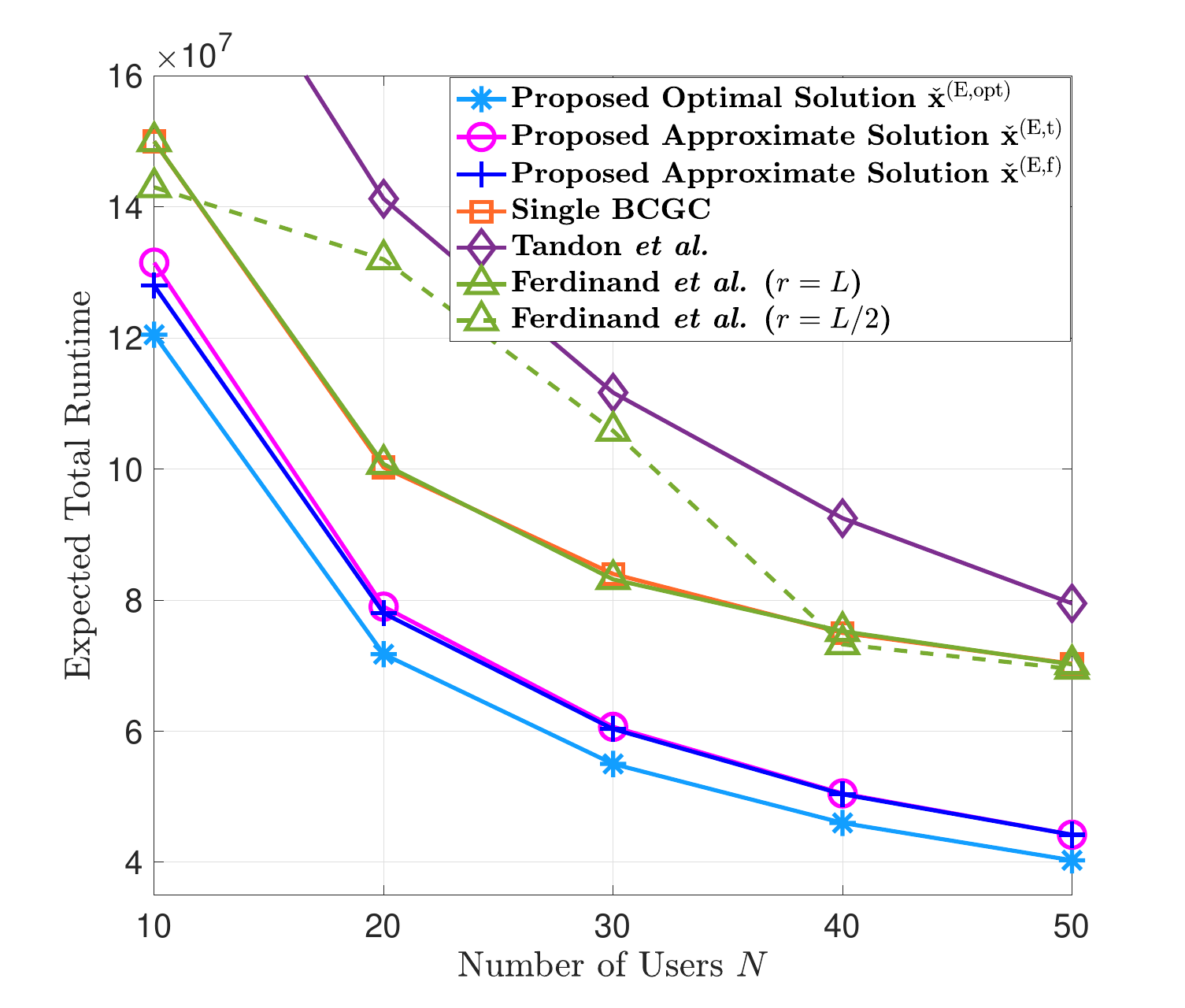}}}
  \ \ 
  \subfigure[\scriptsize{$N=20$ and $L=2\times 10^4$.}\label{subfig:E_vs_mu}]
  {\resizebox{4.35cm}{!}{\includegraphics{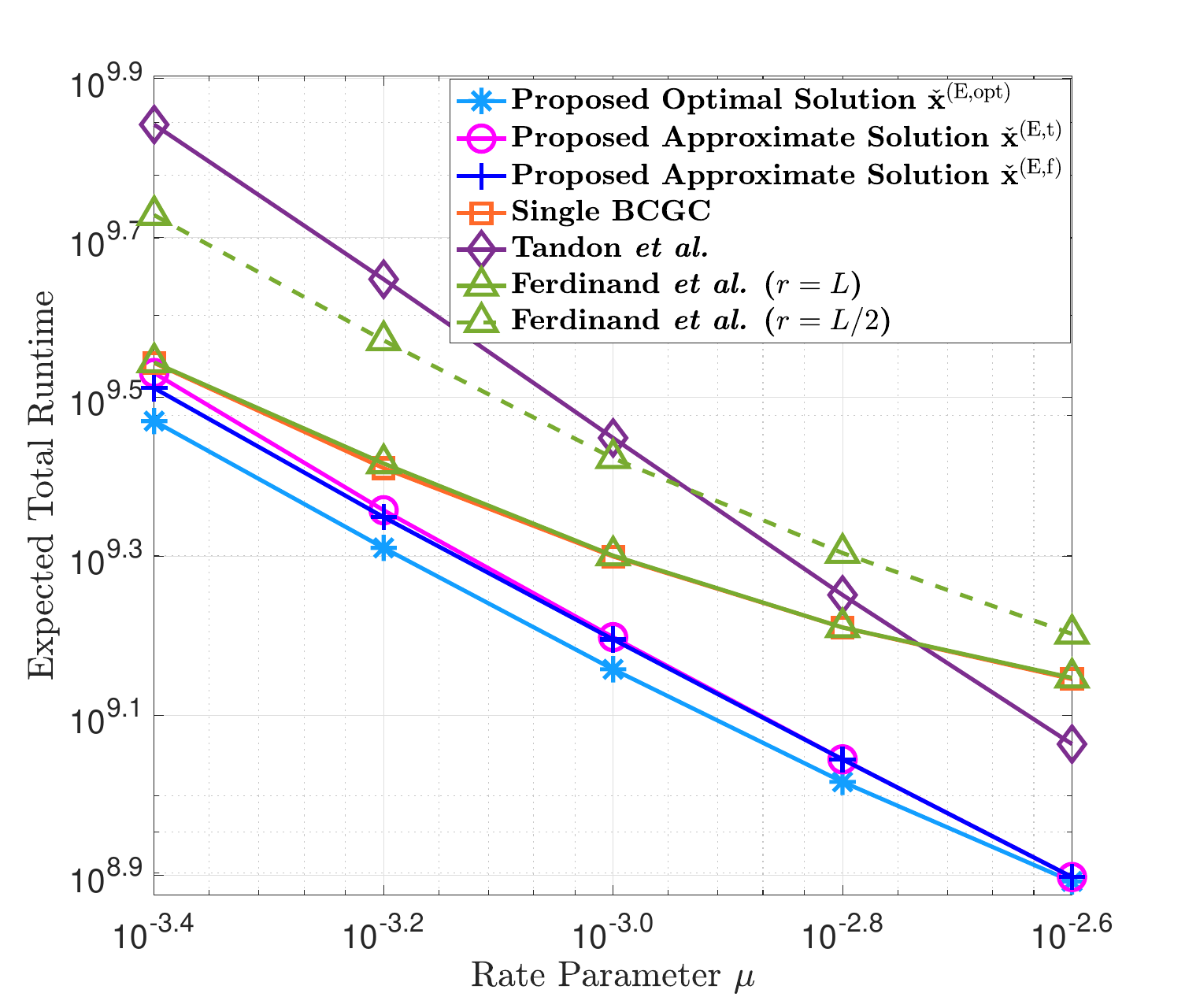}}}
  \ \ 
  \subfigure[\scriptsize{$N=20$ and $\mu=10^{-3}$.}\label{subfig:E_vs_L}]
  {\resizebox{4.35cm}{!}{\includegraphics{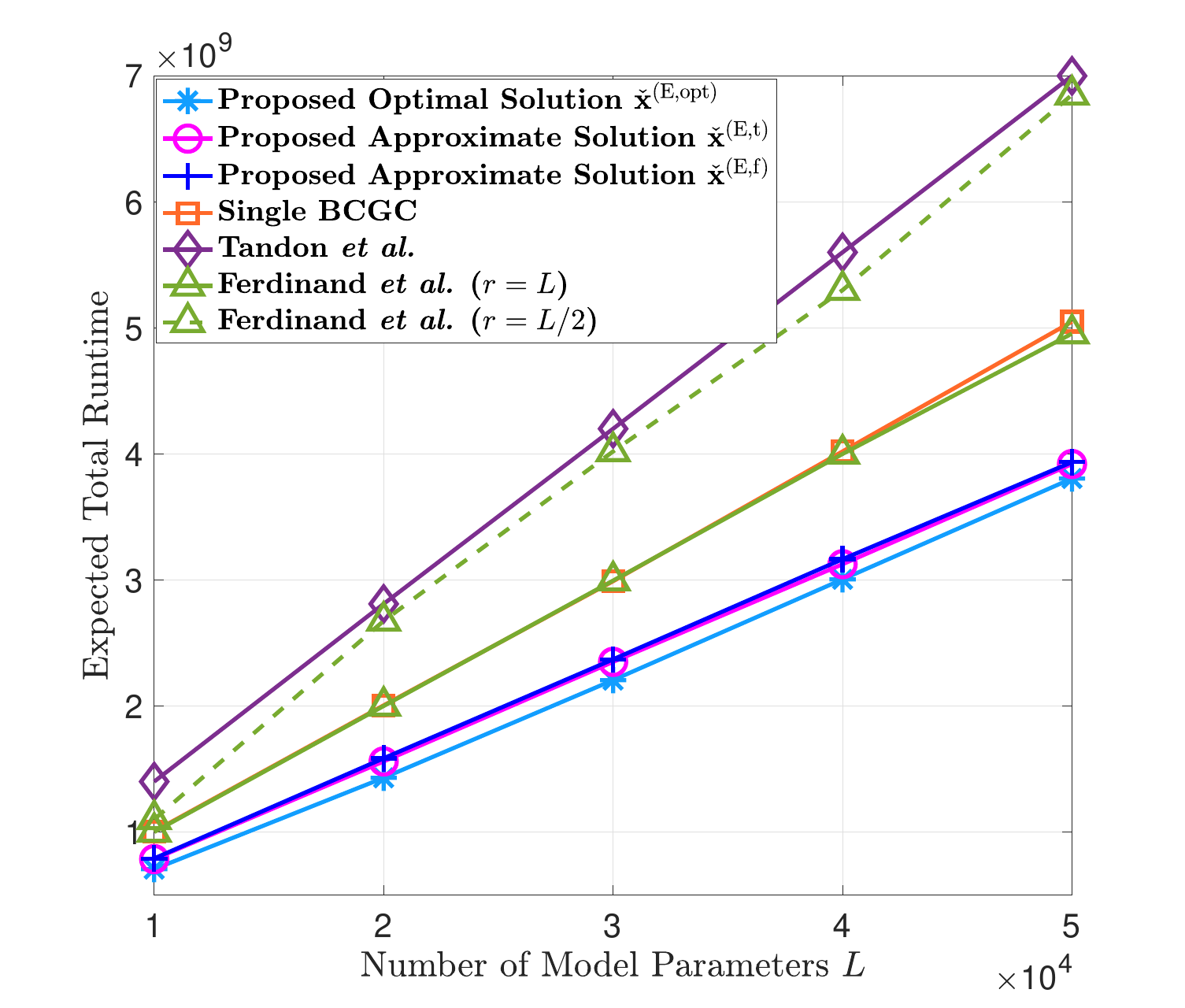}}}
  \end{center}
  \vspace{-0.3cm}
         \caption{Expected total runtime versus $N$, $\mu$, and $L$ at $t_0=100$, respectively. 
         }\label{fig:simulation_E}
         \vspace{-0.3cm}
\end{figure*}

In this section, we numerically evaluate the performance of the integer-valued approximations (obtained using the rounding method in~\cite[pp. 386]{Boyd_cvxbook}) of the proposed solutions. 
Let $\check{\mathbf{x}}^{\rm (E,opt)}$, $\check{\mathbf{x}}^{\rm (E,t)}$, and $\check{\mathbf{x}}^{\rm (E,f)}$ denote the integer-valued approximations of the proposed solutions $\mathbf{x}^{\rm (E,opt)}$, ${\mathbf{x}}^{\rm (E,t)}$, and ${\mathbf{x}}^{\rm (E,f)}$ in \TCOMb{Sec.~\ref{sec:Expected_Overall_Runtime_Minimization}}.
Let $\check{\mathbf{x}}^{\rm (CDF,st)}$ and $\check{\mathbf{x}}^{\rm (CDF,lg)}$ denote the integer-valued approximations of the proposed solutions ${\mathbf{x}}^{\rm (CDF,st)}$ and $\mathbf{x}^{\rm (CDF,lg)}$ in \TCOMb{Sec.~\ref{sec:Completion_Probability_Maximization}}.
We consider the subsequent four baseline schemes.
Single-block coordinate gradient coding (BCGC) corresponds to the integer-valued solution obtained by solving Problem~\ref{prob:equivalent_1} (or Problem~\ref{prob:CDF_equivalent}) with extra constraints $\|\mathbf{x}\|_0=1$ and rounding the optimal solution using the rounding method in~\cite[pp. 386]{Boyd_cvxbook}.
Notice that single-BCGC can be viewed as an optimized version of the gradient coding scheme for full stragglers in~\cite{GC_Tandon_exact}.
Tandon \emph{et al}.'s gradient coding corresponds to the optimal gradient coding scheme in~\cite{GC_Tandon_exact} for $\alpha$-partial stragglers with $\alpha=\frac{ \mathbb{E}[T_n|T_n\le t] } { \mathbb{E}[T_n|T_n> t] }=6$, where $t$ satisfies $\Pr[T_n\le t]=0.5$.
Ferdinand \emph{et al}.'s coded computation ($r=L$) and ($r=L/2$) correspond to the optimal coding scheme in~\cite{mat_vec_Draper_hierarchical_ISIT}, with the optimized coding parameter at the number of layers $r=L$ and $r=L/2$, respectively.
%In the simulation, the distribution of $T_n,n\in[N]$ follows a shifted-exponential distribution with $\mu$ and $t_0$ given in \eqref{equ:E_shifted-exponential_distribution} as in \cite{mat_vec_Lee_speeding_up,mat_vec_Draper_hierarchical_ISIT,mat_mat_Draper_hierarchical_TIT,GC_Min_Ye_communication,GC_WH_improving_RS_Codes}.
We set $M=50$ and $b=1$.\footnote{\REPLYb{This paper focuses on the overall runtime for the master and workers to collaboratively compute the gradient in each iteration of a gradient descent methods for solving a machine learning problem.
Thus, we can evaluate the overall runtime of one iteration without a specific machine learning dataset.}}
\subsection{\REPLYb{Shifted-exponential Distribution}}

%\begin{figure}[t]
%\begin{center}
%	{\resizebox{5.9cm}{!}{\includegraphics{J_fig/fig_E_x_vs_n.eps}}}
%\end{center}
%%  \vspace{-1cm}
%  \caption{Illustration of $\check{\mathbf{x}}^{\rm (E,opt)}$, $\check{\mathbf{x}}^{\rm (E,t)}$, and $\check{\mathbf{x}}^{\rm (E,f)}$ at $N=20$, $L=2\times 10^4$, $\mu=10^{-3}$, and $t_0=100$.}\label{fig:E_x_vs_n}
%%  \vspace{-0.8cm}
%\end{figure}

\REPLYb{In this part, $T_n,n\in[N]$ follow a shifted-exponential distribution (with parameters $\mu$ and $t_0$) given in \eqref{equ:E_shifted-exponential_distribution} as in \cite{mat_vec_Lee_speeding_up,mat_vec_Draper_hierarchical_ISIT,mat_mat_Draper_hierarchical_TIT,GC_Min_Ye_communication,GC_WH_improving_RS_Codes}.}

\REPLYb{First}, we consider expected overall runtime minimization.
%%In the simulation, we set $t_0=100$.
%Fig.~\ref{fig:E_x_vs_n} illustrates the proposed solutions $\check{\mathbf{x}}^{\rm (E,opt)}$, \hspace{-0.1cm} $\check{\mathbf{x}}^{\rm (E,t)}$,\hspace{-0.05cm} and $\check{\mathbf{x}}^{\rm (E,f)}$.\hspace{-0.05cm}
%Fig.~\ref{fig:E_x_vs_n} indicates that in these solutions, the first block (including coordinates $1,\cdots,x_0$) with no redundancy and the last block (including coordinates $L-x_{N-1}+1,\cdots,L$) with redundancy for tolerating $N-1$ stragglers contain most of the $L$ coordinates.
Fig.~\ref{subfig:E_vs_N}, Fig.~\ref{subfig:E_vs_mu}, and Fig.~\ref{subfig:E_vs_L} illustrate the expected runtime versus the number of workers $N$, the rate parameter $\mu$, and the number of model parameters $L$, respectively.
From Fig.~\ref{subfig:E_vs_N}, we see that the expected overall runtime of each scheme decreases with $N$, due to the increase of the overall computation resource with $N$.
From Fig.~\ref{subfig:E_vs_mu}, we see that the expected overall runtime of each scheme decreases with $\mu$ due to the decrease of $\mathbb{E}[T_n]=\frac{1}{\mu}+t_0$ with $\mu$.
From Fig.~\ref{subfig:E_vs_L}, we see that the expected overall runtime of each scheme increases with $L$, due to the increase of computation load with $L$.
Furthermore, from Fig.~\ref{fig:simulation_E}, we can draw the following conclusions.
The proposed solutions significantly outperform the four baseline schemes.
For instance, the proposed solutions can achieve reductions of 44\% in the expected overall runtime over the best baseline scheme at $\mu=10^{-2.6}$ in Fig.~\ref{subfig:E_vs_mu}.
The gains over single BCGC and Tandon \emph{et al.}'s gradient coding are due to the diverse redundancy introduced across partial derivatives.
The gains over Ferdinand \emph{et al.}'s coded computation schemes at $r=L$ and $r=L/2$ indicate that an optimal coded computation scheme for calculating matrix-vector multiplications is no longer effective for calculating a general gradient.
The proposed closed-form approximate solutions are quite close to the proposed optimal solution and hence have significant practical values.
Besides, note that $\check{\mathbf{x}}^{\rm (E,f)}$ slightly outperforms $\check{\mathbf{x}}^{\rm (E,t)}$ which can be drawn from Theorem~\ref{thm:multiplicative_gap_1} and Theorem~\ref{thm:multiplicative_gap_2}.

\begin{figure*}[t]
\begin{center}
  \subfigure[\scriptsize{$L=4\times 10^4$ and $\mu=10^{-1}$.}\label{subfig:CDF_vs_N}]
  {\resizebox{4.35cm}{!}{\includegraphics{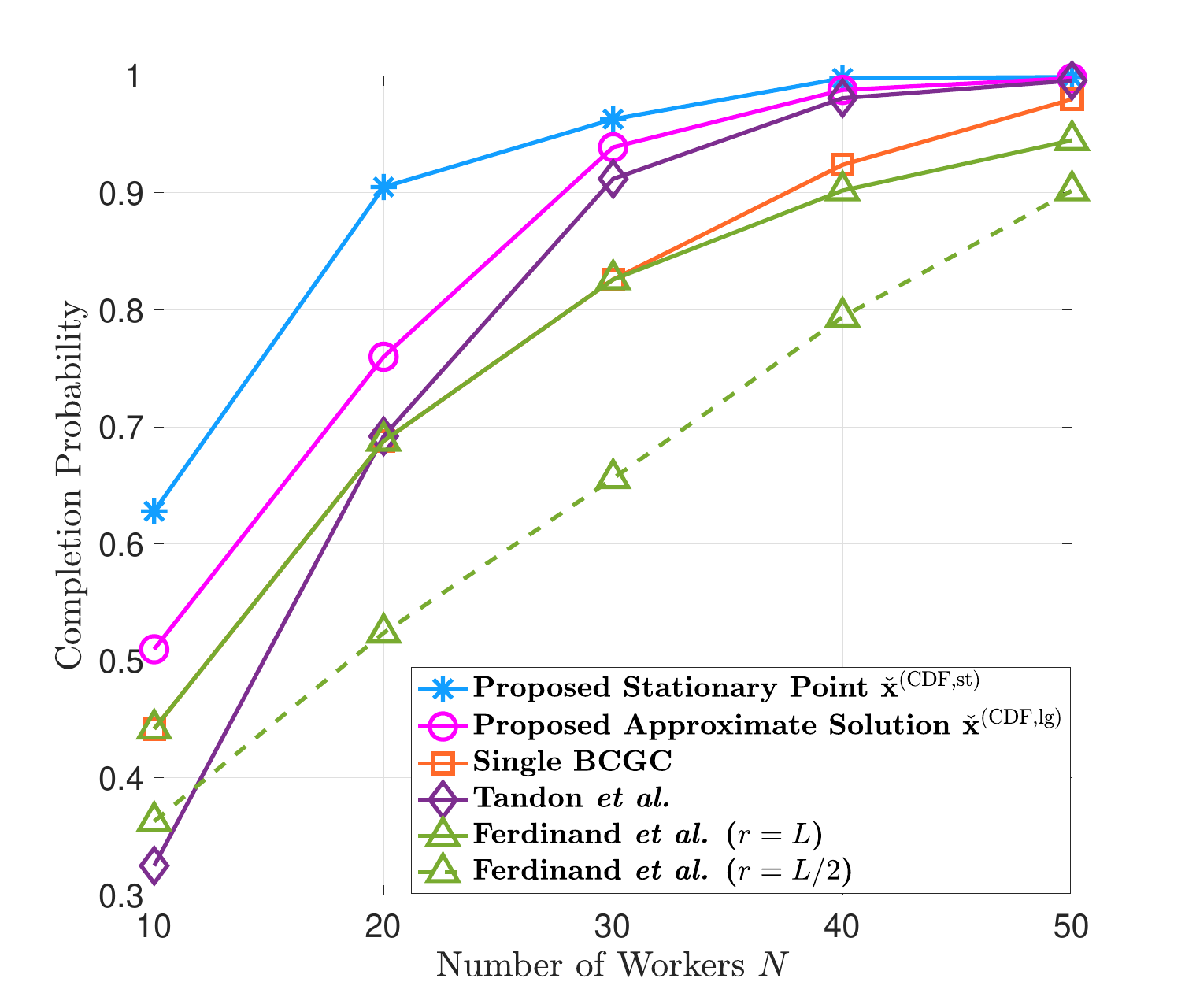}}}
  \ \ 
  \subfigure[\scriptsize{$N=20$ and $L=4\times 10^4$.}\label{subfig:CDF_vs_mu}]
  {\resizebox{4.5cm}{!}{\includegraphics{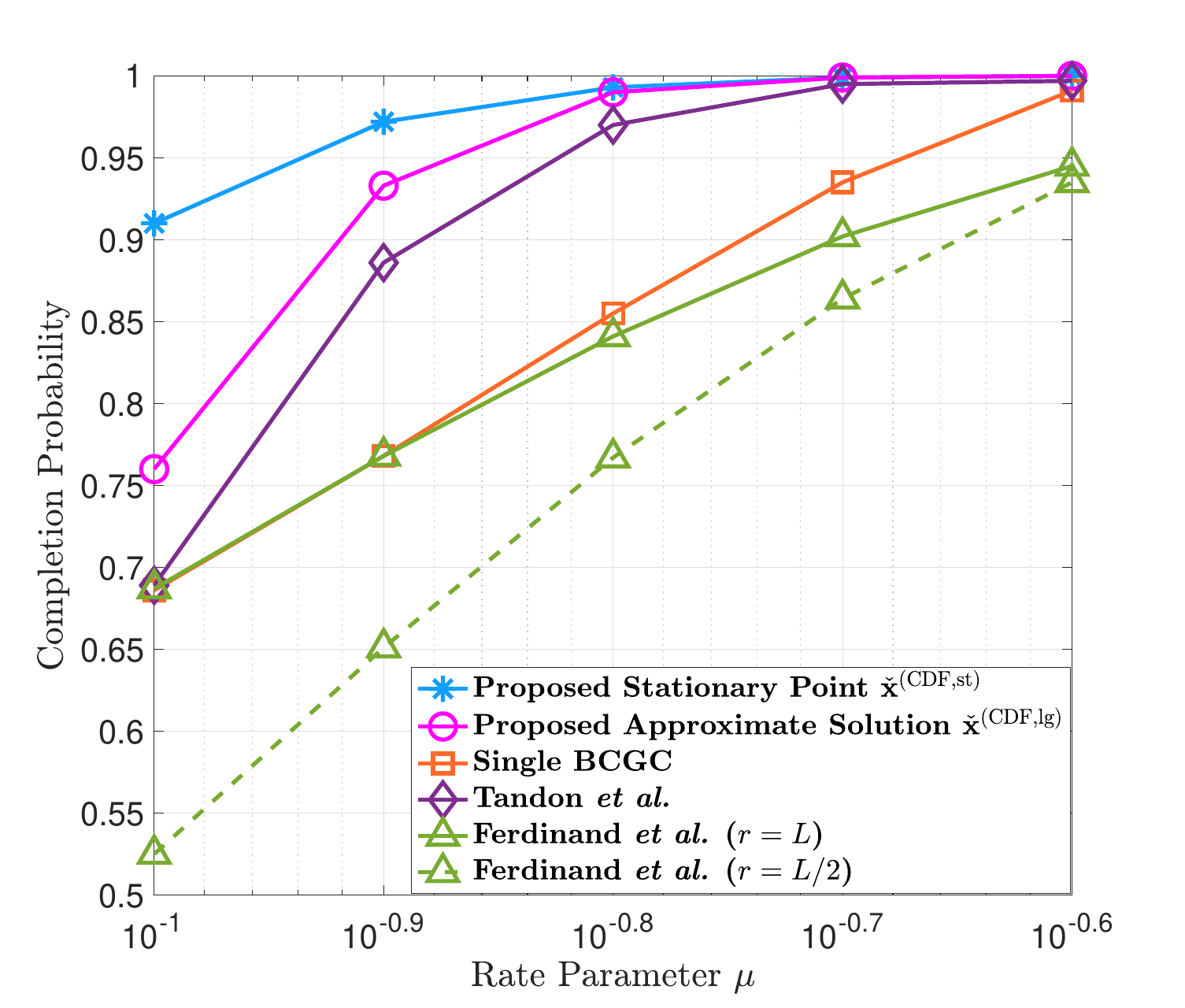}}}
  \ \ ,
  \subfigure[\scriptsize{$N=20$ and $\mu=10^{-1}$.}\label{subfig:CDF_vs_L}]
  {\resizebox{4.5cm}{!}{\includegraphics{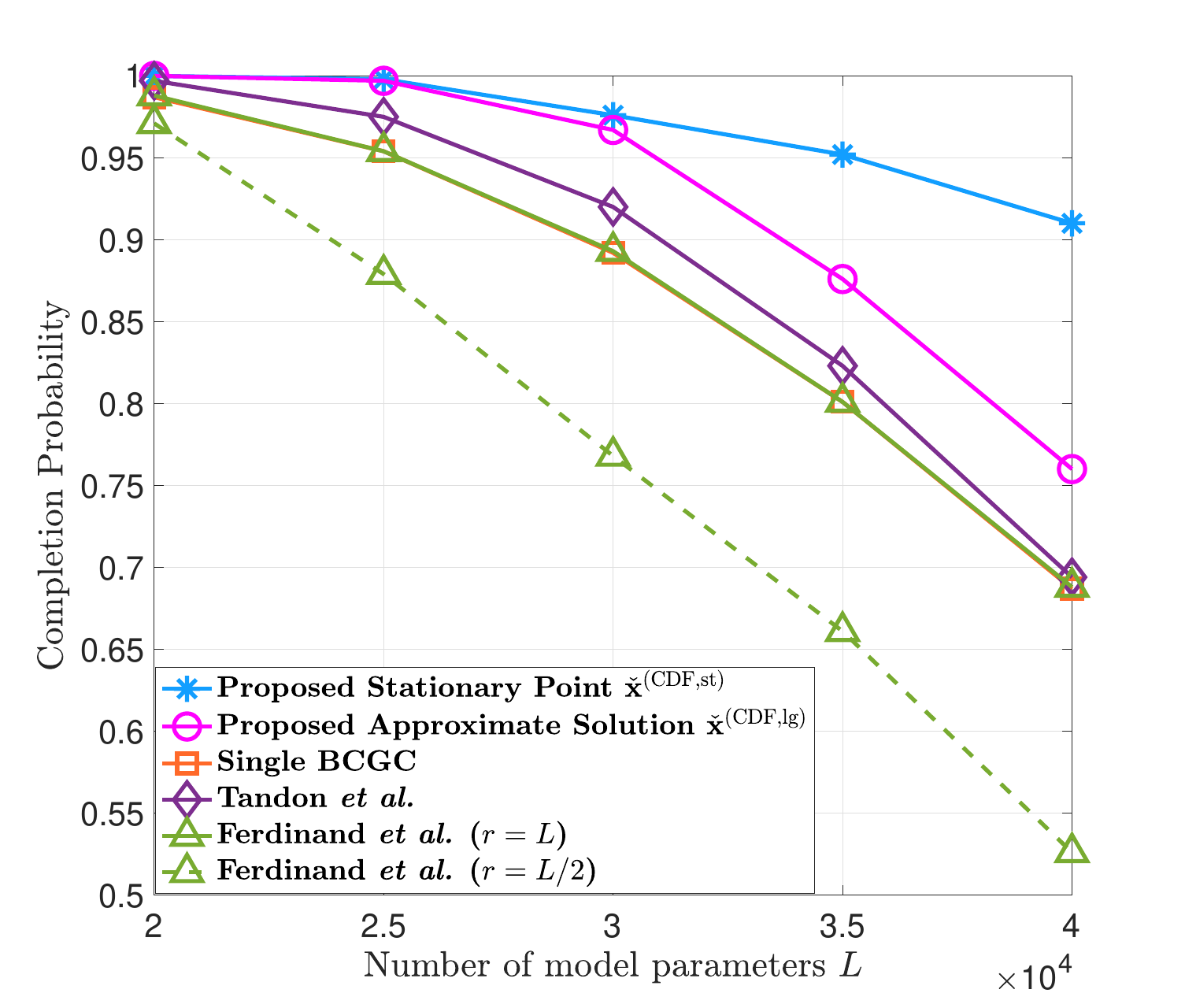}}}
\end{center}
  \vspace{-0.3cm}
         \caption{Completion probability versus $N$, $\mu$ and $L$ at $t= 10^{6.5}$ and $t_0=1$, respectively. 
         }\label{fig:simulation_CDF}
         \vspace{-0.3cm}
\end{figure*}
% ==================================================================

%\subsection{Completion Probability}

%\begin{figure}[t]
%\begin{center}
%	{\resizebox{5.9cm}{!}{\includegraphics{J_fig/fig_CDF_x_vs_n.eps}}}
%\end{center}
%%  \vspace{-1cm}
%  \caption{$\check{\mathbf{x}}^{\rm (CDF,st)}$ and $\check{\mathbf{x}}^{\rm (CDF,lg)}$ at $N=20$, $L=4\times 10^4, \mu=10^{-1}$, and $t_0=1$.}\label{fig:CDF_x_vs_n}
%%  \vspace{-0.8cm}
%\end{figure}

\begin{figure}[h]
\begin{center}
%	{\resizebox{5.9cm}{!}{\includegraphics{J_fig/fig_CDF_vs_t.eps}}}
	{\resizebox{4.5cm}{!}{\includegraphics{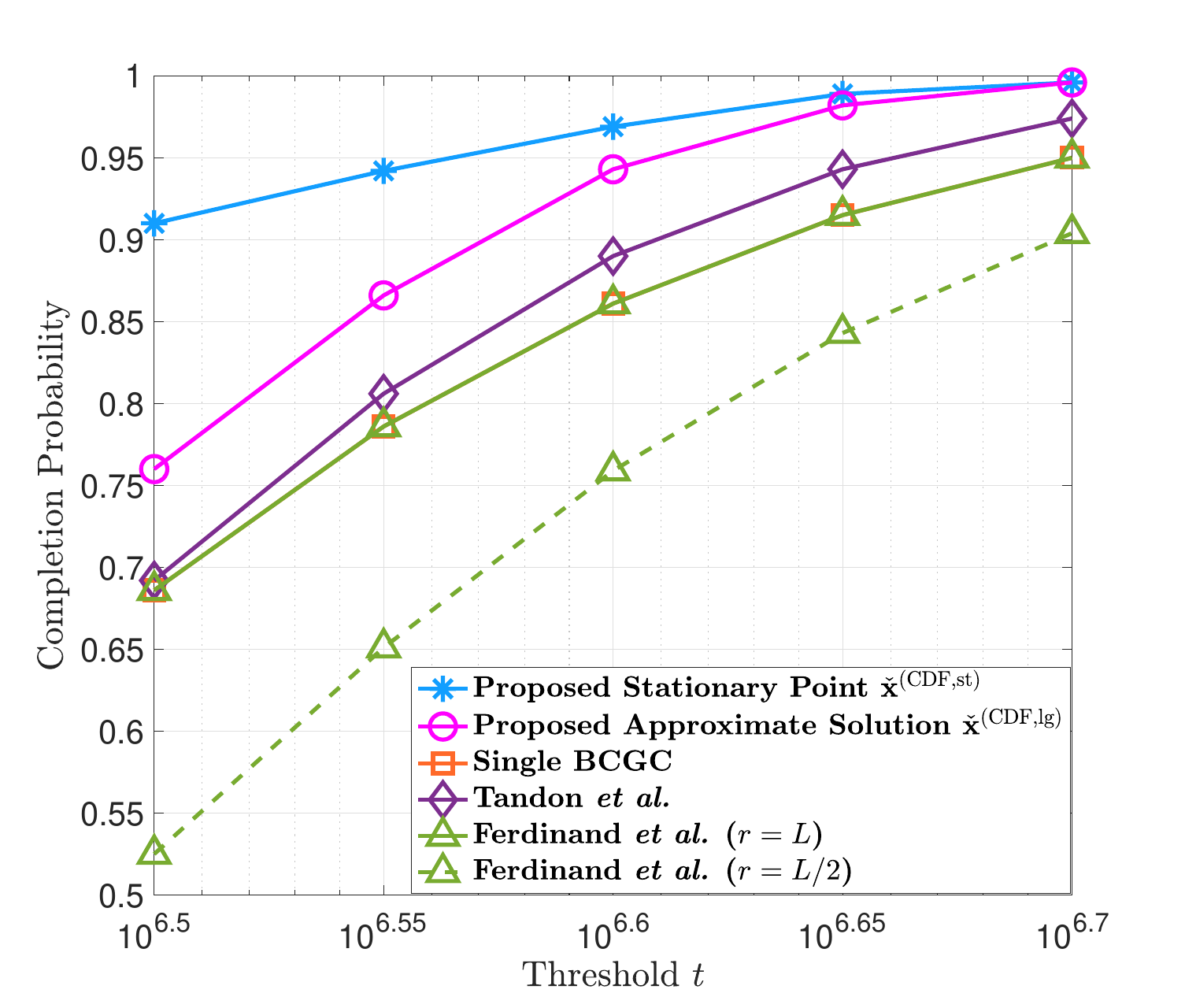}}}
\end{center}
  \vspace{-0.3cm}
  \caption{Completion probability versus $t$ at $N=20$, $L=4\times 10^4$, $\mu=10^{-1}$, and $t_0=1$.}\label{fig:CDF_vs_t}
  \vspace{-0.3cm}
\end{figure}

\REPLYb{Next}, we consider completion probability maximization. 
%Fig.~\ref{fig:CDF_x_vs_n} illustrates the proposed solutions $\check{\mathbf{x}}^{\rm (CDF,st)}$ and $\check{\mathbf{x}}^{\rm (CDF,lg)}$.
%Fig.~\ref{fig:CDF_x_vs_n} indicates that $\check{\mathbf{x}}^{\rm (CDF,st)}$ at $t=10^{8}$ is closer to $\check{\mathbf{x}}^{\rm (CDF,lg)}$ than $\check{\mathbf{x}}^{\rm (CDF,st)}$ at $t=10^{6}$ as expected.
Fig.~\ref{subfig:CDF_vs_N}, Fig.~\ref{subfig:CDF_vs_mu}, and Fig.~\ref{subfig:CDF_vs_L} illustrate the completion probability versus the number of workers $N$, the rate parameter $\mu$, and the number of model parameters $L$, respectively.
From Fig.~\ref{fig:simulation_CDF}, we see that the completion probability of each scheme increases with $N$ and $\mu$, and decreases with $L$.
The explanations are similar to those for the expected overall runtime minimization shown in Fig.~\ref{fig:simulation_E}.
Fig.~\ref{fig:CDF_vs_t} illustrates the completion probability versus  the threshold $t$.
From Fig.~\ref{fig:CDF_vs_t}, we see that the completion probability of each scheme increases with $t$ as expected.
Furthermore, from Fig.~\ref{fig:simulation_CDF} and Fig.~\ref{fig:CDF_vs_t}, we can draw the following conclusions.
The proposed solutions significantly outperform the four baseline schemes.
For instance, in Fig.~\ref{subfig:CDF_vs_N}, the proposed solutions can achieve reductions of 15\% in the completion probability over the best baseline scheme at $N=10$.
Besides, the proposed closed-form approximate solution achieves similar completion probability to the proposed stationary point
with much lower computational complexity and their gap reduces with $t$, indicating the significant practical value of the proposed approximate solution.

%From Fig.~\ref{fig:CDF_vs_t}, we see that the proposed solutions $\check{\mathbf{x}}^{\rm (CDF,st)}$ and $\check{\mathbf{x}}^{\rm (CDF,lg)}$ outperform all the four baseline schemes at all values of $t$ considered in the simulation.
%This indicates that the proposed block coordinate gradient coding scheme can well adapt to changes of threshold $t$.
%Besides, note that the CDF of $\check{\mathbf{x}}^{\rm (CDF,lg)}$ is quite close to that of $\check{\mathbf{x}}^{\rm (CDF,st)}$ at large $t$, indicating that it is a good approximate solution.

\subsection{\REPLYb{Distribution based-on Real-world Platform}}

\begin{figure}[t]
\begin{center}
  \subfigure[\scriptsize{\REPLYb{Expected total runtime versus $N$ at $L=2 \times 10^4$.}}\label{subfig:E_real_vs_N}]
  {\resizebox{4.35cm}{!}{\includegraphics{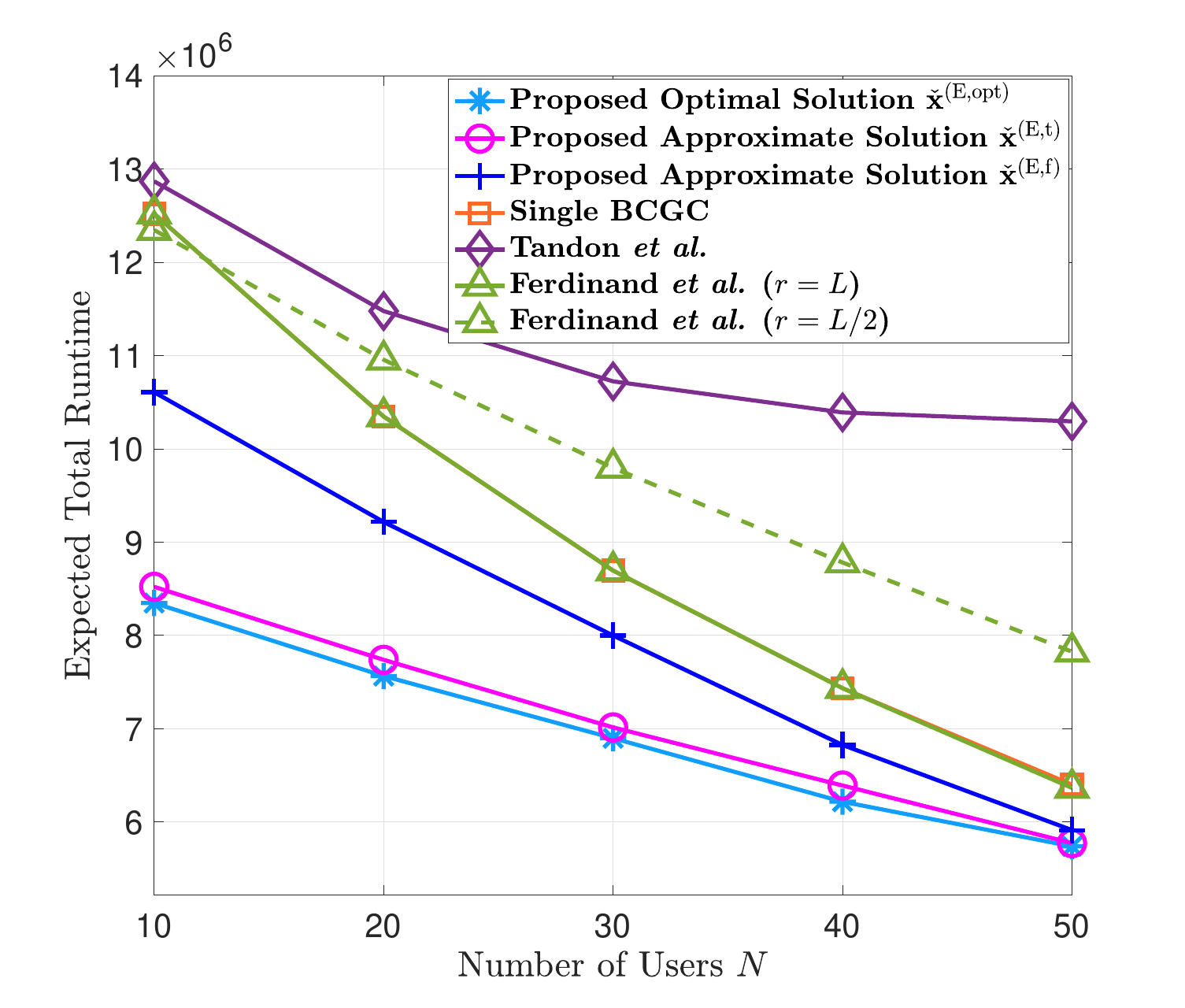}}}
%  \ \
  \subfigure[\scriptsize{\REPLYb{Completion probability versus $N$ at $L=4 \times 10^4$ and $t=10^{6.8}$.}}\label{subfig:CDF_real_vs_N}]
  {\resizebox{4.35cm}{!}{\includegraphics{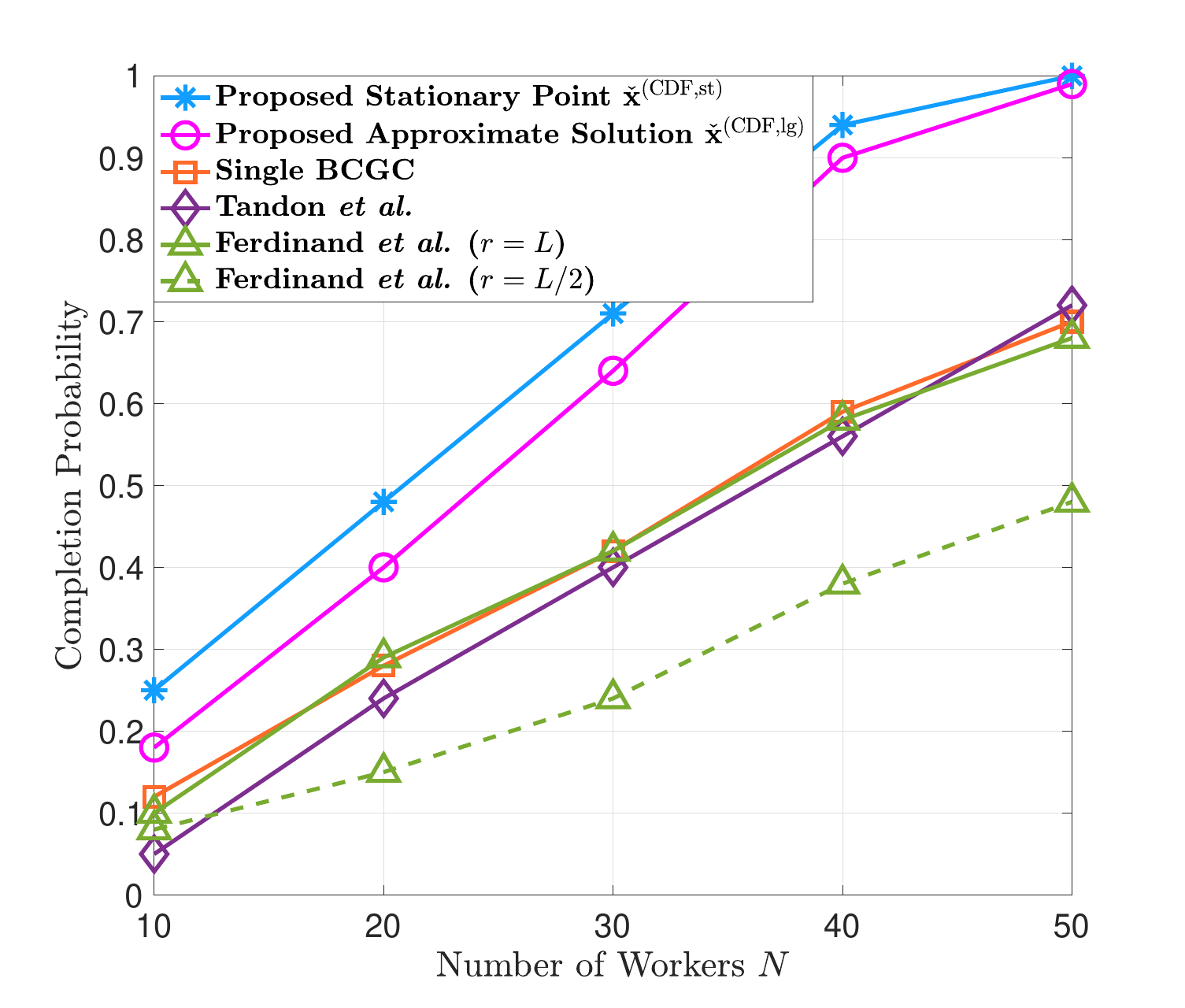}}}
  \end{center}
  \vspace{-0.4cm}
         \caption{\REPLYb{Numerical results under the distribution of $T_n,n\in[N]$ for an Amazon EC2 cluster \cite[Fig. 7]{mat_vec_Lee_speeding_up}.}
         }\label{fig:Amazon}
	\vspace{-0.5cm}
\end{figure}

\REPLYb{In this part, 
$T_n,n\in[N]$ follow the distribution for an Amazon Elastic Compute Cloud (Amazon EC2) cluster, reported in \cite[Fig. 7]{mat_vec_Lee_speeding_up}. 
Specifically, we first approximate the complementary CDF for the Amazon EC2 cluster \cite[Fig. 7]{mat_vec_Lee_speeding_up} with a complementary CDF of a shifted-exponential distribution (with parameters $\mu$ and $t_0$) given in \eqref{equ:E_shifted-exponential_distribution} using moment matching. 
Consequently, we have $t_0=0$, $\mu = \frac{1}{\mathbb{E}[T_n]} = 5\times 10^{-2}$.
Then, we obtain the proposed solutions and baseline solutions under the approximate shifted-exponential distribution.
Finally, we evaluate the expected overall runtime and completion probability using the samples generated from the complementary CDF for the Amazon EC2 cluster \cite[Fig. 7]{mat_vec_Lee_speeding_up}.}
\REPLYb{Fig.~\ref{subfig:E_real_vs_N} and Fig.~\ref{subfig:CDF_real_vs_N} show the corresponding expected overall runtime and completion probability versus the number of workers $N$, respectively. 
%From Fig.~\ref{subfig:E_real_vs_N} and Fig.~\ref{subfig:CDF_real_vs_N}, we see that
%the proposed solutions significantly outperform the four baseline schemes, and
%the proposed approximate solutions are quite close to the proposed optimal solution.
Similar observations can be made.
Thus, we can conclude that the proposed solution framework relying on a shifted-exponential distribution have significant practical values.}

%\begin{figure}[ht]
%\begin{center}
%	{\resizebox{5.9cm}{!}{\includegraphics{reply_fig/fig_E_vs_N_real.eps}}}
%\end{center}
%%  \vspace{-1cm}
%  \caption{\REPLYb{Expected total runtime versus $N$ under the distribution of $T_n,n\in[N]$ for an Amazon EC2 cluster \cite[Fig. 7]{mat_vec_Lee_speeding_up} at $L=2 \times 10^4$.}}\label{subfig:E_real_vs_N}
%%  \vspace{-0.8cm}
%\end{figure}
%
%\begin{figure}[ht]
%\begin{center}
%	{\resizebox{5.9cm}{!}{\includegraphics{reply_fig/fig_CDF_vs_N_real.eps}}}
%\end{center}
%%  \vspace{-1cm}
%  \caption{\REPLYb{Completion probability versus $N$ under the distribution of $T_n,n\in[N]$ for an Amazon EC2 cluster \cite[Fig. 7]{mat_vec_Lee_speeding_up} at $L=4 \times 10^4$ and $t=10^{6.8}$.}}\label{subfig:CDF_real_vs_N}
%%  \vspace{-0.8cm}
%\end{figure}

\addtolength{\topmargin}{0.025in}
\section{Conclusion}
\label{sec:Conclusion}

In this paper, 
%considered a distributed computation system consisting of one master and $N$ workers, characterized by a general partial straggler model and focused on solving a general large-scale machine learning problem with $L$ model parameters \TCOMb{using gradient coding}. 
%Existing gradient coding schemes introduce identical redundancy in coded local partial derivatives corresponding to all $L$ coordinates and hence \TCOMb{cannot effectively utilize} the computing capabilities of partial stragglers.
%To address this limitation, 
we proposed a coordinate gradient coding scheme with $L$ coding parameters, \TCOMb{representing $L$ possibly different diversities for the $L$ coordinates, which generated most gradient coding schemes.}
We considered the minimization of the expected overall runtime and the maximization of the completion probability with respect to the $L$ coding parameters for coordinates.
\TCOMbr{We transformed the coordinate gradient coding scheme to an equivalent but more easily implemented block coordinate gradient coding scheme with $N \ll L$ coding parameters for blocks and adopted continuous relaxation.}
%\TCOMb{To reduce computational complexity}, we equivalently considered a block coordinate gradient coding scheme with $N$ coding parameters for blocks.
%We also adopted continuous relaxation to further reduce \TCOMb{computational complexity}.
For the resulting minimization of expected overall runtime, we 
%developed an iterative algorithm to 
derived an optimal solution and two closed-form approximate solutions.
For the resultant maximization of the completion probability, we 
%developed an iterative algorithm to 
derived a stationary point and a closed-form approximate solution.
Numerical results \TCOMc{demonstrated} that the proposed solutions significantly outperformed existing coded computation schemes and their extensions, and the approximate solutions had close-to-optimal performances.
\TCOMc{To the best of our knowledge, this is the first work that optimizes the redundancies for gradient coding to effectively utilize the computing capabilities of partial stragglers.}

\section*{Appendix A: Proof of Lemma~\ref{lem:E_monotonicity}}

We prove Lemma~\ref{lem:E_monotonicity} by contradiction.
	First, suppose that for the optimal solution $\mathbf{s^*}$, there exists $m\in [L-1]$ such that $s^*_m > s^*_{m+1}$.
	Construct a feasible solution $\tilde{\mathbf{s}} \triangleq (\tilde{s}_l)_{l\in [L]}$, where 
	\begin{small} 
	\begin{alignat}{2}
		\tilde{s}_l &= s^*_l,		&\quad l&\neq m,	\label{equ:lnot=m} \\
		\tilde{s}_l &= s^*_{l+1},	&\quad l&=m.		\label{equ:l=m}
	\end{alignat}
	\end{small}Then, for all $l\in[m-1]$, we have
	\begin{small}
	\begin{align}
	\setlength{\abovedisplayskip}{0pt}
	\setlength{\belowdisplayskip}{0pt}
		\frac{M}{N}b T_{(N-\tilde{s}_l)} \sum_{i=1}^l (\tilde{s}_i + 1) \overset{(a)}{=} \frac{M}{N}b T_{(N-s^*_l)} \sum_{i=1}^l (s^*_i + 1) \le \tau(\mathbf{s}^*,\mathbf{T}).\label{equ:l_case1}
	\end{align}
	\end{small}For all $l\in[N] \backslash [m]$, we have
	\begin{small}
	\begin{align}
	\setlength{\abovedisplayskip}{0pt}
	\setlength{\belowdisplayskip}{0pt}
		\frac{M}{N}b T_{(N-\tilde{s}_l)} \sum_{i=1}^l (\tilde{s}_i + 1) \overset{(b)}{<} \frac{M}{N}b T_{(N-s^*_l)} \sum_{i=1}^l (s^*_i + 1) \le \tau(\mathbf{s}^*,\mathbf{T}).\label{equ:l_case2}
	\end{align}
	\end{small}For $l=m$, we have
	\begin{small}
	\begin{align}
	\setlength{\abovedisplayskip}{0pt}
	\setlength{\belowdisplayskip}{0pt}
		 &\frac{M}{N}b T_{(N-\tilde{s}_m)} \hspace{-0.1cm} \sum_{i=1}^l (\tilde{s}_i + 1) 
		 \hspace{-0.1cm} \overset{(c)}{=} \hspace{-0.1cm} \frac{M}{N}b T_{(N-\tilde{s}_m)} \hspace{-0.1cm} \left( \sum_{i=1}^{m-1} (\tilde{s}_i + 1) \hspace{-0.1cm} + \hspace{-0.1cm} (\tilde{s}_m + 1) \hspace{-0.1cm} \right) \hspace{-0.1cm}\nonumber\\
		 &< \hspace{-0.1cm} \frac{M}{N}b T_{(N-s^*_{m+1})} \hspace{-0.1cm} \sum_{i=1}^{m+1} (s^*_i + 1) \le \tau(\mathbf{s}^* \hspace{-0.05cm} , \hspace{-0.05cm} \mathbf{T}).\label{equ:l_case3}
	\end{align}
	\end{small}Here, $(a),(b)$, and $(c)$ are due to \eqref{equ:lnot=m} and \eqref{equ:l=m}.
	By~\eqref{equ:tau_calculate}, \eqref{equ:l_case1}, \eqref{equ:l_case2} and \eqref{equ:l_case3}, we have $\tau(\tilde{\mathbf{s}},\mathbf{T})\le \tau(\mathbf{s^*,T})$, $\mathbf{T}\in\mathbb{R}_+^N$.
	Besides, $\tau(\tilde{\mathbf{s}},\mathbf{T}) < \tau(\mathbf{s^*,T})$, $\mathbf{T}\in\mathcal{S}\triangleq\{\mathbf{T}:T_n=t, n\in[N], t\in\mathbb{R}_+\}$.
	As the Lebesgue measure is strictly positive on the non-empty open set $\mathcal{S}$,
	$\mathbb{E}\left[\tau(\tilde{\mathbf{s}},\mathbf{T})\right] < \mathbb{E}\left[\tau(\mathbf{s}^*,\mathbf{T})\right]$.
	This indicates that $\mathbf{s}^*$ is not an optimal solution, which contradicts with the assumption.
	Therefore, we complete the proof of Lemma~\ref{lem:E_monotonicity}.

\section*{Appendix B: Proof of Theorem~\ref{thm:E_prob_equvalence}}

	Based on Lemma~\ref{lem:E_monotonicity}, we first equivalently transform Problem~\ref{prob:E_original} to the following  problem.
		\begin{align}
			\min_{\mathbf{s}} \ \  &\mathbb{E}\left[ \tau(\mathbf{s,T}) \right] \label{equ:app_B_prob}\\
		  	\rm{s.t.}\ \  &\eqref{equ:constraint_s_1},\nonumber\\
		  	& s_i \le s_{i+1},\quad i\in[L-1].\label{equ:s_E_monotonicity}
		\end{align}
	Next, we equivalently transform the problem in \eqref{equ:app_B_prob} to Problem~\ref{prob:equivalent_1}. 
	Introduce auxiliary variables $\mathbf{x}\triangleq (x_n)_{n=0,1,\cdots,N-1}$ with $x_n = \sum_{l\in[L]} I(s_l=n) \in\mathbb{N},\ n=0,1,\dots,N-1$, 
	together with \eqref{equ:s_E_monotonicity} implying $s_l = \min\left\{ i : \sum_{n=0}^{i} x_n \ge l \right\},\ l\in[L]$.
	For notation simplicity, treat $\sum_{i=0}^{-1} x_i\triangleq 0$ for ease of illustration and define $\mathcal{S}_n \triangleq \Big\{ \sum_{i=0}^{n-1} x_i+1, \sum_{i=0}^{n-1} x_i+2, \cdots, \sum_{i=0}^{n} x_i \Big\}, n=0,1,\cdots,N-1$.
	Thus, we have 
	\begin{align}
		\sum_{i\in\cup_{j=0}^{n}\mathcal{S}_j} (s_i+1) = \sum_{i=0}^n (i+1)x_i.\label{equ:s_x_sum}
	\end{align}
	Besides, we have
	\begin{small}
	\begin{align*}
		& \tau(\mathbf{s},\mathbf{T})
		 = \frac{M}{N}b \max_{l\in [L]} T_{(N-s_l)} \sum_{i\in[l]}(s_i+1)
		\eqa \frac{M}{N}b \hspace{-0mm} \max_{n=0,\cdots,N-1}\nonumber\\
		& \max_{l\in\mathcal{S}_n} T_{(N-s_l)} \Bigg( \sum_{i\in \cup_{j=0}^{s_l-1} \mathcal{S}_j}(s_i+1) \hspace{-0mm} + \hspace{-0mm} \sum_{i\in\mathcal{S}_{s_l}\cap[l]} \hspace{-0mm} (s_i+1) \Bigg)\\
		&\eqb \frac{M}{N}b \hspace{-1mm} \max_{n=0,\cdots,N-1} \max_{l\in\mathcal{S}_n} T_{(N-n)} \Bigg( \sum_{i\in \cup_{j=0}^{n-1} \mathcal{S}_j}\hspace{-3mm} (s_i+1) \hspace{-0mm} + \hspace{-3mm} \sum_{i\in\mathcal{S}_{s_l}\cap[l]} \hspace{-3mm} (s_i+1) \Bigg)\nonumber\\ 
		& \eqc \frac{M}{N}b \hspace{-1mm} \max_{n=0,\cdots,N-1} \hspace{-1mm} T_{(N-n)} \sum_{i=0}^{n-1}(i+1)x_i
		=\hat{\tau}(\mathbf{x,T}),
	\end{align*}	
	\end{small}where $(a)$ is due to $[L] = \cup_{n=0}^{N-1}\mathcal{S}_n$
	and $[l] = \cup_{j=0}^{s_l-1}\mathcal{S}_j \cup \big\{ \mathcal{S}_{s_l}\cap[l] \big\}$ (by the definition of $\mathcal{S}_n$),
	$(b)$ is due to $s_l = n, l\in\mathcal{S}_n$ (by the expression of $s_l$ and the definition of $\mathcal{S}_n$,
	and $(c)$ is due to \eqref{equ:s_x_sum}. 
	By $x_n = \sum_{l\in[L]} I(s_l=n) \in\mathbb{N},\ n=0,1,\dots,N-1$, we can obtain \eqref{equ:constraint_x_1} and \eqref{equ:constraint_x_2}.
	Thus, the problem in \eqref{equ:app_B_prob} is equivalent to Problem~\ref{prob:equivalent_1}, implying \eqref{equ:change_of_var_x} and \eqref{equ:change_of_var_s} hold and $\tau_{\rm avg}^*=\hat{\tau}_{\rm avg}^*$.
	Therefore, we can show Theorem~\ref{thm:E_prob_equvalence}.

\section*{\REPLYb{Appendix C: Proof of Lemma~\ref{lem:opt_projection}}}

\REPLYb{Let $\bm{\lambda}^{(i)} \triangleq \left( \lambda^{(i)}_n \right)_{n = 0,1,\cdots,N-1}$ denote the Lagrange multipliers corresponding to the inequality constraints in \eqref{equ:constraint_x_3}, and let $\nu$ denote the Lagrange multiplier corresponding to the equality constraint in \eqref{equ:constraint_x_1}.
The KKT conditions are given by
\begin{align}
	& \sum_{n=0}^{N-1}x^{(i)}_n=L,\label{equ_1}\\
	& x^{(i)}_n               \ge 0,\ n=0,1,\cdots,N-1,\label{equ_2}\\
	& \lambda^{(i)}_n         \ge 0,\ n=0,1,\cdots,N-1,\label{equ_3}\\
	& \lambda^{(i)}_n x^{(i)}_n = 0,\ n=0,1,\cdots,N-1,\label{equ_4}\\
	& 2(x^{(i)}_n - \hat{x}^{(i)}_n) - \lambda^{(i)}_n + \nu^{(i)} =0,\label{equ_5}\ 
n=0,1,\cdots,N-1.
\end{align}	
%By the KKT conditions, we can obtain \cite{mydetailedproofarxiv}
%\begin{align}
%	x^{(i)}_n 
%	= \max\left\{ \hat{x}^{(i)}_n - \frac{1}{2}\nu^{(i)},0 \right\},\ n=0,1,\cdots,N-1,\label{equ_9}
%\end{align}
%%Substituting this expression for $x^{(i)}_n$ into \eqref{equ_1}, we obtain
%where $\nu^{(i)}$ satisfies
%\begin{align}
%	\sum_{n=0}^{N-1} \max\left\{ \hat{x}^{(i)}_n -\frac{1}{2}\nu^{(i)},0 \right\} = L.\label{equ_10}
%\end{align}
%%The lefthand side is a piecewise-linear increasing function of $-\frac{1}{2}\nu^{(i)}$, with breakpoints at $-\hat{x}^{(i)}_n$, so
%It can be verified that \eqref{equ_10} has a unique solution which is readily determined by bisection search \SecREPLYc{with initial lower bound and upper bound on $\nu^{(i)}$ given by $\frac{2}{N} (\sum_{n=0}^{N-1} \hat{x}^{(i)}_n - L)$ and $2\underset{n=0,1,\cdots,N-1}{\max} \hat{x}^{(i)}_n$, respectively \cite[pp. 146]{Boyd_cvxbook}}.
%Note that Problem~\ref{prob:projection} is a convex problem with differentiable objective and constraint functions, and it satisfies Slater’s condition. 
%Thus, the KKT conditions provide necessary and sufficient conditions for optimality.
%By \eqref{equ_9}, \eqref{equ_10}, and replacing $-\frac{1}{2}\nu^{(i)}$ with $\nu$ for simplicity, we can show Lemma~\ref{lem:opt_projection}. 
By \eqref{equ_1}, \eqref{equ_2}, \eqref{equ_3}, \eqref{equ_4}, \eqref{equ_5}, we can obtain $\mathbf{x}^{(i)}, \bm{\lambda}^{(i)},$ and $\nu^{(i)}$, which satisfy the KKT conditions as follows.
By \eqref{equ_5}, we have 
\begin{align}
	\lambda^{(i)}_n = 2(x^{(i)}_n - \hat{x}^{(i)}_n) + \nu^{(i)},\ n=0,1,\cdots,N-1.\label{equ_6}
\end{align}
Eliminating \eqref{equ_5} and substituting \eqref{equ_6} into \eqref{equ_3} and \eqref{equ_4}, we have
\begin{align}
	\nu^{(i)} \ge 2(\hat{x}^{(i)}_n - x^{(i)}_n),\ n=0,1,\cdots,N-1,\label{equ_7}\\
	\left( 2(x^{(i)}_n - \hat{x}^{(i)}_n) + \nu^{(i)} \right) x^{(i)}_n = 0,\ n=0,1,\cdots,N-1.\label{equ_8}
\end{align}
If $\nu^{(i)} < 2\hat{x}^{(i)}_n$, \eqref{equ_7} can only hold if $x^{(i)}_n>0$, which by \eqref{equ_8} implies that $x^{(i)}_n = \hat{x}^{(i)}_n - \frac{1}{2}\nu^{(i)}$.
Thus, $x^{(i)}_n = \hat{x}^{(i)}_n - \frac{1}{2}\nu^{(i)}$, if $\nu^{(i)} < 2\hat{x}^{(i)}_n$.
If $\nu^{(i)} \ge 2\hat{x}^{(i)}_n$, then $x^{(i)}_n > 0$ is impossible, because it would imply $2(x^{(i)}_n - \hat{x}^{(i)}_n) + \nu^{(i)} > 0$, which implies $\left( 2(x^{(i)}_n - \hat{x}^{(i)}_n) + \nu^{(i)} \right) x^{(i)}_n > 0$ and violates \eqref{equ_8}.
Thus, $x^{(i)}_n = 0$, if $\nu^{(i)} \ge 2\hat{x}^{(i)}_n$.
Therefore, we have
%By the KKT conditions, we can obtain \cite{mydetailedproof}
\begin{align}
	x^{(i)}_n 
	= \max\left\{ \hat{x}^{(i)}_n - \frac{1}{2}\nu^{(i)},0 \right\},\ n=0,1,\cdots,N-1,\label{equ_9}
\end{align}
Substituting this expression for $x^{(i)}_n$ into \eqref{equ_1}, we obtain
where $\nu^{(i)}$ satisfies
\begin{align}
	\sum_{n=0}^{N-1} \max\left\{ \hat{x}^{(i)}_n -\frac{1}{2}\nu^{(i)},0 \right\} = L.\label{equ_10}
\end{align}
The lefthand side is a piecewise-linear increasing function of $-\frac{1}{2}\nu^{(i)}$, with breakpoints at $-\hat{x}^{(i)}_n$, so
It can be verified that \eqref{equ_10} has a unique solution which is readily determined by bisection search \SecREPLYc{with initial lower bound and upper bound on $\nu^{(i)}$ given by $\frac{2}{N} (\sum_{n=0}^{N-1} \hat{x}^{(i)}_n - L)$ and $2\underset{n=0,1,\cdots,N-1}{\max} \hat{x}^{(i)}_n$, respectively \cite[pp. 146]{Boyd_cvxbook}}.
Note that Problem~\ref{prob:projection} is a convex problem with differentiable objective and constraint functions, and it satisfies Slater’s condition. 
Thus, the KKT conditions provide necessary and sufficient conditions for optimality.
By \eqref{equ_9}, \eqref{equ_10}, and replacing $-\frac{1}{2}\nu^{(i)}$ with $\nu$ for simplicity, we can show Lemma~\ref{lem:opt_projection}. 
}

\section*{Appendix \REPLYb{D}: Proof of Theorem~\ref{thm:E_approx_solution_time}}

First, we prove  $\hat{\tau}(\mathbf{x},\mathbf{t}) \ge \frac{M}{N}b z^{\rm (E,t)}$ for all $\mathbf{x}$ satisfying \eqref{equ:constraint_x_1} and \eqref{equ:constraint_x_3} by contradiction and construction.
	Recall $\hat{\tau}(\mathbf{x},\mathbf{t}) = \frac{M}{N}b\underset{n=0,1,\cdots,N-1}{\max} t_{N-n}\sum_{i=0}^n(i+1) x_i $.
	Suppose $\hat{\tau}(\mathbf{x},\mathbf{t}) < \frac{M}{N}b z^{\rm (E,t)}$.
	Define $\beta_i\triangleq \frac{1}{i(i+1)t_{N+1-i}},\ i\in[N]$.
	We have $L  \eqa \sum_{i=0}^{N-1}x_i 
		\eqb \sum_{i=1}^N \beta_i\Big( t_{N+1-i}\sum_{j=0}^{i-1}(j+1)x_j \Big) 
		\llc z^{\rm (E,t)}\sum_{i=1}^N\beta_i$ 
		$\eqd \frac{L} {\sum_{i=1}^{N-1}\frac{1}{i(i+1)t_{N+1-i}} + \frac{1}{Nt_{1}}} 
		\Big( \sum_{i=1}^{N-1}\frac{1}{i(i+1) t_{N+1-i} } + \frac{1}{N t_{1}} \Big) \hspace{-0.5mm}
	    = \hspace{-0.5mm} L$
%%\begin{small}
%\begin{align}
%	&L  \eqa \sum_{i=0}^{N-1}x_i 
%		\eqb \sum_{i=1}^N \beta_i\bigg( t_{N+1-i}\sum_{j=0}^{i-1}(j+1)x_j \bigg) 
%		\llc z^{\rm (E,t)}\sum_{i=1}^N\beta_i \nonumber\\
%	&\hspace{-1mm} 
%	\eqd \frac{L} {\sum_{i=1}^{N-1}\frac{1}{i(i+1)t_{N+1-i}} + \frac{1}{Nt_{1}}} 
%		\Bigg( \sum_{i=1}^{N-1}\frac{1}{i(i+1) t_{N+1-i} } + \frac{1}{N t_{1}} \Bigg) \hspace{-0.5mm}
%	    = \hspace{-0.5mm} L,\nonumber
%\end{align}
%%\end{small}
which leads to a contradiction, where $(a)$ is due to \eqref{equ:constraint_x_1}, $(b)$ is due the expansion of $\sum_{i=0}^{N-1}x_i$ in terms of $t_{N+1-i} \sum_{j=0}^{i-1}(j+1)x_j, i\in[N]$, $(c)$ is due to the assumption, and $(d)$ is due to the definition of $z^{\rm (E,t)}$ in Theorem~\ref{thm:E_approx_solution_time} and $\beta_i, i\in [N]$.
	Thus, we can show $\hat{\tau}(\mathbf{x},\mathbf{t}) \ge \frac{M}{N}b z^{\rm (E,t)}$.
	Next, it is obvious that ${\mathbf{x}}^{\rm (E,t)}$ is a feasible point and  achieves the minimum, i.e., $\hat{\tau}({\mathbf{x}}^{\rm (E,t)},\mathbf{t}) = z^{\rm (E,t)}$.
	By $\hat{\tau}(\mathbf{x},\mathbf{t}) \ge z^{\rm (E,t)}$ and $\hat{\tau}({\mathbf{x}}^{\rm (E,t)},\mathbf{t}) = z^{\rm (E,t)}$, we know that ${\mathbf{x}}^{\rm (E,t)}$ is an optimal solution of Problem~\ref{prob:approx_solution_time}.
	Therefore, we complete the proof of Theorem~\ref{thm:E_approx_solution_time}.

\section*{Appendix \REPLYb{E}: Proof of Theorem~\ref{thm:multiplicative_gap_1}}

Recall $\hat{\tau}^*_{\rm avg-ct}$, $\mathbf{x}^{\rm (E,opt)}$, and $\mathbf{x}^{\rm (E,t)}$ denote the optimal value of Problem~\ref{prob:relaxed_prob}, an optimal solution of Problem~\ref{prob:relaxed_prob}, and an optimal solution of Problem~\ref{prob:approx_solution_time}, respectively.
First, we have 
\begin{align}
	\hat{\tau}^*_{\rm avg-ct} \gea \hat{\tau}(\mathbf{x}^{\rm (E,opt)},\mathbf{t}) \geb \hat{\tau}({\mathbf{x}}^{\rm (E,t)},\mathbf{t}),\label{equ:APP_E_t}
\end{align}
where $(a)$ is due to Jensen's inequality, and $(b)$ is due to the optimality of ${\mathbf{x}}^{\rm (E,t)}$.
Then, we have
	$\frac{\mathbb{E}\left[\hat{\tau}({\mathbf{x}}^{\rm (E,t)},\mathbf{T})\right]} {\hat{\tau}^*_{\rm avg-ct}} 
	\lec \frac{\mathbb{E}\left[\hat{\tau}({\mathbf{x}}^{\rm (E,t)},\mathbf{T})\right]} {\hat{\tau}({\mathbf{x}}^{\rm (E,t)},\mathbf{t})}$ 
	$\eqd \frac{\frac{M}{N}b\mathbb{E}\left[\underset{n=0,1,\cdots,N-1}{\max}\left\{T_{(N-n)}\sum_{i=0}^n(i+1)x^{\rm (E,t)}_i\right\}\right]} {\frac{M}{N}b\mathbb{E}\left[T_{(N)}\right] x_0^{\rm (E,t)}}$
	$\eqe \mathbb{E}\Big[\underset{n=0,1,\cdots,N-1}{\max} \hspace{-0.1cm} T_{(N-n)} \Big( \hspace{-0.1cm} \frac{1}{\mathbb{E}[T_{(N)}]} \hspace{-0.1cm} + \hspace{-0.1cm} \sum_{i=1}^n\frac{1}{\mu i}\frac{1} {\mathbb{E}[T_{(N-i)}]\mathbb{E}[T_{(N+1-i)}]} \hspace{-0.1cm} \Big) \Big]$
	$\lef \mathbb{E}\Big[\underset{n=0,1,\cdots,N-1}{\max} T_{(N-n)} \Big(\frac{1} {t_0} + \sum_{i=1}^N\frac{1}{\mu i t_0^2} \Big) \Big]$
	$= \Big(\frac{1} {t_0} + \sum_{i=1}^N\frac{1}{\mu i}\frac{1} {t_0^2} \Big) \mathbb{E}\big[T_{(N)} \big]$
	$\eqg \Big(\frac{1} {t_0} + \frac{H_N}{\mu t_0^2}  \Big) \Big( \frac{H_N}{\mu}+t_0 \Big)$
%	= \frac{H_N(H_N+\mu t_0)}{\mu^2 t_0^2} 
	$\eqh \mathcal{O}\left(\log^2(N)\right)$,
where 
$(c)$ is due to \eqref{equ:APP_E_t}, 
$(d)$ is due to the definition of $\hat{\tau}(\cdot)$ in \eqref{equ:tau_hat}, 
$(e)$ is due to Theorem~\ref{thm:E_approx_solution_time} and 
\eqref{equ:Renyi},
%$\mathbb{E}[T_{(N+1-i)}]-\mathbb{E}[T_{(N-i)}] = \frac{1}{\mu i},i\in[n]$,
$(f)$ is due to $t_0 \le T_{(n)} \le T_{(N)}$ for $n\in [N]$,
$(g)$ is due to \eqref{equ:Renyi},
and $(h)$ is due to $H_N = \mathcal{O}\left(\log(N)\right)$.
Therefore, we complete the proof of Theorem~\ref{thm:multiplicative_gap_1}.

\section*{Appendix \REPLYb{F}: Proof of Lemma~\ref{lem:t'}}

For notation simplicity, denote $I(p,q) \triangleq \int_0^1 \frac{x^{p-1} (1-x)^{q-1}}{\log(x)-\mu t_0} dx,\ p,q\in\mathbb{N}_+$.
Before proving Lemma~\ref{lem:t'}, we first show the following lemma.
\begin{Lem}\label{lem:I_p_q}
	$I(p,q) = \sum_{i=0}^{q-1}(-1)^i \binom{q-1}{i} e^{\mu t_0 (p+i)} E_i(-\mu t_0 (p+i)), \ p,q\in\mathbb{N}_+$.
%	\begin{align}
%		I(p,q) = \sum_{i=0}^{q-1}(-1)^i \binom{q-1}{i} f(p+i), \ p,q\in\mathbb{N}_+,\label{equ:APP_D_2}
%	\end{align}
%	where $f(n) \triangleq e^{n\mu t_0} E_i(-n\mu t_0), n\in\mathbb{N}_+$.
\end{Lem}

\begin{IEEEproof}
	Denote $f(n) \triangleq e^{n\mu t_0} E_i(-n\mu t_0), n\in\mathbb{N}_+$, where $E_i(x)$ is the exponential integral.
	We prove Lemma~\ref{lem:I_p_q} by induction on $q$ for any fixed $p\in\mathbb{N}_+$.
%	\emph{(i) Base case:} 
	For $q=1$, we have $I(p,1) = e^{p\mu t_0}\int_{-\infty}^{-\mu t_0} \frac{e^{px}}{x} dx = f(p)$, for $p\in\mathbb{N}_+$.
%	\emph{(ii) Inductive step:} 
	Suppose that for a given $t\ge 2$, Lemma~\ref{lem:I_p_q} holds for $q=t$. 
% 	i.e., $I(p,t) = \sum_{i=0}^{t-1} (-1)^i \binom{t-1}{i} f(p+i)$.
For $q=t+1$, we have
$I(p,t+1) 
	\eqa I(p,t) - I(p+1,t)
	\eqb \sum_{i=0}^{t-1} (-1)^i \binom{t-1}{i} f(p+i) - \sum_{i=0}^{t-1} (-1)^i \binom{t-1}{i} f(p+1+i) 
	= \sum_{i=0}^{t}(-1)^i\binom{t}{i}f(p+i)$,
where $(a)$ is due to $I(p,q)=I(p+1,q)+I(p,q+1)$,
and $(b)$ is due to the assumption.
Thus, Lemma~\ref{lem:I_p_q} also holds for $q=t+1$.
%	\emph{(iii) Conclusion:} 
%Since both the base case and the inductive step have been proved as true, by induction, Lemma~\ref{lem:I_p_q} holds for all $p,q\in\mathbb{N}_+$.
Therefore, by induction, we can show Lemma~\ref{lem:I_p_q}.
\end{IEEEproof}

Now, we prove Lemma~\ref{lem:t'}.
We have $\frac{1}{t'_{n}}
	= \mathbb{E}\Big[ \frac{1}{T_{(n)}} \Big]
	= \mathbb{E}\Big[ \big({\frac{1}{T}}\big)_{(N+1-n)} \Big]
	\eqc -\mu (N+1-n) \binom{N}{n-1} I(N-n+1,n) 
	\eqd -\mu (N+1-n) \binom{N}{n-1} \sum_{i=0}^{n-1}(-1)^i \binom{n-1}{i} e^{\mu t_0(N-n+1+i)} E_i(-\mu t_0 (N-n+1+i))$,
where $(c)$ is due to the PDF of $\frac{1}{T}$, the PDF of $T_{(n)}$~\cite[Proposition 2.2]{Order_Statistics}, and $(d)$ is due to Lemma~\ref{lem:I_p_q}.
Therefore, we complete the proof of Lemma~\ref{lem:t'}.

\section*{Appendix \REPLYb{G}: Proof of Theorem~\ref{thm:multiplicative_gap_2}}
First, by Cauchy-Schwarz inequality, we have $\mathbb{E}\big[T_{(n)}\big]\mathbb{E}\left[\frac{1}{T_{(n)}}\right] \ge 1$, implying 
$\frac{1}{\mathbb{E}\big[T_{(n)}\big]} \le \mathbb{E}\left[\frac{1}{T_{(n)}}\right]$. 
%\begin{align}
%	\frac{1}{\mathbb{E}\big[T_{(n)}\big]} \le \mathbb{E}\left[\frac{1}{T_{(n)}}\right].\label{equ:Cauchy}
%\end{align}
Then, we have
$\frac{\mathbb{E}\left[\hat{\tau}({\mathbf{x}}^{\rm (E,f)},\mathbf{T})\right]} {\hat{\tau}^*_{\rm avg-ct}}
	\lea \frac{\mathbb{E}\left[\hat{\tau}({\mathbf{x}}^{\rm (E,f)},\mathbf{T})\right]}{\hat{\tau}({\mathbf{x}}^{\rm (E,t)},\mathbf{t})}   
	\eqb \frac{\sum_{n=1}^{N-1} \frac{1}{n(n+1)}\frac{1}{\mathbb{E}\left[T_{(N+1-n)}\right]} + \frac{1}{N}\frac{1}{\mathbb{E}\left[T_{(1)}\right]}}{\sum_{n=1}^{N-1} \frac{1}{n(n+1)}\mathbb{E}\left[\frac{1}{T_{(N+1-n)}}\right] + \frac{1}{N}\mathbb{E}\left[\frac{1}{T_{(1)}}\right]}
	\mathbb{E}\bigg[$ $\underset{n\in[N]}{\max}\ T_{(n)}\mathbb{E}\left[\frac{1}{T_{(n)}}\right]\bigg]$
	$\lec \mathbb{E}\left[\underset{n\in[N]}{\max}\ T_{(n)}\mathbb{E}\left[\frac{1}{T_{(n)}}\right]\right]
	\led \frac{1}{t_0} (\frac{H_N}{\mu}+t_0) 
	\eqe \mathcal{O}(\log(N))$,
%\begin{align*}
%	&\frac{\mathbb{E}\left[\hat{\tau}({\mathbf{x}}^{\rm (E,f)},\mathbf{T})\right]} {\hat{\tau}^*_{\rm avg-ct}}
%	\lee \frac{\mathbb{E}\left[\hat{\tau}({\mathbf{x}}^{\rm (E,f)},\mathbf{T})\right]}{\hat{\tau}({\mathbf{x}}^{\rm (E,t)},\mathbf{t})}   
%	\eqf \frac{\sum_{n=1}^{N-1} \frac{1}{n(n+1)}\frac{1}{\mathbb{E}\left[T_{(N+1-n)}\right]} + \frac{1}{N}\frac{1}{\mathbb{E}\left[T_{(1)}\right]}}{\sum_{n=1}^{N-1} \frac{1}{n(n+1)}\mathbb{E}\left[\frac{1}{T_{(N+1-n)}}\right] + \frac{1}{N}\mathbb{E}\left[\frac{1}{T_{(1)}}\right]} \nonumber\\
%	& \times \mathbb{E}\left[\underset{n=0,1,\cdots,N-1}{\max}\left\{T_{(N-n)}\mathbb{E}\left[\frac{1}{T_{(N-n)}}\right]\right\}\right]
%	\leg \mathbb{E}\left[\underset{n=0,1,\cdots,N-1}{\max}\left\{T_{(N-n)}\mathbb{E}\left[\frac{1}{T_{(N-n)}}\right]\right\}\right] \nonumber\\
%	&\leh \frac{H_N}{\mu t_0}+1 \eqi \mathcal{O}\left(\log(N)\right),
%\end{align*}
where $(a)$ is due to \eqref{equ:APP_E_t}, 
$(b)$ is due to the definition of $\hat{\tau}(\cdot)$ in \eqref{equ:tau_hat} and Theorem~\ref{thm:approx_solution_speed}, 
$(c)$ is due to $\frac{1}{\mathbb{E}\big[T_{(n)}\big]} \le \mathbb{E}\left[\frac{1}{T_{(n)}}\right]$, 
$(d)$ is due to $t_0 \le T_{(n)} \le T_{(N)},n\in [N]$, 
and $(e)$ is due to $H_N = \mathcal{O}\left(\log(N)\right)$.
Therefore, we complete the proof of Theorem~\ref{thm:multiplicative_gap_2}.

\section*{Appendix \REPLYb{H}: Proof of Lemma~\ref{lem:CDF_monotonicity}}

For notation simplicity, define $\mathcal{A}_l(\mathbf{s}) \triangleq \Big\{ \mathbf{T}\in\mathbb{R}^N : \frac{M}{N}bT_{(N-s_l)}\sum_{i=1}^l(s_i+1)\le t,T_j\ge t_0,j\in[N] \Big\}$.
It is clear that for all $l\in[L]$, if $\mathbf{T}\in\mathcal{A}_l(\mathbf{s})$, then the master can recover \TCOMr{$\frac{\partial\hat{\ell}(\bm{\theta};\mathcal{D})}{\partial\theta_l}$} by time $t$, 
implying 
\begin{align}
	\Pr[\tau(\mathbf{s},\mathbf{T})\le t] = \Pr[\mathbf{T}\in\cap_{l\in[L]} \mathcal{A}_l(\mathbf{s})].\label{equ:Pr_A_l}
\end{align}
Now, we prove Lemma~\ref{lem:CDF_monotonicity} by contradiction.
First, suppose for any optimal solution $\mathbf{s^*}$, there exists $m\in [L-1]$ such that $s^*_m > s^*_{m+1}$.
Construct a feasible solution $\tilde{\mathbf{s}} \triangleq (\tilde{s}_l)_{l\in [L]}$ given by \eqref{equ:lnot=m} and \eqref{equ:l=m}.
We have
%\begin{small}
\begin{align}
	& \mathcal{A}_m(\mathbf{s}^*) \subset \mathcal{A}_{m+1}(\mathbf{s}^*),\label{equ:event_s^*}\\
	& \mathcal{A}_m(\tilde{\mathbf{s}}) \subset \mathcal{A}_{m+1}(\tilde{\mathbf{s}}),\label{equ:event_tilde_s}\\
	& \mathcal{A}_l(\mathbf{s}^*) = \mathcal{A}_l(\tilde{\mathbf{s}}),\quad l\in[m-1],\label{equ:event_1}\\
	& \mathcal{A}_l(\mathbf{s}^*) \subset \mathcal{A}_l(\tilde{\mathbf{s}}),\quad l\in[L]\backslash[m],\label{equ:event_2}
\end{align}
%\end{small}
where \eqref{equ:event_s^*} is due to the assumption, 
\eqref{equ:event_tilde_s} and \eqref{equ:event_2} are due to the assumption, \eqref{equ:lnot=m}, and \eqref{equ:l=m},
and \eqref{equ:event_1} is due to \eqref{equ:l=m}.
%$\mathcal{S}\triangleq \Big\{\mathbf{T}:T_n=m, n\in[N], t_0 \le m \le \frac{Nt}{Mb\sum_{i=1}^L (s^*_i+1)} \Big\}$
Then, by \eqref{equ:event_s^*}, \eqref{equ:event_tilde_s}, \eqref{equ:event_1}, and \eqref{equ:event_2}, we have
\begin{align}
	\mathcal{S}\triangleq \Bigg\{m^{\prime}\mathbf{1}_N: m^{\prime}\in\Bigg[t_0, \frac{Nt}{Mb\sum_{i=1}^L (s^*_i+1)}\Bigg] \Bigg\}\nonumber\\
	  \in \Bigg( \bigcap_{l\in[L]} \mathcal{A}_l(\tilde{\mathbf{s}}) \Bigg) \backslash \Bigg( \bigcap_{l\in[L]} \mathcal{A}_l(\mathbf{s}^*) \Bigg).\label{equ:S_subseteq}
\end{align}
Finally, we have
$P(\tilde{\mathbf{s}},t) - P(\mathbf{s}^*,t) 
		= \Pr\big[ \tau(\tilde{\mathbf{s}},\mathbf{T})\le t \big] - \Pr\big[ \tau(\mathbf{s}^*,\mathbf{T})\le t \big]
		\eqa \Pr[\mathbf{T}\in\cap_{l\in[L]} \mathcal{A}_l(\tilde{\mathbf{s}})] - \Pr[\mathbf{T}\in\cap_{l\in[L]} \mathcal{A}_l(\mathbf{s}^*)]
		\eqb \Pr\big[\mathbf{T}\in \big( \bigcap_{l\in[L]} \mathcal{A}_l(\tilde{\mathbf{s}}) \big)
		\backslash \big( \bigcap_{l\in[L]} \mathcal{A}_l(\mathbf{s}^*) \big) \big] 
		\gec \Pr[\mathbf{T}\in\mathcal{S}] 
		\gld 0$,
where $(a)$ is due to \eqref{equ:Pr_A_l},
$(b)$ is due to $\cap_{l\in[L]} \mathcal{A}_l(\mathbf{s}^*) \subseteq \cap_{l\in[L]} \mathcal{A}_l(\tilde{\mathbf{s}})$ (by \eqref{equ:event_s^*}, \eqref{equ:event_tilde_s}, \eqref{equ:event_1}, and \eqref{equ:event_2}),
$(c)$ is due to \eqref{equ:S_subseteq}, and 
$(d)$ is due to that the Lebesgue measure is strictly positive on the non-empty open set $\mathcal{S}$.
	This indicates that $\mathbf{s}^*$ is not an optimal solution, which contradicts with the assumption.
	Therefore, we complete the proof of Lemma~\ref{lem:CDF_monotonicity}.
	
\section*{Appendix \REPLYb{I}: Proof of Theorem~\ref{thm:CDF_prob_equvalence}}

Based on Lemma~\ref{lem:CDF_monotonicity}, we first equivalently transform Problem~\ref{prob:CDF_original} to the following problem.
		\begin{align}
			\max_{\mathbf{s}} \ \  &P(\mathbf{s},t) \label{equ:app_H_prob}\\
		  	\rm{s.t.}\ \  &\eqref{equ:constraint_s_1},\eqref{equ:s_E_monotonicity}.\nonumber
		\end{align}
Next, we equivalently transform the problem in \eqref{equ:app_H_prob} to Problem~\ref{prob:CDF_equivalent}.
Introduce auxiliary variables $\mathbf{x}\triangleq (x_n)_{n=0,1,\cdots,N-1}$ with $x_n = \sum_{l\in[L]} I(s_l=n)\in\mathbb{N},\ n=0,1,\cdots,N-1$, together with \eqref{equ:s_E_monotonicity} implying $s_l = \min\left\{ i : \sum_{n=0}^{i} x_n \ge l \right\},\ l\in[L]$.
For notation simplicity, treat $\hat{k}_{-1}\triangleq 0$ for ease of illustration, and
%$\sum_{j=0}^{-1}x_j \triangleq  0$
%\TCOMg{Recall $\mathcal{U}_n = \Big\{ \sum_{i=0}^{n-1} x_i+1, \sum_{i=0}^{n-1} x_i+2, \cdots, \sum_{i=0}^{n} x_i \Big\}, n=0,1,\cdots,N-1$}.
define 
$\times$ as the Cartesian product of two sets, 
$\tilde{\mathbf{k}}_n(\mathbf{x})\triangleq (k_l)_{l=\sum_{j=0}^{n-1} x_j,\cdots,\sum_{j=0}^n x_j-1} \in\mathbb{N}^{x_n},n=0,1,\cdots,N-1$, 
$\tilde{\mathbf{k}}_N(\mathbf{x})\triangleq k_L \in\mathbb{N}$,
%$\mathcal{U}_n(\mathbf{x},\hat{\mathbf{k}}) \triangleq \Big\{ \tilde{\mathbf{k}}_n(\mathbf{x})\in\mathbb{N}^{x_n} : \big|\tilde{\mathbf{k}}_n(\mathbf{x})\big|_1 =\hat{k}_{n-1} \Big\},n=0,1,\cdots,N-1$
%and $\mathcal{U}(1,\hat{k}_{N-1}) \triangleq \Big\{\hat{k}_{N-1}\Big\}$.
\TCOMg{$\mathcal{U}(x,\hat{k}) \triangleq \Big\{ \tilde{\mathbf{k}}\in\mathbb{N}^{x} : \big|\tilde{\mathbf{k}}\big|_1 =\hat{k} \Big\}$.}
%and $\mathcal{U}(1,\hat{k}_{N-1}) \triangleq \Big\{\hat{k}_{N-1}\Big\}$.
%We treat $\hat{k}_{-1}\triangleq 0$ and $\sum_{j=0}^{-1}x_j \triangleq  0$ for ease of illustration.
Recall $\mathcal{K}(\mathbf{s},N,L)\triangleq \left\{ \mathbf{k}\in\mathbb{N}^{L+1} : \sum_{i=0}^{l-1} k_i\le s_{l}, l\in[L], \sum_{i=0}^{L}k_i=N \right\}$.
Thus, we have 
\begin{align}
	\mathcal{K}(s,N,L) 
	= \underset{\hat{\mathbf{k}}\in\hat{\mathcal{K}}(N)}{\bigcup} \Big( \mathcal{U}(x_0,\hat{k}_{-1}) \times \mathcal{U}(x_1,\hat{k}_{0}) \times \cdots \nonumber\\
	\times \mathcal{U}(x_{N-1},\hat{k}_{N-2}) \times \mathcal{U}(1,\hat{k}_{N-1}) \Big).\label{equ:K_decomposition}
\end{align}
Besides, we have 
%$P(\mathbf{s},t)
%	= \underset{\mathbf{k}\in\mathcal{K}(\mathbf{s},N,L)}{\sum} \binom{N} {k_0,k_1,\cdots,k_L} \overset{L}{\underset{l=0}{\prod}}\big( F_l(\mathbf{s},t)-F_{l+1}(\mathbf{s},t) \big)^{k_l}$
\begin{small}
\begin{align*}
&P(\mathbf{s},t)
	= \underset{\mathbf{k}\in\mathcal{K}(\mathbf{s},N,L)}{\sum} \binom{N} {k_0,k_1,\cdots,k_L} \overset{L}{\underset{l=0}{\prod}}\big( F_l(\mathbf{s},t)-F_{l+1}(\mathbf{s},t) \big)^{k_l}\nonumber\\
	& \eqa \underset{\hat{\mathbf{k}}\in\hat{\mathcal{K}}(N)}{\sum} \
  		\underset{\substack{\tilde{\mathbf{k}}_0(\mathbf{x}) \in \mathcal{U}(x_0,\hat{k}_{-1})}}{\sum} \
  		\underset{\substack{\tilde{\mathbf{k}}_1(\mathbf{x}) \in \mathcal{U}(x_1,\hat{k}_{0})}}{\sum} \
  		\cdots \hspace{-0.15cm}
  		\underset{\substack{\tilde{\mathbf{k}}_{N-1}(\mathbf{x}) \in \mathcal{U}(x_{N-1},\hat{k}_{N-2})}}{\sum} \nonumber\\ 
  	& \underset{\substack{\tilde{\mathbf{k}}_{N}(\mathbf{x}) \in \mathcal{U}(1,\hat{k}_{N-1})}}{\sum}
  		\binom{N} {k_0,k_1,\cdots,k_L} \overset{L}{\underset{l=0}{\prod}}\big( F_l(\mathbf{s},t)-F_{l+1}(\mathbf{s},t) \big)^{k_l}\\
	& \eqb \underset{\hat{\mathbf{k}}\in\hat{\mathcal{K}}(N)}{\sum} \
		\underset{\substack{\tilde{\mathbf{k}}_{N}(\mathbf{x}) \in \mathcal{U}(1,\hat{k}_{N-1})}}{\sum} \
		\underset{\substack{\tilde{\mathbf{k}}_{N-1}(\mathbf{x}) \in \mathcal{U}(x_{N-1},\hat{k}_{N-2})}}{\sum} \ \hspace{-0.15cm}
		\cdots \hspace{-0.15cm}
		\underset{\substack{\tilde{\mathbf{k}}_2(\mathbf{x}) \in \mathcal{U}(x_1,\hat{k}_{0})}}{\sum} \nonumber\\
	& \underset{\substack{\tilde{\mathbf{k}}_1(\mathbf{x}) \in \mathcal{U}(x_0,\hat{k}_{-1})}}{\sum} 
		\binom{N} {k_{x_0+x_1},k_{x_0+x_1+1},\cdots,k_L,\hat{k}_0} \\
	& \binom{\hat{k}_0} {k_{x_0},k_{x_0+1},\cdots,k_{x_0+x_1-1}} 
		\Bigg(\prod_{l=x_0+x_1}^{L-1}\bigg(F_l(\mathbf{s},t)-F_{l+1}(\mathbf{s},t)\bigg)^{k_l} \Bigg)\nonumber\\
	& \times \bigg(F_l(\mathbf{s},t)-F_{l+1}(\mathbf{s},t)\bigg)^{k_L}
		\prod_{l=x_0}^{x_0+x_1-1}\hspace{-0.0cm}\bigg(F_l(\mathbf{s},t)-F_{l+1}(\mathbf{s},t)\bigg)^{k_l}\\
	& = \underset{\hat{\mathbf{k}}\in\hat{\mathcal{K}}(N)}{\sum} \
		\underset{\substack{\tilde{\mathbf{k}}_{N}(\mathbf{x}) \in \mathcal{U}(1,\hat{k}_{N-1})}}{\sum} \
		\underset{\substack{\tilde{\mathbf{k}}_{N-1}(\mathbf{x}) \in \mathcal{U}(x_{N-1},\hat{k}_{N-2})}}{\sum} \hspace{-0.2cm}
		\cdots \hspace{-0.2cm}
		\underset{\substack{\tilde{\mathbf{k}}_2(\mathbf{x}) \in \mathcal{U}_2(\mathbf{x},\hat{\mathbf{k}})}}{\sum} \nonumber\\
	& \binom{N} {k_{x_0+x_1},k_{x_0+x_1+1},\cdots,k_L,\hat{k}_0} \hspace{-0.1cm}
		\Bigg(\prod_{l=x_0+x_1}^{L-1} \hspace{-0.2cm} \bigg(F_l(\mathbf{s},t)-F_{l+1}(\mathbf{s},t)\bigg)^{k_l}\hspace{-0.1cm}\Bigg)\\
	& \times \bigg(F_l(\mathbf{s},t)-F_{l+1}(\mathbf{s},t)\bigg)^{k_L} \hspace{-0.5cm}
	    \underset{\substack{\tilde{\mathbf{k}}_1(\mathbf{x}) \in \mathcal{U}(x_1,\hat{k}_{0})}}{\sum} \hspace{-0.1cm}
		\binom{\hat{k}_0} {k_{x_0},k_{x_0+1},\cdots,k_{x_0+x_1-1}} \hspace{-0.0cm} \\
	& \prod_{l=x_0}^{x_0+x_1-1}\hspace{-0.0cm}\bigg(F_l(\mathbf{s},t)-F_{l+1}(\mathbf{s},t)\bigg)^{k_l}
	  \eqc  \sum_{\hat{\mathbf{k}}\in\hat{\mathcal{K}}(N)} \hspace{-0.1cm} 
		\binom{N} {\hat{k}_0,\hat{k}_1,\cdots,\hat{k}_{N-1}}\\
	& \Bigg(\prod_{n=0}^{N-2}\hspace{-0.1cm}\bigg( F_{\sum_{i=0}^n x_i}(\mathbf{s},t)-F_{\sum_{i=0}^{n+1} x_i}(\mathbf{s},t)  \bigg)^{\hat{k}_n} \hspace{-0.1cm} \Bigg)
		\bigg( F_{\sum_{i=0}^{N-1} x_i}(\mathbf{s},t) \\
	& - F_{L+1}(\mathbf{s},t) \bigg)^{\hat{k}_{N-1}} \hspace{-0.1cm}
	  \eqd \hat{P}(\mathbf{x},t),
\end{align*}
\end{small}where 
	$(a)$ is due to \eqref{equ:K_decomposition}, 
%	$(b)$ is due to changing the order of summation, 
	$(b)$ is due to $\binom{n}{k_0,k_1,\cdots,k_{r-1}}=\binom{n}{\sum_{i=0}^{m-1}k_i,k_m,\cdots,k_{r-1}} \binom{\sum_{i=1}^{m-1}k_i}{k_0,\cdots,k_{m-1}},m\in[r-1]$ and the definition of $\mathcal{U}(x_0,\hat{k}_{-1})$,
	$(c)$ is due to the multinomial theorem, 
	and $(d)$ is due to \eqref{equ:s_x_sum}
	and the definition of $\hat{P}(\mathbf{x},t)$ in \eqref{equ:CDF_fobj_x}.
%	Besides, it is obvious that the constraints of the above problem is equivalent to those of Problem~\ref{prob:CDF_equivalent}.
	By $x_n = \sum_{l\in[L]} I(s_l=n) \in\mathbb{N},\ n=0,1,\dots,N-1$, we can obtain \eqref{equ:constraint_x_1} and \eqref{equ:constraint_x_2}. 
	Thus, the problem in \eqref{equ:app_H_prob} is equivalent to Problem~\ref{prob:CDF_equivalent}, implying that \eqref{equ:change_of_var_x} and \eqref{equ:change_of_var_s} hold and $P^*(t) = \hat{P}^*(t)$.
	Therefore, we can show Theorem~\ref{thm:CDF_prob_equvalence}.
	
\section*{Appendix \REPLYb{J}: Proof of Lemma~\ref{lem:CDF_equi_stoch_prob}}

First, we show
\begin{small} 
\begin{align}
	f(m,n) 
	&\triangleq \hspace{-0.1cm} \sum_{i=0}^{m+1} \hspace{-0.1cm} \frac{i+2}{n+i+1}\binom{2n+i-1}{n-1} \hspace{-0.1cm} 
	- \hspace{-0.1cm} \frac{m+2}{n+m+2} \hspace{-0.05cm} \binom{2n+m+1}{n} \hspace{-0.05cm}\nonumber\\
	 &= \hspace{-0.05cm}0, n\in\mathbb{N}_+, m\in\mathbb{N},\label{equ:induction}
\end{align} 
\end{small}by induction on $m$ for any fixed $n\in\mathbb{N}_+$.
For $m=0$, $f(m,n) = \frac{2}{n+1}\binom{2n-1}{n-1}+\frac{3}{n+2}\binom{2n}{n-1}
	- \frac{2}{n+2}\binom{2n+1}{n} \eqa 0$, where $(a)$ is due to $\binom{n}{m} = \frac{n!}{m!(n-m)!}$.
Suppose that for a given $t\ge 1$, $f(m,n)=0$ holds for $m=t$.
Then, for $m=t+1$, $f(m,n)= \sum_{i=0}^{t+2} \frac{i+2}{n+i+1}\binom{2n+i-1}{n-1} - \frac{t+3}{n+t+2}\binom{2n+t+2}{n}
	\eqb  \frac{t+2}{n+t+2}\binom{2n+t+1}{n} + \frac{t+4}{n+t+3}\binom{2n+t+1}{n-1} - \frac{t+3}{n+t+2}\binom{2n+t+2}{n}
	\eqc 0$,
where $(b)$ is due to the assumption, and $(c)$ is due to $\binom{n}{m} = \frac{n!}{m!(n-m)!}$.
Thus, $f(m,n)=0$ also holds for $m=t+1$.
	Therefore, by induction, we can show \eqref{equ:induction}.

Next, we show 
\begin{align}
	\Big| \hat{\mathcal{K}}_m(N) \Big| = \frac{m+2}{N+m+1}\binom{2N+m-1}{N-1},\ m\in\mathbb{N},\ N\in\mathbb{N}_+,\label{equ:hat_mathcal_K_m}
\end{align}
where $\hat{\mathcal{K}}_m(N) \triangleq \Big\{\hat{\mathbf{k}}\in\mathbb{N}^{N} : \sum_{i=0}^{n-1} \hat{k}_i \le n+m, n\in[N-1], \sum_{i=0}^{N-1} \hat{k}_i=N+m \Big\}$, by induction on $N$ for any fixed $m\in\mathbb{N}$.
For $N=1$, $\big| \hat{\mathcal{K}}_m(N) \big|=1=\frac{m+2}{N+m+1}\binom{2N+m-1}{N-1}$.
Suppose that for a given $n \ge 1$, \eqref{equ:hat_mathcal_K_m} holds for $N=n$.
Then, for $N=n+1$, 
	$\big| \hat{\mathcal{K}}_{m}(n+1) \big| 
		\eqd \sum_{i=0}^{m+1} \big| \hat{\mathcal{K}}_{m}(n+1) \cap \big\{\hat{\mathbf{k}}\in\mathbb{N}^n : \hat{k}_0=i \big\} \big|
		  = \sum_{i=0}^{m+1} \big| \hat{\mathcal{K}}_{m+1-i}(n) \big|
		  = \sum_{i=0}^{m+1} \big| \hat{\mathcal{K}}_{i}(n) \big|
		\eqe \sum_{i=0}^{m+1} \frac{i+2}{n+i+1}\binom{2n+i-1}{n-1}
		\eqf \frac{m+2}{n+m+2}\binom{2n+m+1}{n}$,
where $(d)$ is due to $\hat{\mathcal{K}}_{m}(n+1) = \overset{m+1}{\underset{i=0}{\bigcup}} \Big( \hat{\mathcal{K}}_{m}(n+1) \cap \big\{\hat{\mathbf{k}}\in\mathbb{N}^n : \hat{k}_0=i \big\} \Big)$, 
	$(e)$ is due to the assumption, 
	and $(f)$ is due to \eqref{equ:induction}.
Thus, \eqref{equ:hat_mathcal_K_m} also holds for $N=n+1$.
	Therefore, by induction, we can show \eqref{equ:hat_mathcal_K_m}.
	
Finally, we have 
$\big| \hat{\mathcal{K}}(N) \big| 
= \big| \hat{\mathcal{K}}_{0}(N) \big| 
\eqg \frac{2}{N+1}\binom{2N-1}{N-1}
= \prod_{n=2}^N \frac{n+N+1}{n}
\ge \prod_{n=2}^N 2
= 2^{N-1}$, where $(g)$ is due to \eqref{equ:hat_mathcal_K_m}. 

Therefore, we complete the proof of Lemma~\ref{lem:CDF_equi_stoch_prob}.

\section*{Appendix \REPLYb{K}: Proof of Theorem~\ref{thm:CDF_conditional_probability}}

First, by $\Pr\Big[\bm\xi = \hat{\mathbf{k}}\Big] = \frac{1}{\left|\hat{\mathcal{K}}(N)\right|}$, we obtain \eqref{equ:CDF_conditional_probability_1} and \eqref{equ:CDF_conditional_probability_2}.
We have
$\Pr[\xi_n=\hat{k}_n | \xi_0 = \hat{k}_0,\cdots,\xi_{n-1}=\hat{k}_{n-1}] \hspace{-0.1cm}
	\eqa \hspace{-0.1cm} \frac{\big| \hat{\mathcal{K}}(N) \cap \big\{{\bm\zeta}\in\mathbb{N}^N : \zeta_0=\hat{k}_0,\cdots,\zeta_{n-1}=\hat{k}_{n-1},\zeta_n=\hat{k}_n \big\} \big|} {\big| \hat{\mathcal{K}}(N) \cap \big\{{\bm\zeta}\in\mathbb{N}^N : \zeta_0=\hat{k}_0,\cdots,\zeta_{n-1}=\hat{k}_{n-1} \big\} \big|} \hspace{-0.1cm}
	\eqb \hspace{-0.1cm} \frac{\big| \hat{\mathcal{K}}_{n+1}(N-n-1) \big|} {\big| \hat{\mathcal{K}}_{n}(N-n) \big|} \hspace{-0.1cm}
	\eqc \hspace{-0.1cm} \frac{\left(n+3-\sum_{i=0}^n\hat{k}_i\right)\left(N+n+1-\sum_{i=0}^{n-1}\hat{k}_i\right) } {\left(n+2-\sum_{i=0}^{n-1}\hat{k}_i\right)\left(N+n+2-\sum_{i=0}^n\hat{k}_i\right)}$ 
	$\times\frac{\binom{2N+n-\sum_{i=0}^n\hat{k}_i}{N-1}} {\binom{2N+n-1-\sum_{i=0}^{n-1}\hat{k}_i}{N-1}}$,
where
$(a)$ is due to the definition of conditional probability,
$(b)$ is due to $\hat{\mathcal{K}}(N) \cap \big\{{\bm\zeta}\in\mathbb{N}^N : \zeta_0=\hat{k}_0,\cdots,\zeta_n=\hat{k}_n \big\} = \hat{\mathcal{K}}_{n+1}(N-n-1)$, and 
$(c)$ is due to \eqref{equ:hat_mathcal_K_m}.
Next, by \eqref{equ:CDF_conditional_probability_1} and \eqref{equ:CDF_conditional_probability_2}, we obtain $\Pr\Big[\bm\xi = \hat{\mathbf{k}}\Big] = \frac{1}{\left|\hat{\mathcal{K}}(N)\right|}$.
We have
$\Pr\Big[\bm\xi=\hat{\mathbf{k}}\Big]
	\eqd \overset{N-1}{\underset{n=0}{\prod}} \Pr[\xi_n=\hat{k}_n | \xi_0 = \hat{k}_0,\cdots,\xi_{n-1}=\hat{k}_{n-1}]
	\eqe \overset{N-1}{\underset{n=0}{\prod}} \frac{\big| \hat{\mathcal{K}}_{n+1}(N-n-1) \big|} {\big| \hat{\mathcal{K}}_{n}(N-n) \big|}
	= \frac{1}{\big| \hat{\mathcal{K}}(N) \big|}$,
where $(d)$ is due to the chain rule of probability and $(e)$ is due to \eqref{equ:CDF_conditional_probability_1}, \eqref{equ:CDF_conditional_probability_2} and \eqref{equ:hat_mathcal_K_m}.
Therefore, we complete the proof of Theorem~\ref{thm:CDF_conditional_probability}.

\section*{Appendix \REPLYb{L}: \TCOMg{Proof of Lemma~\ref{lem:CDF_large_t_asymp_completion_probability}}}

For notation simplicity, define $\Omega(N)\triangleq \Big\{\hat{\mathbf{k}}\in\mathbb{N}^N : \sum_{n=0}^{N-1}\hat{k}_n=N \Big\}$, 
	$\Upsilon_1(N) \triangleq \Big\{\hat{\mathbf{k}}\in\mathbb{N}^N:\hat{k}_n=0,n\in\{0,1,\cdots,N-2\} \backslash \big(\mathcal{N}_1(\mathbf{x})-1\big) \Big\}$.
	Recall $\hat{\mathcal{K}}(N)$ given in \eqref{equ:hat_K} and $\mathcal{N}_1(\mathbf{x}) \triangleq \left\{n\in[N-1] : x_n\neq 0\right\}$.
Before proving Lemma~\ref{lem:CDF_large_t_asymp_completion_probability}, we first show the following lemma.
%Recall that $\hat{\mathbf{k}} \triangleq (k_n)_{n=0,1,\cdots,N-1}$, 
%$\hat{\mathcal{K}}(N) \triangleq \Big\{ \hat{\mathbf{k}}\in \mathbb{N}^{N} : \sum_{i=0}^{n-1} \hat{k}_i \le n, n\in[N], \sum_{i=0}^{N-1} \hat{k}_i=N \Big\}$,
%$\mathcal{N}_1(\mathbf{x}) \triangleq \left\{n\in[N-1] : x_n\neq 0\right\}$,
%and $\mathcal{N}_2(\mathbf{x}) \triangleq \underset{n\in\mathcal{N}_1(\mathbf{x})}{\argmin} \frac{n+1}{\sum_{i=0}^{n} (i+1)x_i}$.
\begin{Lem}\label{lem:app_I_opt_solution}
	The optimal value of $\underset{\hat{\mathbf{k}}\in\Upsilon_2(N)}{\min} \ \  f_{\rm IP}(\mathbf{\hat{\mathbf{k}},x})$, 
	where $f_{\rm IP}(\mathbf{\hat{\mathbf{k}},x}) \hspace{-0.0cm} \triangleq \hspace{-0.0cm} \underset{n\in\mathcal{N}_1(\mathbf{x})-1}{\sum}\frac{\hat{k}_n}{\sum_{i=0}^{n+1}(i+1)x_i}$ and
	$\Upsilon_2(N) \triangleq \Omega(N) \backslash \hat{\mathcal{K}}(N) \cap \Upsilon_1(N)$,
	denoted by $f_{\rm IP}^*(\mathbf{x})$, is $f_{\rm IP}^*(\mathbf{x}) = \underset{{n\in\mathcal{N}_1(\mathbf{x})}}{\min} \frac{n+1}{\sum_{i=0}^n (i+1)x_i}$. 
\end{Lem}

\begin{IEEEproof}
	For notation simplicity, let 
%	$\Omega(N)\triangleq \Big\{\hat{\mathbf{k}}\in\mathbb{N}^N : \sum_{n=0}^{N-1}\hat{k}_n=N \Big\}$ 
%	and $\Upsilon_1(N) \triangleq \Big\{\hat{\mathbf{k}}\in\mathbb{N}^N:\hat{k}_n=0,n\notin\mathcal{N}_1(\mathbf{x})-1\cup\{N-1\} \Big\}$, implying $\Upsilon_2(N) \triangleq \Omega(N) \backslash \hat{\mathcal{K}}(N) \cap \Upsilon_1(N)$ 
	$\mathcal{T}_n(N) \triangleq \Big\{ \hat{\mathbf{k}}\in\Upsilon_2(N) : \Big\|\big(\hat{k}_{i}\big)_{i=0,\cdots,N-2}\Big\|_0=n \Big\}$, $n\in[|\mathcal{N}_1(\mathbf{x})|]$,
	$\mathcal{T}_{\mathcal{S}}(N) \triangleq \Big\{ \hat{\mathbf{k}}\in\mathcal{T}_{|\mathcal{S}|}(N) : \hat{k}_n\neq 0,n\in\mathcal{S} \Big\}$, $\mathcal{S}\subseteq \mathcal{N}_1(\mathbf{x})-1$, $\mathcal{S}\neq\emptyset$,
	implying $\Upsilon_2(N) = \underset{n\in[|\mathcal{N}_1(\mathbf{x})|]}{\bigcup} \mathcal{T}_n (N)$, 
	$\mathcal{T}_i(N) = \underset{|\mathcal{S}|=i,\mathcal{S}\subseteq \mathcal{N}_1(\mathbf{x})-1}{\bigcup} \mathcal{T}_{\mathcal{S}}(N)$,
	and $\mathcal{T}_i(N)\cap\mathcal{T}_j(N)=\emptyset$, $i \neq j$.
	First, we decompose the problem in Lemma~\ref{lem:app_I_opt_solution} (i.e., $\underset{\hat{\mathbf{k}}\in\Upsilon_2(N)}{\min} f_{\rm IP}(\hat{\mathbf{k}},\mathbf{x})$) into $|\mathcal{N}_1(\mathbf{x})|$ subproblems, \TCOMb{i.e.,
%	\begin{align}
%		f_{\rm IP}^*(\mathbf{x}) 
%		= \underset{\hat{\mathbf{k}}\in\Upsilon_2(N)}{\min} f_{\rm IP}(\hat{\mathbf{k}},\mathbf{x}) 
%		= \underset{m\in[|\mathcal{N}_1(\mathbf{x})|]}{\min}\  \underset{\hat{\mathbf{k}}\in\mathcal{T}_m(N)}{\min} f_{\rm IP}(\hat{\mathbf{k}},\mathbf{x}),\label{equ:subprob_partition}
%	\end{align}
	\begin{align}
		f_{\rm IP}^*(\mathbf{x}) 
		= \underset{m\in[|\mathcal{N}_1(\mathbf{x})|]}{\min} f^*_m,\label{equ:subprob_partition}
	\end{align}
	where $f^*_m \triangleq \underset{\hat{\mathbf{k}}\in\mathcal{T}_m(N)}{\min} f_{\rm IP}(\hat{\mathbf{k}},\mathbf{x}),m\in[|\mathcal{N}_1(\mathbf{x})|]$.}
%	implying that subproblem $m$ is $\underset{\hat{\mathbf{k}}\in\mathcal{T}_m(N)}{\min} f_{\rm IP}(\hat{\mathbf{k}},\mathbf{x}),m\in[|\mathcal{N}_1(\mathbf{x})|]$.
%  	Let $f^*_m \triangleq \underset{\hat{\mathbf{k}}\in\mathcal{T}_m(N)}{\min} f_{\rm IP}(\hat{\mathbf{k}},\mathbf{x})$, $m\in[|\mathcal{N}_1(\mathbf{x})|]$ be the optimal value of subproblem $m$.
  	Then, we analyze $f^*_m,m\in[|\mathcal{N}_1(\mathbf{x})|]$.
  	
	(i) \TCOMb{We analyze $f^*_1$.}
  	\begin{small}
  	\begin{align}
  		&f^*_1 
  		\hspace{-0.0cm}=\hspace{-0.0cm} \underset{\hat{\mathbf{k}}\in\mathcal{T}_1(N)}{\min} f_{\rm IP}(\hat{\mathbf{k}},\mathbf{x})
  		\hspace{-0.0cm}=\hspace{-0.0cm} \underset{\substack{n\in\\\mathcal{N}_1(\mathbf{x})-1}}{\min}  \underset{\hat{\mathbf{k}}\in \mathcal{T}_{\{n\}}(N)}{\min} f_{\rm IP}(\hat{\mathbf{k}},\mathbf{x})\nonumber\\
  		\hspace{-0.1cm}&\overset{(a)}{=}\hspace{-0.3cm} \underset{\substack{n\in\\\mathcal{N}_1(\mathbf{x})-1}}{\min}\ \underset{\substack{\hat{\mathbf{k}}\in\mathbb{N}^N:n+1<\hat{k}_n\le N,\\ \hat{k}_n+\hat{k}_{N-1}=N,  \hat{k}_i=0, \\ i\in\{0,\cdots,N-2\}\backslash\{n\}}}{\min} \hspace{-0.5cm} f_{\rm IP}(\hat{\mathbf{k}},\mathbf{x})
  		\hspace{-0.1cm}\overset{}{=}\hspace{-0.3cm} \underset{n\in\mathcal{N}_1(\mathbf{x})}{\min} \frac{n+1}{\sum_{i=0}^n(i+1)x_i},\label{equ:f_star_1}
  	\end{align}
  	\end{small}where $(a)$ is due to $\mathcal{T}_{\{n\}}(N) = \Upsilon_2(N) \cap \Big\{ \hat{\mathbf{k}}\in\mathbb{N}^N : \hat{k}_n\neq 0,\hat{k}_i=0,i\in\{0,\cdots,N-2\}\backslash\{n\} \Big\}
	= \Omega(N) \cap \Upsilon_1(N) \backslash \hat{\mathcal{K}}(N)
	\cap \Big\{ \hat{\mathbf{k}}\in\mathbb{N}^N : \hat{k}_n\neq 0,\hat{k}_i=0,i\in\{0,\cdots,N-2\}\backslash\{n\} \Big\}
	= \Big( \Omega(N) \cap \Upsilon_1(N) \cap \Big\{ \hat{\mathbf{k}}\in\mathbb{N}^N : \hat{k}_n\neq 0,\hat{k}_i=0,i\in\{0,\cdots,N-2\}\backslash\{n\} \Big\} \Big) \backslash \Big( \hat{\mathcal{K}}(N) \cap \Big\{ \hat{\mathbf{k}}\in\mathbb{N}^N : \hat{k}_n\neq 0,\hat{k}_i=0,i\in\{0,\cdots,N-2\}\backslash\{n\} \Big\} \Big)
	= \Big\{\hat{\mathbf{k}}\in\mathbb{N}^N : \hat{k}_n+\hat{k}_{N-1}=N, \hat{k}_i=0,i\in\{0,\cdots,N-2\}\backslash\{n\} \Big\} 
	  \backslash 
	  \Big\{ \hat{\mathbf{k}}\in\mathbb{N}^N: \hat{k}_n\le n+1, \hat{k}_n+\hat{k}_{N-1}=N, \hat{k}_i=0,i\in\{0,\cdots,N-2\}\backslash\{n\} \Big\}
	= \Big\{ \hat{\mathbf{k}}\in\mathbb{N}^N: n+1<\hat{k}_n\le N, \hat{k}_n+\hat{k}_{N-1}=N, \hat{k}_i=0,i\in\{0,\cdots,N-2\}\backslash\{n\} \Big\}$.
	
	(ii) \TCOMb{We show $f^*_m>f^*_1,m\in\{2,3,\cdots,|\mathcal{N}_1(\mathbf{x})|\}$.}
	\begin{small}
	\begin{align}
		&f^*_m 
		= \underset{\hat{\mathbf{k}}\in\mathcal{T}_m(N)}{\min} f_{\rm IP}(\hat{\mathbf{k}},\mathbf{x})
		= \underset{\substack{\mathcal{S}\subseteq \mathcal{N}_1(\mathbf{x})-1}}{\min} \  \underset{\hat{\mathbf{k}}\in\mathcal{T}_{\mathcal{S}}(N)}{\min} f_{\rm IP}(\hat{\mathbf{k}},\mathbf{x})\nonumber\\
		& \overset{(c)}{=} \underset{\substack{\mathcal{S}\subseteq \mathcal{N}_1(\mathbf{x})-1}}{\min}\ \underset{\substack{i\in[m]}}{\min} \  \underset{\hat{\mathbf{k}}\in\mathcal{U}_{i,\mathcal{S}}(N)}{\min} f_{\rm IP}(\hat{\mathbf{k}},\mathbf{x}),\nonumber\\
		& = \underset{\substack{\mathcal{S}\subseteq \mathcal{N}_1(\mathbf{x})-1}}{\min}\ \underset{\substack{i\in[m]}}{\min} \ 
		\underset{\hat{\mathbf{k}}\in\mathcal{U}_{i,\mathcal{S}}(N)}{\min} \sum_{n\in\{a_1,\cdots,a_m\}} \frac{\hat{k}_n}{\sum_{j=0}^{n+1}(j+1)x_j}\nonumber\\
		& \gld \underset{\substack{\mathcal{S}\subseteq \mathcal{N}_1(\mathbf{x})-1}}{\min}\ \underset{\substack{i\in[m]}}{\min} \ 
		\underset{\hat{\mathbf{k}}\in\mathcal{U}_{i,\mathcal{S}}(N)}{\min} \sum_{n\in\{a_1,\cdots,a_{i}\}} \frac{\hat{k}_n}{\sum_{j=0}^{a_{i}+1} (j+1)x_j}\nonumber\\
		& = \underset{\substack{\mathcal{S}\subseteq \mathcal{N}_1(\mathbf{x})-1}}{\min}\ \underset{\substack{i\in[m]}}{\min} \ 
		\underset{\hat{\mathbf{k}}\in\mathcal{U}_{i,\mathcal{S}}(N)}{\min} \frac{\sum_{n\in\{a_1,\cdots,a_{i}\}} \hat{k}_n} {\sum_{j=0}^{a_{i}+1} (j+1)x_j}\nonumber\\
		& \gee \underset{\substack{\mathcal{S}\subseteq \mathcal{N}_1(\mathbf{x})-1}}{\min}\ \underset{\substack{i\in[m]}}{\min} \ 
		 \frac{a_{i}+2}{\sum_{j=0}^{a_{i}+1} (j+1)x_j}
		\eqf f^*_1, \TCOMb{m\in\{2,\cdots,|\mathcal{N}_1(\mathbf{x})|\}},\label{equ:f_star_m_ge_f_star_1}
%		&\overset{(d)}{>} f^*_1,
	\end{align}
	\end{small}where $\mathcal{S}\triangleq\{a_1,\cdots,a_m\}$ with $0 \le a_1<a_2<\cdots<a_m \le N-2$, 
	$\mathcal{U}_{i,\mathcal{S}}(N) \triangleq \Big\{ \hat{\mathbf{k}}\in\mathcal{T}_{\mathcal{S}}(N) : \hat{k}_{a_1},\cdots\hat{k}_{a_m}>0, \hat{k}_{a_1}\le a_1+1,\hat{k}_{a_1}+\hat{k}_{a_2}\le a_2+1,\cdots,\hat{k}_{a_1}+\cdots+\hat{k}_{a_{i-1}}\le a_{i-1}+1,\hat{k}_{a_1}+\cdots+\hat{k}_{a_{i}}>a_{i}+1 \Big\}$, 
	$(c)$ is due to $\mathcal{T}_{\mathcal{S}}(N) = \bigcup_{i\in[m]}\mathcal{U}_{i,\mathcal{S}}(N)$ (by the definition of $\mathcal{T}_{\mathcal{S}}(N)$ and $\mathcal{U}_{i,\mathcal{S}}(N)$),
	$(d)$ is due to $\{a_1,\cdots,a_i\} \subseteq \{a_1,\cdots,a_m\}$ and $x_{a_i+1}>0$,
	$(e)$ is due to the definition of $\mathcal{U}_{i,\mathcal{S}}(N)$, and
	$(f)$ is due to \eqref{equ:f_star_1}.
%	Then, we have
%	\begin{align}
%		&\underset{\hat{\mathbf{k}}\in\mathcal{U}_{i,\mathcal{S}}(N)}{\min} f_{\rm IP}(\hat{\mathbf{k}},\mathbf{x})
%		= \underset{\hat{\mathbf{k}}\in\mathcal{U}_{i,\mathcal{S}}(N)}{\min} \sum_{n\in\{a_1,\cdots,a_m\}} \frac{\hat{k}_n}{\sum_{i=0}^{n+1}(i+1)x_i}
%		> \underset{\hat{\mathbf{k}}\in\mathcal{U}_{i,\mathcal{S}}(N)}{\min} \sum_{n\in\{a_1,\cdots,a_{i+1}\}} \frac{\hat{k}_n}{\sum_{i=0}^{a_{i+1}+1} (i+1)x_i}\nonumber\\
%		&= \underset{\hat{\mathbf{k}}\in\mathcal{U}_{i,\mathcal{S}}(N)}{\min} \frac{\sum_{n\in\{a_1,\cdots,a_{i+1}\}} \hat{k}_n}{\sum_{i=0}^{a_{i+1}+1} (i+1)x_i} \overset{(d)}{\ge} \frac{a_{i+1}+2}{\sum_{i=0}^{a_{i+1}+1} (i+1)x_i}
%		\ge f^*_1,\label{equ:f_star_m_due_1}
%	\end{align}
%	where $(d)$ is due to the definition of $\mathcal{U}_{i,\mathcal{S}}(N)$.
%	Next, we have 
%	\begin{align}
%		\underset{\substack{\mathcal{S}=\{a_1,\cdots,a_m\}: \\ \mathcal{S}\subseteq\{0\}\cup[|\mathcal{N}_1(\mathbf{x})|]}}{\min} \underset{i\in\{0\}\cup[m-1]}{\min} \  \underset{\hat{\mathbf{k}}\in\mathcal{U}_{i,\mathcal{S}}(N)}{\min} f_{\rm IP}(\hat{\mathbf{k}},\mathbf{x})
%		\overset{(e)}{>} f^*_1,\label{equ:f_star_m_due_2}
%	\end{align}
%	where $(e)$ is due to \eqref{equ:f_star_m_due_1}.
%	Finally, by \eqref{equ:f_star_m} and \eqref{equ:f_star_m_due_2}, we have 
%	\begin{align}
%		f^*_m > f^*_1,\ m\in\{2,3,\cdots,|\mathcal{N}_1(\mathbf{x})|\}.\label{equ:f_star_m_ge_f_star_1}
%	\end{align}

	By \eqref{equ:subprob_partition}, \eqref{equ:f_star_1}, and \eqref{equ:f_star_m_ge_f_star_1}, we have 
	$f_{\rm IP}^*(\mathbf{x}) \hspace{-0.0cm} = \hspace{-0.2cm}\underset{n\in\mathcal{N}_1(\mathbf{x})}{\min} \frac{n+1}{\sum_{i=0}^n(i+1)x_i}$.
	Thus, we can show Lemma~\ref{lem:app_I_opt_solution}.
\end{IEEEproof} 
Now, we prove Lemma~\ref{lem:CDF_large_t_asymp_completion_probability} \TCOMb{based on Lemma \ref{lem:app_I_opt_solution}}.
%First, recall that $\Omega(N)\triangleq \Big\{\hat{\mathbf{k}}\in\mathbb{N}^N : \sum_{n=0}^{N-1}\hat{k}_n=N \Big\}$,
%$\Upsilon_1(N) \triangleq \Big\{\hat{\mathbf{k}}:\hat{k}_n=0,n\notin\mathcal{N}_2(\mathbf{x})\cup\{N-1\} \Big\}$,
%$\mathcal{N}_1(\mathbf{x}) \triangleq \left\{n\in[N-1] : x_n\neq 0\right\}$,
%$\mathcal{N}_2(\mathbf{x}) \triangleq \underset{n\in\mathcal{N}_1(\mathbf{x})}{\argmin} \frac{n+1}{\sum_{i=0}^{n} (i+1)x_i}$,
%and $\mathcal{N}_1(\mathbf{x})-1 \triangleq \Big\{ n\in\{0\}\cup[N-2]:x_{n+1}\neq 0 \Big\}$.
We have
$1-\hat{P}(\mathbf{x},t) 
	\eqg \underset{\hat{\mathbf{k}}\in\Omega(N) \backslash \hat{\mathcal{K}}(N)}{\sum} \hspace{-0.0cm} \binom{N} {\hat{k}_0,\hat{k}_1,\cdots,\hat{k}_{N-1}}
\overset{N-2}{\underset{n=0} {\prod}}\Bigg( F_T\bigg( \frac{t} {\frac{M}{N}b\sum_{i=0}^n (i+1)x_i}  \bigg) - F_T\left( \frac{t} {\frac{M}{N}b\sum_{i=0}^{n+1} (i+1)x_i} \right) \Bigg)^{\hat{k}_n} \hspace{-0.3cm}
F_T\left( \frac{t} {\frac{M}{N}b\sum_{i=0}^{N-1} (i+1)x_i} \right)^{\hat{k}_{N-1}}
		\hspace{-0.1cm} 
	\overset{(h)}{\sim} \hspace{-0.5cm} \underset{\hat{\mathbf{k}}\in \Upsilon_2(N)}{\sum}$ $\hspace{-0.0cm} \binom{N} {\hat{k}_0,\hat{k}_1,\cdots,\hat{k}_{N-1}} 
\overset{N-2}{\underset{n=0}{\prod}} 
		\Bigg(I(x_{n+1} \hspace{-0.1cm} > \hspace{-0.1cm} 0) e^{-\mu\Big(\frac{t}{\frac{M}{N}b\sum_{i=0}^{n+1}(i+1)x_i}-t_0\Big)} \Bigg)^{\hat{k}_n} \hspace{-0.4cm}
	=\hspace{-0.4cm} \underset{\hat{\mathbf{k}}\in\Upsilon_2(N)}{\sum} \binom{N} {\hat{k}_0,\hat{k}_1,\cdots,\hat{k}_{N-1}}e^{\mu t_0 \underset{n\in\mathcal{N}_1(\mathbf{x})-1}{\sum}\hat{k}_n} 
		e^{-\frac{\mu N t}{Mb} \underset{n\in\mathcal{N}_1(\mathbf{x})-1}{\sum} \frac{\hat{k}_n}{\sum_{i=0}^{n+1}(i+1)x_i} }$ 
$\eqi E_{\rm lg} (\mathbf{x},t)$,
%\begin{small}
%\begin{align*}
%	& 1-\hat{P}(\mathbf{x},t) 
%	= \underset{\hat{\mathbf{k}}\in\Omega(N) \backslash \hat{\mathcal{K}}(N)}{\sum}\binom{N} {\hat{k}_0,\hat{k}_1,\cdots,\hat{k}_{N-1}}
%	\overset{N-2}{\underset{n=0}{\prod}}\Bigg( F_T\bigg( \frac{t} {\frac{M}{N}b\sum_{i=0}^n (i+1)x_i} \bigg)
%	- F_T\left( \frac{t} {\frac{M}{N}b\sum_{i=0}^{n+1} (i+1)x_i} \right) \Bigg)^{\hat{k}_n}\\
%	& F_T\left( \frac{t} {\frac{M}{N}b\sum_{i=0}^{N-1} (i+1)x_i} \right)^{\hat{k}_{N-1}}
%	\overset{(f)}{\sim} \underset{\hat{\mathbf{k}}\in \Omega(N)\backslash\hat{\mathcal{K}}(N)}{\sum} \binom{N} {\hat{k}_0,\hat{k}_1,\cdots,\hat{k}_{N-1}} \overset{N-2}{\underset{n=0}{\prod}} 
%	\Bigg(I(x_{n+1}>0)e^{-\mu\Big(\frac{t}{\frac{M}{N}b\sum_{i=0}^{n+1}(i+1)x_i}-t_0\Big)} \Bigg)^{\hat{k}_n}\\
%	&= \underset{\hat{\mathbf{k}}\in\Upsilon(N)}{\sum} \binom{N} {\hat{k}_0,\hat{k}_1,\cdots,\hat{k}_{N-1}}e^{\mu t_0 \underset{n\in\mathcal{N}_1(\mathbf{x})-1}{\sum}\hat{k}_n} 
%	e^{-\frac{\mu N t}{Mb} \underset{n\in\mathcal{N}_1(\mathbf{x})-1}{\sum} \frac{\hat{k}_n}{\sum_{i=0}^{n+1}(i+1)x_i} }
%	\eqg 1-\hat{P}_{\rm lg} (\mathbf{x},t),
%\end{align*}
%\end{small}
where \hspace{0.05cm}
$(g)$ \hspace{0.05cm} is \hspace{0.05cm} due \hspace{0.05cm} to \hspace{0.05cm} $\hat{\mathcal{K}}(N)\subseteq \Omega(N)$,
$(h)$ \hspace{0.05cm} is \hspace{0.05cm} due \hspace{0.05cm} to $\underset{t\rightarrow\infty}{\lim} F_T(\frac{t}{\frac{M}{N}b\sum_{i=0}^n (i+1)x_i}) = 1$, 
$\underset{t\rightarrow\infty}{\lim} e^{-\mu t \Big(\frac{1}{\frac{M}{N}b\sum_{i=0}^n(i+1)x_i} - \frac{1}{\frac{M}{N}b\sum_{i=0}^{n+1}(i+1)x_i}\Big)} = 0$, and $0^{\hat{k}_n}=I(\hat{k}_n=0)$, and 
$(i)$ is due to Lemma~\ref{lem:app_I_opt_solution} and \eqref{equ:CDF_large_completion_probability}.
Therefore, we complete the proof of Lemma~\ref{lem:CDF_large_t_asymp_completion_probability}.

%\setlength{\intextsep}{-10pt}
%\setlength{\textfloatsep}{10pt}
%\renewcommand{\thealgorithm}{3}
%\begin{algorithm}[h]
%	\caption{\hspace{-0.12cm} Algorithm \hspace{-0.12cm} for \hspace{-0.12cm} Obtaining \hspace{-0.12cm} Encoding \hspace{-0.12cm} Matrix \hspace{-0.12cm} $\mathbf{B}_l$ \hspace{-0.2cm} \cite{GC_Tandon_exact}}
%\begin{small}
%\REPLYb{
%	\begin{algorithmic}[1]
%		\STATE \textbf{input}: Number of workers $N$ and number of full stragglers $s_l$.
%		\STATE \textbf{output}: Encoding matrix $\mathbf{B}_l\in\mathbb{R}^{N\times N}$.
%		\STATE $\mathbf{H}=\text{randn}(s_l,N)$;
%		\STATE $\mathbf{H}(:,N)=-\text{sum}(H(:,1:N-1),2)$;
%		\STATE $\mathbf{B}_l=\text{zeros}(N)$;
%		\STATE \textbf{for} $i=1:N$ \textbf{do}
%		\STATE \quad $j=\text{mod}(i-1:s_l+i-1,N)+1$;
%		\STATE \quad $\mathbf{B}_l(i,j)=[1;-\mathbf{H}(:,j(2:s_l+1))\backslash \mathbf{H}(:,j(1))]$;
%		\STATE \textbf{end for}
%	\end{algorithmic}\label{alg:encoding_mat}}
%\end{small} 
%\end{algorithm}

\section*{Appendix \REPLYb{M}: Proof of Theorem~\ref{thm:CDF_large_t_approx_closedform}}

First, it is obvious that Problem \ref{prob:CDF_large_t_approx} is equivalent to the following problem.
%\begin{Prob}[Approximate Problem of Problem \ref{prob:CDF_large_t_approx}]\label{prob:CDF_large_t_approx_equivalent}
	\begin{align}
			\min_{\mathbf{x}} \ \  &\underset{n=0,1,\cdots,N-1}{\max} \frac{\sum_{i=0}^n (i+1)x_i} {n+1} \nonumber\\
		  	{\rm s.t.}\ \  &\eqref{equ:constraint_x_1}, \eqref{equ:constraint_x_3}. \nonumber
	\end{align}
%\end{Prob}
%Then, by using methods similar to those in the proof of Theorem~\ref{thm:E_approx_solution_time}, 
\TCOMb{Then, by Theorem~\ref{thm:E_approx_solution_time} and by letting $t_n=\frac{1}{N-n+1},n\in[N]$,
we obtain an optimal solution of the above problem given by \eqref{equ:CDF_large_t_approx_closedform}.}
Therefore, we complete the proof of Theorem \ref{thm:CDF_large_t_approx_closedform}.

\section*{\REPLYb{Appendix N: Encoding and decoding matrix for computing $\frac{\partial\hat{\ell}(\bm{\theta};\mathcal{D}_i)}{\partial\theta_l}$ where $l\in[L]$}}
%For any $l\in[L]$, the encoding matrix $\mathbf{B}_l$ and decoding matrix $\mathbf{A}_l$ are generated according to \cite{GC_Tandon_exact} with $s=s_l$, which are given by Alg.~\ref{alg:encoding_mat} and Alg.~\ref{alg:decoding_mat}, respectively.

\renewcommand{\thealgorithm}{3}
\begin{algorithm}[h]
	\caption{\hspace{-0.12cm} Algorithm \hspace{-0.12cm} for \hspace{-0.12cm} Obtaining \hspace{-0.12cm} Encoding \hspace{-0.12cm} Matrix \hspace{-0.12cm} $\mathbf{B}_l$ \hspace{-0.2cm} \cite{GC_Tandon_exact}}
\begin{small}
\REPLYb{
	\begin{algorithmic}[1]
		\STATE \textbf{input}: Number of workers $N$ and number of full stragglers $s_l$.
		\STATE \textbf{output}: Encoding matrix $\mathbf{B}_l\in\mathbb{R}^{N\times N}$.
		\STATE $\mathbf{H}=\text{randn}(s_l,N)$;
		\STATE $\mathbf{H}(:,N)=-\text{sum}(H(:,1:N-1),2)$;
		\STATE $\mathbf{B}_l=\text{zeros}(N)$;
		\STATE \textbf{for} $i=1:N$ \textbf{do}
		\STATE \quad $j=\text{mod}(i-1:s_l+i-1,N)+1$;
		\STATE \quad $\mathbf{B}_l(i,j)=[1;-\mathbf{H}(:,j(2:s_l+1))\backslash \mathbf{H}(:,j(1))]$;
		\STATE \textbf{end for}
	\end{algorithmic}\label{alg:encoding_mat}}
\end{small} 
\end{algorithm}

\setlength{\intextsep}{-2pt}
\setlength{\textfloatsep}{2pt}
\renewcommand{\thealgorithm}{4}
\begin{algorithm}[h]
	\caption{\hspace{-0.12cm} Algorithm \hspace{-0.12cm} for \hspace{-0.12cm} Obtaining \hspace{-0.12cm} Decoding \hspace{-0.12cm} Matrix \hspace{-0.12cm} $\mathbf{A}_l$ \hspace{-0.2cm} \cite{GC_Tandon_exact}}
\begin{small}
\REPLYb{
	\begin{algorithmic}[1]
		\STATE \textbf{input}: Number of workers $N$, number of full stragglers $s_l$, and encoding matrix $\mathbf{B}_l$.
%		\cite{GC_Tandon_exact}.
		\STATE \textbf{output}: Decoding matrix $\mathbf{A}_l\in\mathbb{R}^{\binom{N}{s_l} \times N}$.
%		 such that $\mathbf{A}_l \mathbf{B}_l=\mathbf{1}_{\binom{N}{s} \times N}$.
		\STATE $f=\text{nchoosek}(N,s_l)$;
		\STATE $\mathbf{A}_l=\text{zeros}(f,N)$;
		\STATE \textbf{for} $\mathcal{I}\subseteq [N], |\mathcal{I}|=N-s_l$ \textbf{do}
		\STATE \quad $\mathbf{x}=\text{zeros}(1,N)$;
		\STATE \quad $\mathbf{y}=\text{ones}(1,k)/\mathbf{B}_l(\mathcal{I},:)$;
		\STATE \quad $\mathbf{x}(\mathcal{I})=\mathbf{y}$;
		\STATE \quad $\mathbf{A}_l=[\mathbf{A}_l;\mathbf{x}]$;
		\STATE \textbf{end for}
	\end{algorithmic}\label{alg:decoding_mat}}
\end{small} 
\end{algorithm}

%\bibliographystyle{IEEEtran}
%%\bibliography{refs_wq_J,refs_wq_isit.bib}
%\bibliography{refs_wq.bib, refs_supplementary.bib, refs_book.bib, refs_reply1.bib, refs_reply2.bib}

% Generated by IEEEtran.bst, version: 1.14 (2015/08/26)

\end{document}